\documentclass[letterpaper,twocolumn,10pt]{article}
\usepackage{usenix}
\usepackage{amsmath}
\usepackage{textcomp,hyperref}
\usepackage{pifont}
\usepackage{xspace}
\usepackage{bbding}
\usepackage{multirow}
\usepackage{threeparttable}
\usepackage{amssymb}
\usepackage{xspace}
\usepackage{booktabs}
\usepackage{subcaption}
\usepackage{threeparttable}
\usepackage{graphicx}
\usepackage{amsthm}
\theoremstyle{plain}
\newtheorem{theorem}{Theorem}
\newtheorem{example}{Example}
\newtheorem{definition}{Definition}
\usepackage{colortbl}
\usepackage{color, xcolor}
\usepackage{booktabs}
\usepackage{balance}
\usepackage{enumerate}
\usepackage{bm}
\usepackage{url}

\newcommand{\heading}[1]{{\vspace{1pt}\noindent{\textbf{#1}}}}

\newcommand{\figref}[1]{Figure~\ref{#1}\xspace}
\newcommand{\secref}[1]{Sec.~\ref{#1}\xspace}
\newcommand{\tabref}[1]{\mbox{Table~\ref{#1}}}
\newcommand{\appref}[1]{Appendix~\ref{#1}\xspace}
\newcommand{\thmref}[1]{Theorem~\ref{#1}\xspace}

\usepackage{tikz}
\usepackage{amsmath}
\usepackage{filecontents}

\begin{document}

\date{}

\title{\Large \bf Timestamp Manipulation: Incentive Attacks on Timestamp-Based Proof-of-Work Blockchains with Minimal Additional Risk}

\author{
{\rm Junjie Hu}\\
Shanghai Jiao Tong University\\
nakamoto@sjtu.edu.cn
\and
{\rm Na Ruan}\\
Shanghai Jiao Tong University\\
naruan@sjtu.edu.cn
\and
{\rm Sisi Duan}\\
Tsinghua University\\
duansisi@tsinghua.edu.cn
} 

\maketitle

\begin{abstract}
Nakamoto consensus is the most widely adopted decentralized consensus mechanism in cryptocurrency systems. Since its proposal in 2008, many studies have investigated incentive attacks, where the adversary aims to earn more rewards than its fair share but the security properties are not violated. 
One notable example is the recent Riskless Uncle Maker (RUM) attack on timestamp-based Nakamoto blockchains (CCS'23) such as Ethereum 1.x (deprecated in Sep 2022) and Ethereum Classic (still active) and EthereumPoW (still active). Specifically, the RUM attack shows that an adversary with 25\% relative hash power can achieve a normalized profit share of 26.12\%.

In this work, we propose two new incentive attacks on timestamp-based Nakamoto consensus: Unrestricted Uncle Maker (UUM) and Staircase-Unrestricted Uncle Maker (SUUM). Compared with RUM, both attacks achieve higher profit without risking reward loss. We demonstrate through theoretical analysis and experimental validation that if an adversary controls 25\% relative hash power, it can achieve a normalized profit share of 28.41\% and 33.30\% in UUM and SUUM, respectively. We further conduct an empirical analysis on three major ETH 1.x-based blockchains, namely Ethereum 1.x, Ethereum Classic, and EthereumPoW. Our on-chain observations reveal persistent SUUM-style behavior in four large mining pools. Compared to honest mining, the estimated additional revenues gained by these mining pools are nearly three million USD (\$2,961,078 to be specific). 
\end{abstract}

\section{Introduction}\label{Introduction}
\heading{Timestamp-based Nakamoto-style Blockchains.}
Timestamp-based Nakamoto-style blockchains denote ones that use Proof-of-Work (PoW)-based consensus~\cite{BitcoinAPeer-to-PeerElectronicCashSystem, LayDowntheCommonMetricsEvaluatingProofofWorkConsensusProtocols'Security, Inclusiveblockchainprotocols, Fruitchain, HybridConsensus, Bitcoin-NG, Monoxide} and incorporate physical timestamps into the \textit{difficulty adjustment} rule, an important criterion for block proposer selection. As of writing, Nakamoto-style blockchains have a market cap of over \$1.9 trillion~\cite{coinmarketcap.com}. The goal of earlier designs such as  Ethereum 1.x~\cite{EthereumWhitepaper, Anext-generationsmartcontractanddecentralizedapplicationplatform} was to improve the performance of conventional PoW-based blockchains such as Bitcoin. Namely, the block time was reduced from 10 minutes in Bitcoin to 13 seconds. Although Ethereum 1.x was deprecated in Sep 2022 due to the transition from PoW to Proof-of-Stake (PoS), several derivatives of Ethereum 1.x are still active today, including EthereumPoW \cite{EthereumPoW} with a total market capitalization of \$33.24 million, Ethereum Fair \cite{EthereumFair} with \$1.03 million, and Ethereum Classic \cite{EthereumClassic} with over \$1.382 billion market cap\footnote{Date source (accessed in Mar 2026): \url{https://coinmarketcap.com/}.}.



\heading{Existing Incentive Attacks.} 
Incentive attacks denote attacks where the adversary seeks to gain higher profit rather than breaking the security properties of the system. A notable example in conventional Nakamoto consensus is the selfish mining attack~\cite{Majorityisnotenough}. In selfish mining, the adversary first withholds newly mined blocks. After gaining a leading advantage, the blocks are released, after which the withheld chain becomes the longest chain, and blocks by honest miners are discarded. Since the adversary gains extra profit, honest miners are discouraged and the computing power of honest miners is wasted. As summarized in \tabref{tab:comp}, if the adversary controls $\alpha=25\%$ of the computational power, it can gain 25.49\% expected profit, higher than $25\%$ in honest mining, where each miner honestly follows the protocol. 

 
Timestamp-based Nakamoto-style blockchains relax the use of timestamps in difficulty adjustment to gain higher performance. The cost is that soft forks~\cite{softFork} become more frequent, where multiple conflicting chains are formed. Although honest miners can eventually merge to the same chain, such soft forks create more opportunities for incentive attacks, where miners can manipulate block difficulty to gain an unfair advantage~\cite{ShortPaperRevisitingDifficultyControlforBlockchainSystems, Highfrequencyvolatilitycomovementsincryptocurrencymarkets, CorrectCryptocurrencyASICPricingAreMinersOverpaying, MindtheMining, EnergyEquilibriainProofofWorkMining, UnstableThroughputWhentheDifficultyAlgorithmBreaks, BlockchainStretchingSqueezingManipulatingTimeforYourBestInterest}.

Note that selfish mining is not risk-free, i.e., there is a certain probability that the withheld chain does not become the longest chain in the end. The recently proposed Riskless Uncle Maker (RUM) attack~\cite{Unclemakertimestampingoutthecompetitioninethereum} to timestamp-based Nakamoto-style blockchains presented the first attack where the withheld chain will become the longest chain. 
This is achieved by introducing a new attack vector called timestamp manipulation. In RUM, the adversary strategically sets block timestamps under specific time conditions, artificially elevating the block difficulty to exceed that of honest blocks. In this way, the adversary obtains advantages in the forking race (i.e., competition to become the longest chain among soft forks). In RUM for $\alpha=25\%$, the adversary can gain a 26.12\% profit.

\begin{table}[t]
\caption{Comparison of different mining strategies.}
\label{tab:comp}
\centering
\resizebox{\columnwidth}{!}{%
\begin{tabular}{@{}|c|c|c|c|c|@{}}
\toprule
\shortstack[c]{Attack\\Strategy} &
\shortstack[c]{Normalized\\Reward Share} &
\shortstack[c]{Attack-state\\Probability} &
\shortstack[c]{Fork\\Rate} &
\shortstack[c]{On-chain\\Evidence} \\ \midrule
Honest Mining \cite{BitcoinAPeer-to-PeerElectronicCashSystem} & 25.00\% & 0 & $\approx 0$ & \Checkmark \\ \midrule
Selfish Mining \cite{Majorityisnotenough} & 25.49\% & 0.25 & 0.36 & \ding{55} \\ \midrule
RUM \cite{Unclemakertimestampingoutthecompetitioninethereum} & 26.12\% & 0.17 & 0.02 & \Checkmark \\ \midrule
UUM [\secref{The design of uum attack}] & \colorbox{gray!40}{28.41\%} & \colorbox{gray!40}{0.47} & \colorbox{gray!40}{0.11} & \Checkmark \\ \midrule
SUUM [\secref{The design of suum attack}] & \colorbox{gray!60}{33.30\%} & \colorbox{gray!60}{0.61} & \colorbox{gray!60}{0.13} & \Checkmark \\ \bottomrule
\end{tabular}%
}
\begin{tablenotes}
\footnotesize
\setlength{\itemindent}{0pt}
\setlength{\rightskip}{10pt}
\item* All results assume the adversary controls $25\%$ of the mining power. The reward share with other $\alpha$ can be obtained from \figref{Exp}~(a).
\end{tablenotes}
\end{table}

\begin{figure}[t]
  \centering
  \includegraphics[width=1\columnwidth]{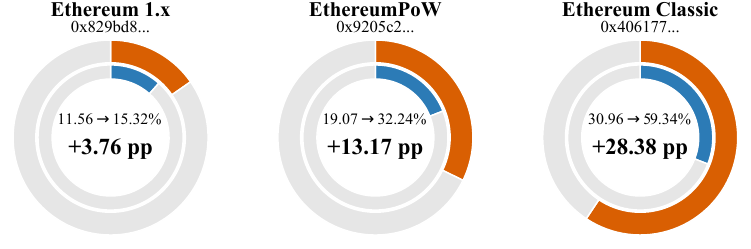}
  \caption{External hash share (inner ring) and realized main-chain share (outer ring) for three flagged pools; the centers report the corresponding surplus.}
  \label{fig:hash-share-gap}
\end{figure}

\heading{Our Approach.} Our empirical analysis reveals a complementary observation: the realized main-chain block shares of the flagged pools substantially exceed their external hash-share benchmarks. Across the three pools with available external benchmarks, the surplus ranges from $3.76$ to $28.38$ percentage points (\figref{fig:hash-share-gap} and Table~\ref{tab:unified_pool_statistics}). Under honest mining, and assuming that the external benchmarks accurately track the pools' physical mining capacity, a pool's long-run main-chain block share should remain close to its relative hash power. A persistent surplus of this magnitude is therefore difficult to reconcile with honest mining alone: it requires a mechanism that makes the pool's blocks disproportionately survive on the main chain while displacing competing honest blocks. Strategic private mining and selective release provide exactly such a mechanism---the adversary creates a competing branch and releases it when its timestamp-calibrated difficulty can replace honest blocks.

This observation motivates us to study how timestamp manipulation can turn selective block replacement into a reliable and profitable strategy. Our crucial technical finding is that manipulating the timestamp directly affects the probability that subsequent honest miners generate a new block per unit time, because timestamp manipulation can also raise the subsequent mining difficulty. If the difficulty rises excessively, the attack cannot be launched frequently enough and is thus less profitable. To capture this tradeoff, we refine the definition of \textit{risk} into \textit{pre-mining risk} and \textit{post-mining risk}. Pre-mining risk denotes the reduction in block-generation rate caused by the difficulty selected before an attack block is mined, whereas post-mining risk denotes the subsequent difficulty increase caused by the attack. Eliminating pre-mining risk ensures that the adversarial branch can win the forking race; reducing post-mining risk prevents the attack from suppressing future block production and eroding its long-term profit. From this perspective, RUM primarily controls pre-mining risk but does not fully account for post-mining risk.


Guided by these two insights, we consider both pre-mining and post-mining risk and introduce \textit{minimal risk control}, which seeks to minimize the cost of creating a competing branch while limiting its effect on subsequent difficulty. We then present two new incentive attacks: Unrestricted Uncle Maker (UUM) and Staircase-Unrestricted Uncle Maker (SUUM). UUM relaxes RUM's triggering condition and thereby creates more opportunities for single-block replacement. SUUM further generalizes this strategy to multi-block withholding and staircase release, allowing a private branch to displace multiple honest blocks while controlling the timestamp of each released block. We show that SUUM incurs only negligible additional pre-mining risk while keeping post-mining risk as low as possible. As summarized in \tabref{tab:comp}, SUUM can be launched with 61\% probability, significantly higher than RUM's 17\%, and both proposed attacks achieve substantially higher normalized reward shares than existing approaches.

\begin{center}
\fbox{
\parbox{0.95\linewidth}{
\smallskip
$\noindent\textbf{Responsible disclosure}$. For ethical reasons, we responsibly disclosed our findings to the Ethereum Foundation and relevant representatives of the Ethereum Classic and EthereumPoW communities in December 2025.
}}
\end{center}

\heading{Experimentation and Real-Data Analysis.} 
We validate our results via both simulation and on-chain data analysis. For our on-chain data analysis, we use three mainstream ETH 1.x-style blockchains: Ethereum 1.x (block heights 15424591--15535776), Ethereum Classic (23353196--23453195), and EthereumPoW (22989157--23089156). As mentioned earlier, Ethereum 1.x has been deprecated. We thus collected 111,186 sampled main-chain blocks from the pre-merge Ethereum 1.x period. For Ethereum Classic and EthereumPoW, we collected the data for 100,000 main-chain blocks in October 2025. In total, we have collected 311,186 main-chain blocks for our analysis, and we summarize our results below. 

\begin{enumerate}
    \item[$\bullet$] We observe persistent timestamp manipulation patterns similar to our SUUM attack in four large mining pools, including Ethereum 1.x pool \texttt{0x829bd8...}, EthereumPoW pool \texttt{0x9205c2...}, and Ethereum Classic pools \texttt{0x406177...} and \texttt{0x35aa26...}. 
    \item[$\bullet$] Among these four flagged pools, the evidence that the mining pool is taking SUUM attack strategies is the strongest for the Ethereum Classic pool \texttt{0x406177...} (publicly labeled as F2Pool)~\cite{Thepool0x406177...4e95ispubliclylabeledasF2Pool}. Specifically, in our dataset of 100,000 main-chain blocks, the pool withheld and released a maximum of 21 consecutive blocks, by controlling $\alpha=30.96\%$ of the computational power \cite{ethereumclassichashrate}. Our case study on this mining pool shows that the profit of the mining pool is highly close to our theoretical analysis. 
\end{enumerate}

\heading{Our Contributions.} We make the following contributions. 
\begin{enumerate}

\item[$\bullet$] We propose UUM and SUUM attacks on the incentive mechanism of timestamp-based Nakamoto-style blockchains. Under the guidance of our refined notion of post-mining risks and a minimum post-mining risk control approach, both of our attacks have higher profit than known attacks. 

\item[$\bullet$] Our experimental results show that an adversary controlling 25\% relative hash power achieves 28.41\% and 33.30\% expected profit, respectively. Both are higher than 26.12\% for RUM. 

\item[$\bullet$] We analyze 311,186 main-chain blocks from three mainstream Ethereum 1.x-style blockchains and find strong evidence that four mining pools exhibit behavior consistent with our SUUM attack. Among them, the Ethereum Classic pool \texttt{0x406177...} provides the strongest evidence in our datasets.
\end{enumerate}

\section{Threat Model}\label{Threat model}
\heading{Participant Composition.}
We assume a decentralized protocol, denoted as $\Gamma$, which is collectively executed by a set of participants, $\mathcal{P}$, across consecutive time epochs. The set of participants, $\mathcal{P}$, can be subdivided into two subsets: the set of honest participants, $\mathcal{P}_{\mathcal{H}} = \left\{ p_{h}^{1},~p_{h}^{2},\ldots,p_{h}^{n} \right\}$, who strictly adhere to the protocol $\Gamma$, and the set of adversaries, $\mathcal{P}_{\mathcal{A}}$. Thus, $\mathcal{P} = \mathcal{P}_{\mathcal{H}} \cup \mathcal{P}_{\mathcal{A}}$.

It is noteworthy that the scope of honest mining participants, $\mathcal{P}_{\mathcal{H}}$, is not limited to a single entity. Its composition can be extended to mining pools and even consortia formed by multiple mining pools. Correspondingly, the adversary set, $\mathcal{P}_{\mathcal{A}}$, may also exist in the form of mining pools or mining pool alliances, exhibiting considerable complexity and diversity. This paper follows the common assumptions in the blockchain literature \cite{Majorityisnotenough, EthereumWhitepaper, Fruitchain} that during the attack period, the hash power of each participant remains constant, all mining hardware is preconfigured, and relevant costs (including but not limited to electricity expenses and mining hardware acquisition cost) are prepaid. In practice, historical data indicate that the active hashing rate in the network remains relatively stable over short periods, with minor fluctuations \cite{BitInfoCharts}. Our assumptions are thus reasonable.

In protocol $\Gamma$, each participant $p \in \mathcal{P}$ is associated with a numerical value $\mu_{p}$ that lies between $0$ and $1$, such that $\sum_{{p} \in \mathcal{P}}\mu_{p} = 1$.
Here, $\mu_{p}$ represents the participation power ratio of participant $p$ in protocol $\Gamma$, which can be embodied specifically as hash power, stake power, or other relevant metrics.

Regarding the settings of timestamps and block structures, we assume that the timestamp of the genesis block $\mathcal{B}_{0}$ is $t_{0} = 0$. Starting from the genesis block, the timestamp of block $\mathcal{B}_{i}$, a block with height $i$ on the main chain, is denoted as $t_{i}$. We use $\mathcal{D}_{i}$ to denote the difficulty of block $\mathcal{B}_{i}$. We denote the timestamp of the parent block $\mathcal{B}_{i-1}$ by $t_i^p$, i.e., $t_i^p = t_{i-1}$. Additionally, in Ethereum-style timestamp-based Nakamoto blockchains, an uncle block refers to a valid stale block that is not included in the main-chain but can still be referenced by a later main-chain block for partial reward. The miner of the referenced stale block receives an uncle reward, while the miner of the later main-chain block that references it receives a nephew reward. A variable ${pu}_{i} \in \left\{ 0, 1 \right\}$ indicates whether the parent block of block $\mathcal{B}_{i}$ references any uncle blocks. If it does, then ${pu}_{i} = 1$; otherwise, ${pu}_{i} = 0$.

\heading{Adversary's Target.}
Ideally, participants with an expected hash rate proportion of $\mu_{p}$ should be able to mine a corresponding proportion of $\mu_{p}$ blocks and thus obtain $\mu_{p}$ of the mining rewards~\cite{Majorityisnotenough, Fruitchain}. In this study, the target of the adversary $\mathcal{P}_{\mathcal{A}}$ is to surpass its fair share based on hash rate, specifically by mining more blocks than their proportional share $\mu_{\mathcal{P}_{\mathcal{A}}}$ to gain higher rewards.

\heading{Honest Participant's Strategy Space.}
The definition of honest mining protocol in this paper refers to the relevant academic literature in the field of blockchain \cite{Majorityisnotenough, OntheSecurityandPerformanceofProofofWorkBlockchains} and specifically follows the rules established by the Ethereum 1.x White Paper \cite{EthereumWhitepaper}. Accordingly, honest participants always strive to mine the blocks with the highest total difficulty and strictly follow the protocol requirements. 

\heading{Adversary's Strategy Space.}  To gain extra profit, the adversary may adjust the timestamps of the blocks they mine, but must ensure that these timestamps fall within a valid range to ensure that the blocks are valid, i.e., the timestamp of a block to be mined by adversaries must be strictly greater than that of its parent block on the main-chain. Furthermore, the adversary can choose which blocks to mine.

To clearly distinguish the attack models proposed in this study from the attack strategies presented in previous literature \cite{SoKToolsforGameTheoreticModelsofSecurityforCryptocurrencies, SoKResearchPerspectivesandChallengesforBitcoinandCryptocurrencies, Majorityisnotenough, OntheSecurityandPerformanceofProofofWorkBlockchains, Unclemakertimestampingoutthecompetitioninethereum}, we introduce two specific types of adversaries: UUM and SUUM. The UUM adversary increases the probability of launching the attack, but does not withhold blocks. Once a valid block is found by the UUM adversary, it must be immediately released. The SUUM adversary allows the adversary to withhold their blocks to further improve the attack surface and improve the probability of launching the attack. 

\section{Preliminaries}\label{The design of rum attack}

In this section, we formally define the fork selection rule, fairness, block difficulty adjustment rule, and block reward classification of Nakamoto-style blockchains. Next, we systematically introduce the design principle of the RUM attack.

The honest participants $\mathcal{P}_{\mathcal{H}}$ select the block with higher difficulty in the presence of fork competition in the Nakamoto-style blockchains.

We denote the timestamp and difficulty of the $i$-th block $
\mathcal{B}_{{i}}^{{p}_{{h}}}$ on the honest branch during a fork competition as ${t}_{{i}}^{{p}_{{h}}}$ and $
\mathcal{D}_{{i}}^{{p}_{{h}}}$, respectively. Similarly, the timestamp and difficulty of the $i$-th block $\mathcal{B}_{{i}}^{{p}_{{a}}}$ on the adversary's branch during a fork competition are denoted as $
{t}_{{i}}^{{p}_{{a}}}$ and $\mathcal{D}_{{i}}^{{p}_{{a}}}$.

\begin{example}[\textbf{Fork Selection}]\label{example1}
The blockchain comprises three blocks, denoted as $\mathcal{B}_{0}$, $\mathcal{B}_{1}^{{p}_{{a}}}$, and $\mathcal{B}_{1}^{{p}_{{h}}}$. The timestamp of block $\mathcal{B}_{0}$ is represented by ${t}_{0}^{{p}_{{h}}}$ or ${t}_{0}^{{p}_{{a}}}$, and its difficulty is denoted by $
\mathcal{D}_{0}^{{p}_{{h}}}$ or $
\mathcal{D}_{0}^{{p}_{{a}}}$. For block $\mathcal{B}_{1}^{{p}_{{a}}}$, the timestamp is $
{t}_{1}^{{p}_{{a}}}$ and its difficulty is $
\mathcal{D}_{1}^{{p}_{{a}}}$. Similarly, for block $\mathcal{B}_{1}^{{p}_{{h}}}$, the timestamp is $
{t}_{1}^{{p}_{{h}}}$ and its difficulty is $
\mathcal{D}_{1}^{{p}_{{h}}}$. According to the Ethereum 1.x protocol, each honest miner selects the block with a higher difficulty. Specifically:

\begin{enumerate}[(1)]
\item Case 1: Select block $\mathcal{B}_{1}^{{p}_{{h}}}$ if $
\mathcal{D}_{1}^{{p}_{{h}}} = \max \left\{ \mathcal{D}_{1}^{{p}_{{h}}},\mathcal{D}_{1}^{{p}_{{a}}} \right\}$.

\item Case 2: Select block $\mathcal{B}_{1}^{{p}_{{a}}}$ if $
\mathcal{D}_{1}^{{p}_{{a}}} = \max \left\{ \mathcal{D}_{1}^{{p}_{{h}}},\mathcal{D}_{1}^{{p}_{{a}}} \right\}$.

\item Case 3: Select either block arbitrarily if $
\mathcal{D}_{1}^{{p}_{{h}}} = \mathcal{D}_{1}^{{p}_{{a}}}$.
\end{enumerate}
\end{example}

\begin{definition}[\textbf{Fairness}]\label{def:fairness}
    Let the time be partitioned into discrete intervals $t = 1, 2, \dots, T$, where each interval $t$ generates $L_t$ blocks. A blockchain system satisfies fairness if, for any participant $p_i \in \mathcal{P}$, the following condition holds: 
    \begin{align}
        \lim_{T \to \infty} \frac{1}{T} \sum_{t=1}^T \frac{R_i^{(t)}}{L_t \cdot R} =
\lim_{T \to \infty} \frac{1}{T} \sum_{t=1}^T \mu_i^{(t)},
    \end{align}
    where $\mu_i^{(t)}$ is the power proportion of participant $p_i$ during interval $t$, $R_i^{(t)}$ is the reward earned by participant $p_i$ during interval $t$. $R$ is the total reward per block, i.e., coinbase rewards, transaction fees, etc.
\end{definition}

Definition~\ref{def:fairness} formally defines the fairness of blockchains. Over the long term, in accordance with the law of large numbers, the reward distribution among participants strictly converges to their resource input (power/stake). 

\begin{definition}[\textbf{Block Difficulty}]\label{def:difficulty}
    The difficulty $\mathcal{D}_{i}$ of block $\mathcal{B}_{i}$ satisfies the following equation:
\begin{align}
    \mathcal{D}_{i} \overset{def}{=} {\max\left\{ {2^{17},\mathcal{D}_{i}^{p} + f \cdot \lfloor \frac{\mathcal{D}_{i}^{p}}{2048} \rfloor} \right\}},
\end{align}
where $f={\max\left\{ {1 + {pu}_{i} - \lfloor \frac{t_{i} - t_{i - 1}}{9} \rfloor, - 99} \right\}}$. 
\end{definition} 

In the definition, $2^{17}$ denotes the minimal difficulty to ensure the difficulty never falls below this value. The value $2048$ is a system parameter in Ethereum 1.x's difficulty adjustment mechanism. $-99$ is the lower bound of the difficulty adjustment coefficient $f$. The parameter aims to prevent excessive difficulty reduction when block generation times are excessively long. 
All Ethereum 1.x-based blockchains, such as ETH 1.0, ETH Fair, ETH Classic, and ETH PoW, adopt these parameters. In contrast, non-Ethereum 1.x-based blockchains, such as Bitcoin and Bitcoin Cash, use different parameters. It is worth noting that our proposed attacks apply to all of the timestamp-based, per-block difficulty adjustment Nakamoto-style blockchains, as we later show in \secref{Generality}. 

It is worth mentioning that in Ethereum 1.x, the difficulty of blocks after height $15$ exceeds $2^{17}$. Consequently, in subsequent analysis concerning block difficulty, the difficulty $
\mathcal{D}_{i}$ of block $\mathcal{B}_{i}$ is determined by $\mathcal{D}_{i} \overset{def}{=} \mathcal{D}_{i}^{p} + {\max\left\{ {1 + {pu}_{i} - \lfloor (t_{i} - t_{i - 1})/{9} \rfloor, - 99} \right\}} \cdot \lfloor  \mathcal{D}_{i}^{p} / 2048 \rfloor$. When the timestamp difference $\Delta t = t_i - t_{i-1}$ exceeds 900 seconds, the adjustment term $\max\{1 - \lfloor\Delta t / 9\rfloor, -99\}$ hits its lower bound of $-99$. This causes the block difficulty to drop by a large, fixed amount $(-99 \cdot \lfloor \mathcal{D}_{i}^{p} / 2048\rfloor)$, regardless of how much larger $\Delta t$ becomes.

\begin{definition}[\textbf{Pre-mining Risk}]\label{def:pre_mining_risk}
Let $\mathcal P_{\mathcal{A}}(\mathcal{D})$ denote the adversary's block generation rate when mining on a branch of difficulty $\mathcal{D}$. The pre-mining risk induced by an attack in a given round is defined as the reduction in the adversary's block generation rate caused by mining on the selected branch during the attack, relative to the honest mining baseline before the round's reward is realized. Formally, $\mathbb{AC} \overset{def}{=} \mathcal P_{\mathcal{A}}(\mathcal{D}_{0}) - \mathcal P_{\mathcal{A}}(\mathcal{D}_{1})$,
where $\mathcal{D}_{0}$ is the counterfactual difficulty the adversary would face under honest mining from the same parent state, and $\mathcal{D}_{1}$ is the difficulty of the branch on which the adversary actually mines under the attack strategy. An attack is \textit{pre-mining risk-free} if and only if $\mathbb{AC}=0$.
\end{definition}

\begin{definition}[\textbf{Post-mining Risk}]\label{def:post_mining_risk}
The \textit{post-mining risk} $R$ induced by an attack in a given round is defined as the increase in the difficulty of the subsequent main-chain block, relative to the difficulty during honest mining. Formally, $R \overset{def}{=} \mathcal{D}_{Attack}-\mathcal{D}_{Honest}$, where $\mathcal{D}_{Attack}$ is the difficulty of the first block mined on the adversary's winning chain after the attack round, and $\mathcal{D}_{Honest}$ is the counterfactual difficulty had the honest block won that round.
\end{definition}

The unit of $R$ is the base unit of Ethereum difficulty adjustment, i.e., an integer multiple of $\lfloor \mathcal D_{\mathrm{parent}} / 2048 \rfloor$. \textit{Minimal post-mining risk} is achieved when the adversary's timestamp calibration minimizes $R$ while still ensuring victory in the forking race (Theorems~\ref{theorem3}, \ref{theorem7}).

\begin{figure}[t]
  \centering
  \includegraphics[width=\linewidth]{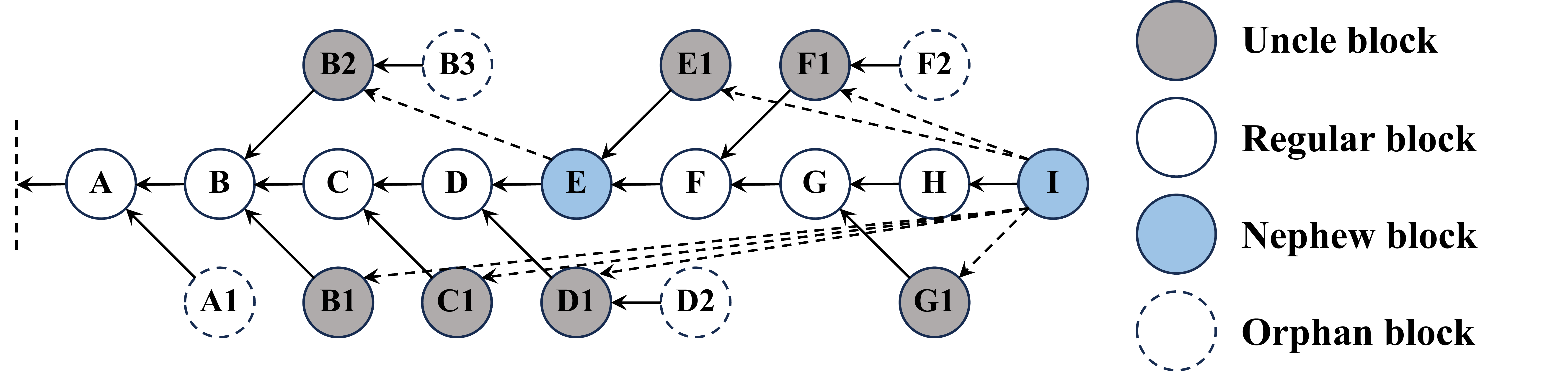}
  \caption{Different block types. Blocks A to I contain coinbase rewards. Blocks B1 to G1 contain uncle rewards. Blocks E and I contain nephew rewards (Definition \ref{Block Reward}).}
  \label{fig:1}
\end{figure}

Each participant $p$ who generates a block has the potential to receive a reward, and different block types correspond to different types of rewards. We formally define block rewards in \appref{Different types of rewards} (an example is provided in \figref{fig:1}).

Unless otherwise stated, all theoretical reward comparisons and the primary experimental reward plots use \emph{coinbase-only realized main-chain rewards}. Under this accounting, we distinguish two reward notions.

First, let $\widehat{\mathbb{R}}_{\mathcal{P}}$ denote participant $\mathcal{P}$'s 
normalized realized main-chain coinbase reward share, i.e., the fraction of realized coinbase rewards obtained by $\mathcal{P}$ under the stationary distribution induced by the corresponding mining strategy. Second, let $\mathbb{R}_{\mathcal{P}}$ denote the corresponding absolute realized main-chain coinbase reward rate per unit physical time.

Moreover, timestamp-based Ethereum 1.x-style protocols adjust difficulty toward the target block interval $T$. Hence, the realized main-chain block generation rate converges to $1/T$ in steady state. Let $R_c$ denote the coinbase reward of each realized main-chain block. Then
\(
\mathbb{R}_{\mathcal{P}}
=
\frac{R_c}{T}\widehat{\mathbb{R}}_{\mathcal{P}}.
\)

Therefore, a strict increase in normalized realized main-chain coinbase reward share immediately implies a strict increase in absolute realized main-chain coinbase reward rate per unit physical time. Uncle/nephew rewards further amplify the attacker's realized gains.

\section{Overview of Our Attacks}

\subsection{Review of RUM}\label{Overview of RUM}
The main workflow of RUM is similar to selfish mining and consists of three phases. 
\begin{enumerate}
    \item[$\bullet$] Upon an \textit{initiation condition}, the adversary withholds a block it is supposed to propose. 
    \item[$\bullet$] Since the honest miners cannot see the withheld block, they propose blocks according to their view and construct a \textit{public chain}. By closely monitoring the public chain, the adversary may continue to withhold more blocks and construct a \textit{private chain}. 
    \item[$\bullet$] Upon the \textit{success condition}, the entire private chain of withheld blocks is released simultaneously. If it wins the forking race, the honest competing blocks are discarded and the adversary obtains more than its fair share reward.
\end{enumerate}

\begin{figure}[t]
  \centering
  \includegraphics[width=\linewidth]{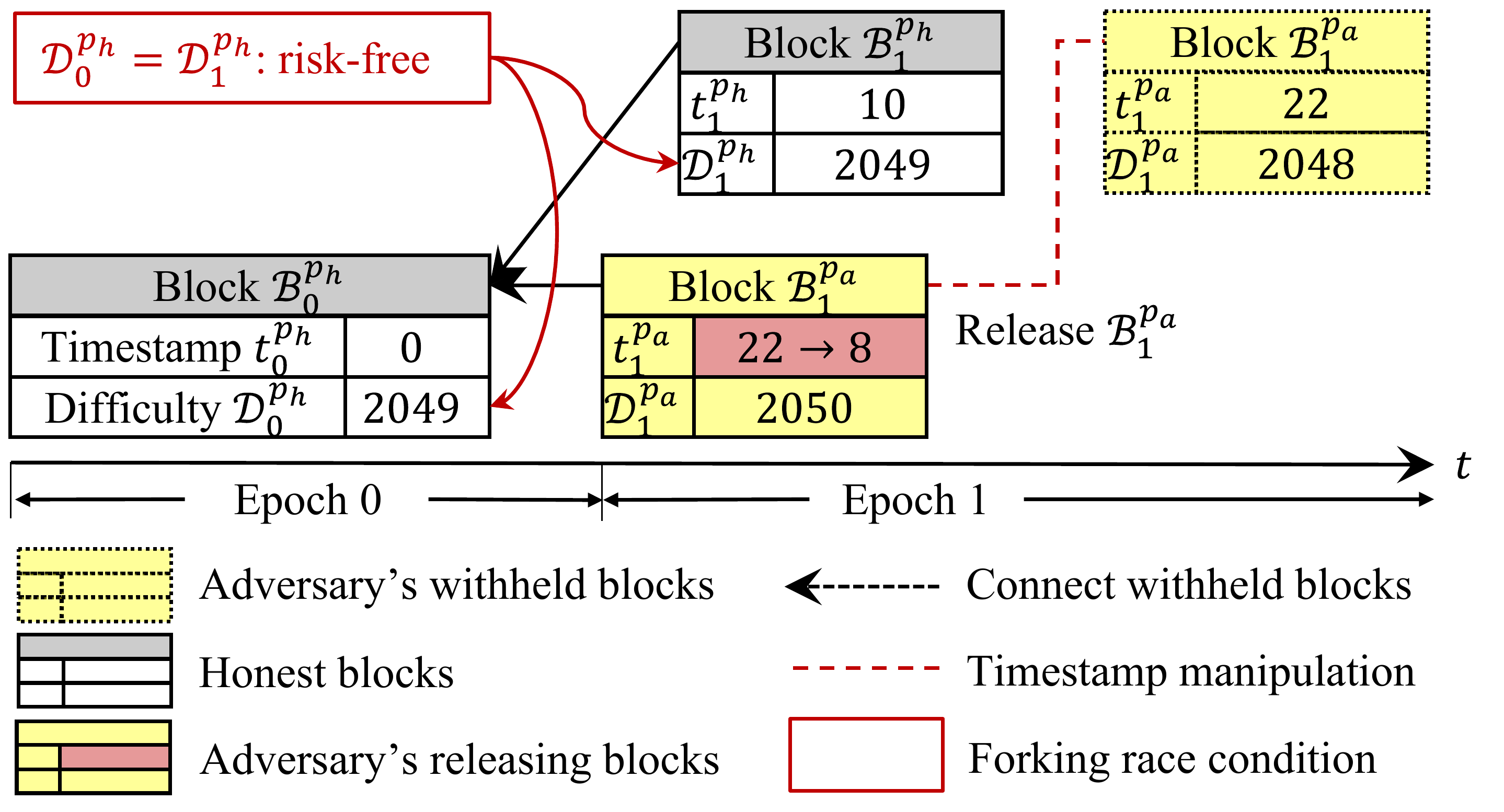}
  \caption{An Example of the RUM Attack. The RUM attack is risk-free since $\mathcal{D}_{0}^{p_h}=\mathcal{D}_{1}^{p_h}$.}
  \label{A Case of RUM}
\end{figure}

To make the attack pre-mining risk-free, RUM manipulates the timestamp so that the private branch maintains a difficulty advantage. Here, $\mathcal{B}_{i}^{p_h}$ and $\mathcal{B}_{i}^{p_a}$ denote the $i$-th block on the honest branch and the adversarial branch, respectively; $t_{i}^{p_h}$ and $t_{i}^{p_a}$ are their timestamps; and $\mathcal{D}_{i}^{p_h}$ and $\mathcal{D}_{i}^{p_a}$ are their block difficulties. By Definition~\ref{def:difficulty}, the difficulty of a child block depends on its parent difficulty and the timestamp gap to its parent. RUM requires $t_{1}^{p_h}-t_{0}^{p_h}\in[9,18)$, exactly the interval in which $\lfloor (t_{1}^{p_h}-t_{0}^{p_h})/9 \rfloor = 1$. By adjusting the timestamp in this way, the honest child keeps the same difficulty as its parent, i.e., $\mathcal{D}_{1}^{p_h}=\mathcal{D}_{0}^{p_h}$. In the forking race, the private chain is guaranteed to win. We show an example in \figref{A Case of RUM}.

We review the attack in detail in \appref{Methods of RUM Attacks}. The adversary's profit grows as the controlled hash power increases. An adversary with 25\% relative hash power gains 26.12\% normalized reward share. However, RUM can be launched only in the narrow window $[9,18)$.

\subsection{Overview of Our Approach}\label{Our Insight} 
Our key insight is that timestamp calibration affects not only the current forking race but also the difficulty of subsequent blocks. We call the latter effect \textit{post-mining risk}. Without controlling this risk, the attack remains profitable only in a narrow set of timestamp conditions and cannot be launched frequently enough. In particular, RUM is restricted to $t_{1}^{p_h}-t_{0}^{p_h}\in[9,18)$.


In our work, we seek to control both pre-mining risk and post-mining risk to increase the attack opportunities and improve profitability. Specifically, we introduce a minimal risk control approach to guide the adversary not only to exploit timestamp manipulation for short-term advantages but also to minimize the difficulty growth risk (i.e., post-mining risk). Based on this minimal risk control approach, we propose a non-trivial extension of the RUM attack, as summarized below. 


\begin{enumerate}
    \item [$\bullet$] UUM (\secref{The design of uum attack}) relaxes RUM's initiation window from $[9,18)$ to $t_1^{p_h}-t_0^{p_h}\in[9,+\infty)$, and calibrates $t_1^{p_a}$ so that the adversarial block obtains the smallest difficulty advantage that still guarantees victory in the forking race. As summarized in \tabref{tab:comp}, when the adversary controls $25\%$ of the mining power, the attack-state probability increases from $0.17$ under RUM to $0.47$ under UUM.
    \item [$\bullet$] SUUM (\secref{The design of suum attack}) extends UUM from a one-round race to a multi-round withholding strategy. In this extension, the adversary withholds a sequence of private blocks and calibrates their timestamps to preserve the smallest difficulty advantage needed for eventual victory while controlling post-mining risk. Once the release condition is met, the withheld private chain is published and wins the forking race, allowing the adversary to reclaim multiple public blocks in one attack episode. When the adversary controls $25\%$ of the mining power, the attack-state probability increases from $0.47$ under UUM to $0.61$ under SUUM, and the normalized reward share increases from $28.41\%$ to $33.30\%$.
\end{enumerate}


\section{The Design of UUM Attack}\label{The design of uum attack}
\begin{figure}[h]
  \centering
  \includegraphics[width=\linewidth]{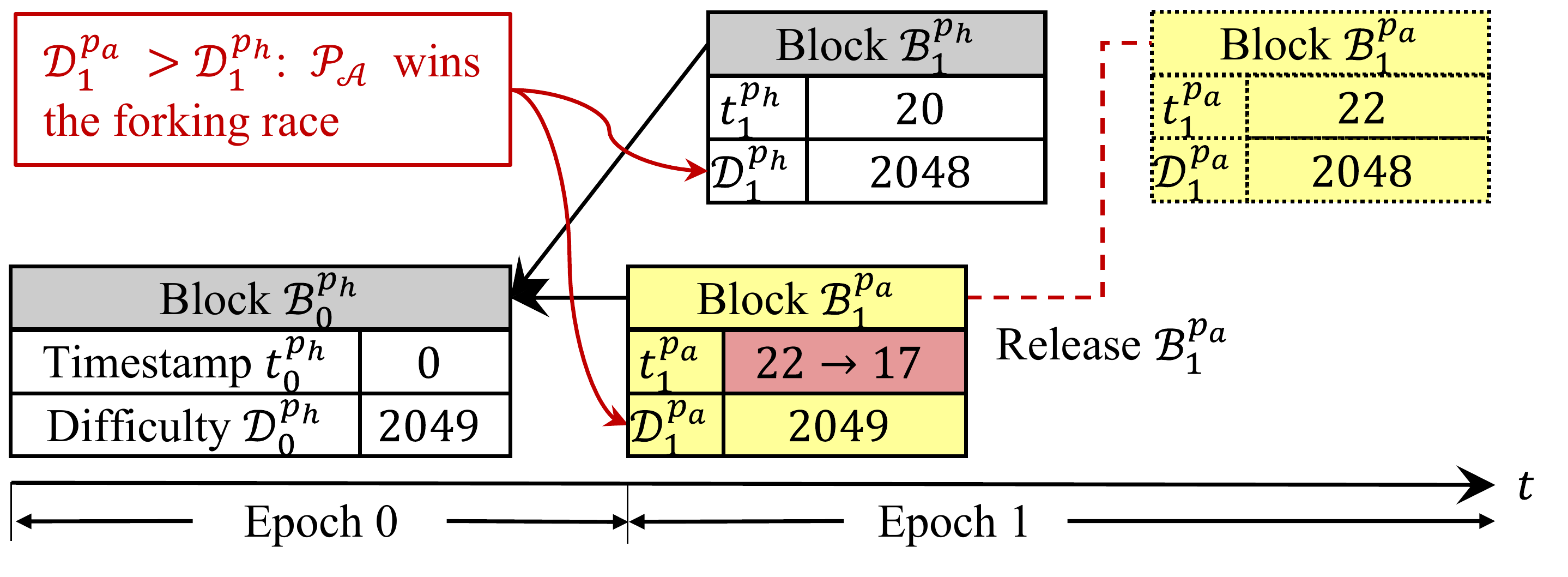}
  \caption{An example of the UUM attack. Before mining $\mathcal{B}_{1}^{p_a}$, the adversary pre-selects $t_{1}^{p_a}=17$; although the block is found at time $t=22$, this calibration raises its difficulty from 2048 to 2049. The block therefore obtains the smallest advantage needed to defeat $\mathcal{B}_{1}^{p_h}$ while minimizing post-mining risk (Theorem~\ref{theorem3}).}
  \label{A Case of UUM}
\end{figure}

We are now ready to present the UUM attack. We illustrate an example in \figref{A Case of UUM}. 

\subsection{The UUM Attack}

Relative to RUM, UUM makes two changes. First, RUM can be triggered only when the
honest competing block satisfies $t_1^{p_h}-t_0^{p_h}\in[9,18)$, under which its
difficulty remains the same as its parent, i.e., $\mathcal{D}_1^{p_h}=\mathcal{D}_0^{p_h}$. UUM relaxes this condition to $t_1^{p_h}-t_0^{p_h}\in[9,+\infty)$, which includes
all cases where the honest competing block does not become harder than its parent,
i.e., $\mathcal{D}_1^{p_h}\leq \mathcal{D}_0^{p_h}$. Second, once triggered, UUM
calibrates the adversarial timestamp so that the adversarial block wins the first
forking race with the smallest difficulty advantage compatible with success. Specifically, the main differences include: 

\begin{enumerate}
    \item[$\bullet$] UUM can be initiated whenever $\lfloor (t_1^{p_h} - t_0^{p_h}) / 9 \rfloor \geq 1$.
    \item[$\bullet$] To win the first forking race while controlling post-mining risk, UUM chooses a valid adversarial timestamp in $[1,900)$ and calibrates it to be exactly one 9-second bucket earlier than the honest competing block. By contrast, RUM consistently sets $1 \le t_{1}^{p_a}-t_{0}^{p_a}<9$.
\end{enumerate}

We summarize the key conclusions and proof sketches below, and defer full details to \appref{section4.1}. Briefly speaking, \thmref{theorem1} identifies the necessary and sufficient initiation condition for UUM; \thmref{theorem2} characterizes when UUM succeeds; \thmref{theorem3} shows how to achieve such success while minimizing the post-mining risk; and \thmref{theorem4} characterizes the exact condition under which UUM downgrades to RUM.




\begin{theorem}[\textbf{UUM Initiation Condition}]\label{theorem1}
    The initiation condition for the UUM attack is given by $\lfloor \frac{{t}_{1}^{{p}_{{h}}} - {t}_{0}^{{p}_{{h}}}}{9} \rfloor \in \left\lbrack 1, + {\infty} \right)$.
\end{theorem}

\thmref{theorem1} characterizes exactly when the relaxed UUM trigger is compatible with our minimal post-mining risk objective. Specifically, we require that the first honest competing block lies in the non-increasing difficulty region, i.e., $\mathcal D_1^{p_h}\le \mathcal D_0^{p_h}$. Substituting the difficulty formula into this inequality yields
\(
\max\left\{1-\left\lfloor (t_1^{p_h}-t_0^{p_h})/9 \right\rfloor,-99\right\}\cdot
\left\lfloor \mathcal D_0^{p_h}/2048 \right\rfloor \le 0.
\)
Since $\left\lfloor \mathcal D_0^{p_h}/2048 \right\rfloor > 0$, this is equivalent to
\(
1-\left\lfloor (t_1^{p_h}-t_0^{p_h})/9 \right\rfloor \le 0,
\)
namely,
\(
\left\lfloor (t_1^{p_h}-t_0^{p_h})/9 \right\rfloor \ge 1.
\)
Thus, $\left\lfloor (t_1^{p_h}-t_0^{p_h})/9 \right\rfloor \ge 1$ is precisely the necessary and sufficient initiation condition for UUM under this objective. The detailed proof is given in \appref{Proof of Theorem 1}.


\begin{theorem}[\textbf{UUM Success Condition}]\label{theorem2}
    The UUM attack is successful if and only if $\lfloor \frac{{t}_{1}^{{p}_{{h}}} - {t}_{0}^{{p}_{{h}}}}{9} \rfloor-\lfloor \frac{{t}_{1}^{{p}_{{a}}} - {t}_{0}^{{p}_{{h}}}}{9} \rfloor \in \left\lbrack 1, + {\infty} \right)$ and ${t}_{1}^{{p}_{{a}}} - {t}_{0}^{{p}_{{a}}} \in \lbrack 1,900)$.
\end{theorem}

\thmref{theorem2} characterizes exactly when the first adversarial block can defeat the honest competing block by difficulty dominance. Starting from the success requirement $\mathcal{D}_1^{p_a} > \mathcal{D}_1^{p_h}$ and using the fact that the two blocks share the same parent, the difficulty comparison reduces to the timestamp adjustment terms alone, which yields the one-bucket advantage condition above. The interval $[1,900)$ is the valid range for the adversarial timestamp. Together, these two conditions are necessary and sufficient for UUM to win the first forking race. The detailed proof is given in \appref{Proof of Theorem 2}.

\begin{figure*}[t]
  \centering
  \includegraphics[width=0.90\linewidth]{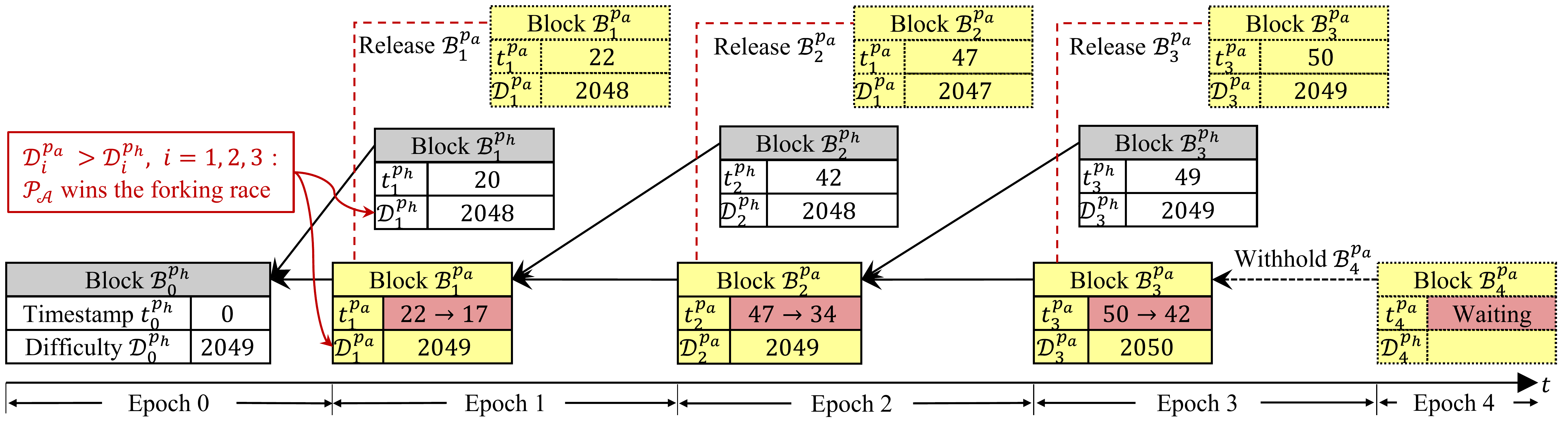}
  \caption{An example of the SUUM attack. The adversary builds a private chain by withholding blocks and pre-calibrating their timestamps. In epoch~1, setting $t_1^{p_a}=17$ gives $\mathcal{B}_1^{p_a}$ difficulty 2049, compared with 2048 for $\mathcal{B}_1^{p_h}$; epochs~2 and~3 repeat this calibration, and $\mathcal{B}_4^{p_a}$ is withheld in epoch~4. Releasing the calibrated private chain wins the forking race while controlling post-mining risk.
  }
  \label{A Case of SUUM}
\end{figure*}


\begin{theorem}[\textbf{UUM Successful Condition with Minimal Risk}]\label{theorem3}
    The UUM attack is successful with minimal risk if and only if $
\lfloor \frac{{t}_{1}^{{p}_{{h}}} - {t}_{0}^{{p}_{{h}}}}{9} \rfloor-\lfloor \frac{{t}_{1}^{{p}_{{a}}} - {t}_{0}^{{p}_{{h}}}}{9} \rfloor = 1$ and ${t}_{1}^{{p}_{{a}}} - {t}_{0}^{{p}_{{a}}} \in \lbrack 1,900)$.
\end{theorem}

Among all timestamp choices that make UUM succeed, minimal post-mining risk is obtained by taking the smallest bucket advantage that still guarantees victory, namely equality in the first condition. The validity interval $t_1^{p_a}-t_0^{p_a}\in[1,900)$ remains unchanged. We also assume $pu=0$ for the adversarial block, since not referencing uncles minimizes the difficulty of the adversarial chain. The analysis of victory conditions considers the case where honest blocks may have $pu=0$ or $pu=1$; the derived calibration ensures the adversary’s advantage holds in both cases, preserving the heaviest chain victory guarantee. The detailed proof of \thmref{theorem3} can be found in \appref{Proof of Theorem 3}.

\heading{UUM Downgrades to RUM.} We show below that if \(\lfloor (t_1^{p_h} - t_0^{p_h}) / 9 \rfloor = 1\), the UUM attack downgrades to the RUM attack. 

\begin{theorem}[\textbf{UUM Downgrades to RUM}]\label{theorem4}
    The downgrade of UUM to RUM occurs if and only if $
\lfloor \frac{t_{1}^{p_{h}} - t_{0}^{p_{h}}}{9} \rfloor = 1$.
\end{theorem}

The proof of \thmref{theorem4} starts from the defining RUM property that the honest competing block keeps the same difficulty as its parent, i.e., $\mathcal{D}_1^{p_h}=\mathcal{D}_0^{p_h}$. Substituting the difficulty formula into this equality yields $\lfloor (t_1^{p_h}-t_0^{p_h})/9 \rfloor = 1$, which is exactly the original RUM initiation window. Hence, UUM reduces to RUM precisely in this case. The detailed proof is given in \appref{Proof of Theorem 4}.


\subsection{Reward Analysis}
UUM increases the adversary's normalized realized main-chain coinbase reward share by reducing the honest participants' realized main-chain coinbase reward mass. By the steady-state reward relation established in the Preliminaries, this also increases the adversary's absolute realized main-chain coinbase reward rate per unit physical time.

\begin{theorem}[\textbf{UUM Adversary Rewards}]\label{theorem5}
Under UUM, the adversary's normalized realized main-chain coinbase reward share strictly exceeds its honest-mining baseline $\mu_A$. By the steady-state reward relation established in the Preliminaries, its absolute realized main-chain coinbase reward rate per unit physical time also strictly exceeds the honest-mining baseline. Correspondingly, the honest participants' normalized and absolute realized main-chain coinbase rewards both decrease.
\end{theorem}

The proof of \thmref{theorem5} is deferred to \appref{Proof of theorem 5}.

\section{The Design of SUUM Attack}\label{The design of suum attack}
We further extend UUM to SUUM. The main differences include: 
\begin{enumerate}
    \item[$\bullet$] Unlike UUM where an adversarial block is always released immediately, SUUM allows the adversary to withhold multiple blocks and release them as a private chain.
    \item[$\bullet$] SUUM applies minimal post-mining risk control block by block: the first withheld block obtains a one-bucket advantage, meaning that its timestamp gap falls exactly one 9-second difficulty-adjustment bucket below that of the competing honest block. Each later withheld block is then calibrated with the smallest additional advantage compatible with eventual success.
    \item[$\bullet$] This enlarged strategy space lets the adversary achieve higher profit than UUM because multiple blocks in the public chain are discarded.
\end{enumerate}

\subsection{Method}
\label{SUUM Method}
\begin{figure}[t]
  \centering
  \includegraphics[width=\linewidth]{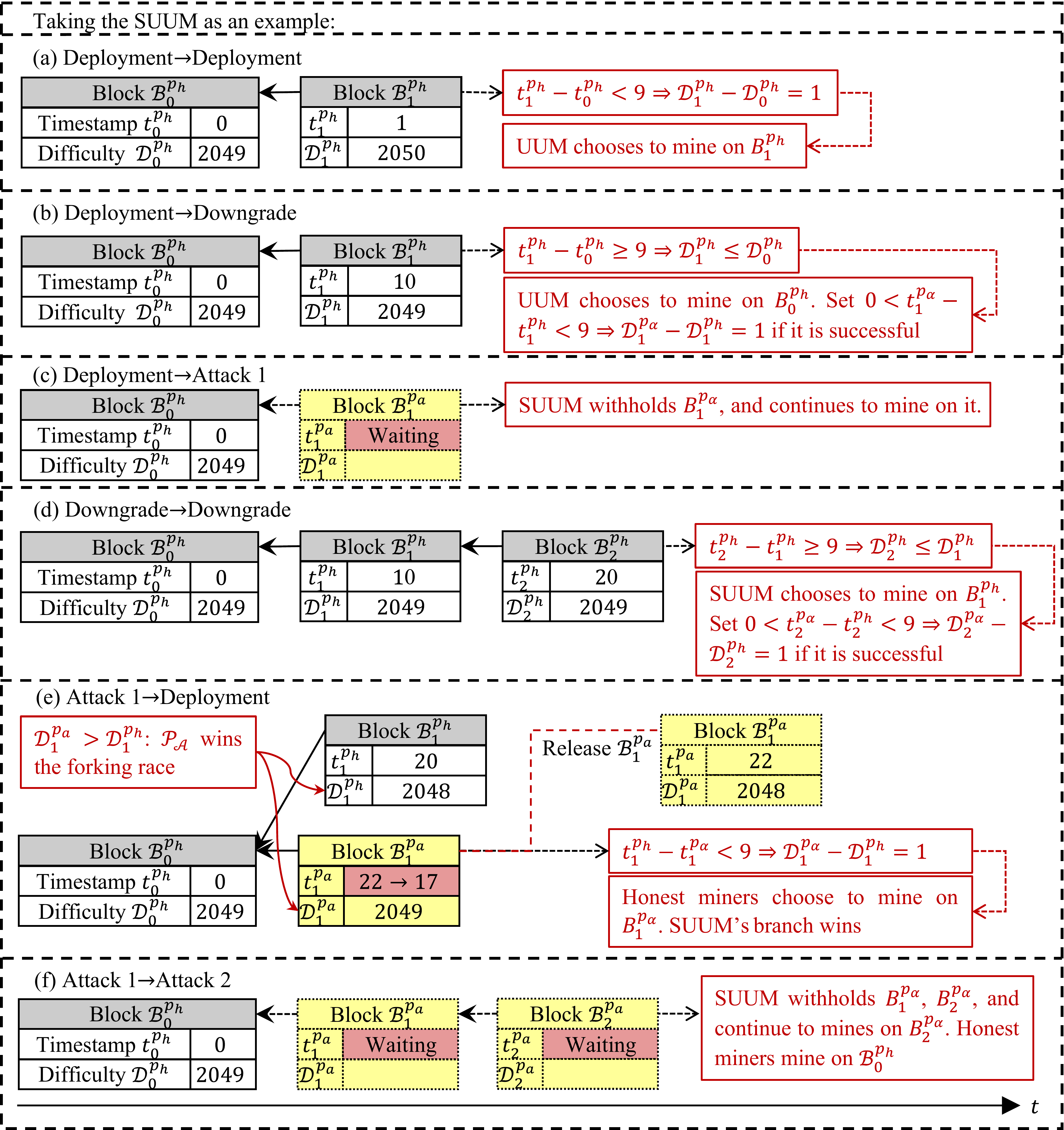}
  \caption{SUUM state transitions. From Deployment, a valid trigger moves the adversary either to Downgrade, where SUUM reduces to UUM, or to Attack, where blocks are withheld and later released as a private chain. Minimal-risk timestamp calibration preserves the required difficulty advantage; after release, the adversary returns to Deployment or continues withholding.}
  \label{SUUM State Transition Process}
\end{figure}

We show a case of the SUUM attack in \figref{A Case of SUUM}. We decompose SUUM into five logical phases: Downgrade, Deployment, Withholding, Releasing, and Minimal Risk Control. The corresponding state transitions are shown in \figref{SUUM State Transition Process}; full algorithmic details are deferred to \appref{Methods of SUUM Attacks}.

\heading{SUUM Successful Condition.}
\thmref{SUUM Successful Condition} defines the precise success conditions for the SUUM attack. For the first adversarial block ($i = 1$), it requires that \(\lfloor ( t_i^{p_h} - t_{i-1}^{p_h})/9 \rfloor - \lfloor(t_i^{p_a} - t_{i-1}^{p_a}) \big / 9 \rfloor \in [1, +\infty)\), and that the adversary’s own block timestamp satisfies \(t_1^{p_a} - t_0^{p_a} \in [1, 900)\). For subsequent blocks ($i \geq 2$), the condition becomes \(\lfloor ( t_i^{p_h} - t_{i-1}^{p_h})/9 \rfloor - \lfloor(t_i^{p_a} - t_{i-1}^{p_a}) \big / 9 \rfloor \in [0, +\infty)\) together with the same validity range \(t_i^{p_a} - t_{i-1}^{p_a} \in [1, 900)\). These criteria ensure that each adversarial block in the private chain achieves a higher difficulty than the corresponding honest block, guaranteeing its eventual inclusion in the main-chain.

\begin{theorem}[\textbf{SUUM Successful Condition}]\label{SUUM Successful Condition}
    The SUUM attack is successful if and only if the following conditions are met:

\begin{enumerate}[(1)]
\item For $i = 1$, the conditions $\lfloor \frac{t_{i}^{p_{h}} - t_{i - 1}^{p_{h}}}{9} \rfloor - \lfloor \frac{t_{i}^{p_{a}} - t_{i - 1}^{p_{a}}}{9} \rfloor \in \lbrack 1, + \infty)$ and $t_{i}^{p_{a}} - t_{i - 1}^{p_{a}} = t_{1}^{p_{a}} - t_{0}^{p_{a}} \in \lbrack 1,900)$ must be satisfied.

\item For $i \geq 2$, the conditions $\lfloor \frac{t_{i}^{p_{h}} - t_{i - 1}^{p_{h}}}{9} \rfloor - \lfloor \frac{t_{i}^{p_{a}} - t_{i - 1}^{p_{a}}}{9} \rfloor \in \lbrack 0, + \infty)$ and $t_{i}^{p_{a}} - t_{i - 1}^{p_{a}} \in \lbrack 1,900)$ must be satisfied.
\end{enumerate}
\end{theorem}

The proof of \thmref{SUUM Successful Condition} starts from the requirement that each adversarial block in the withheld chain must dominate its honest counterpart in difficulty. For the first withheld block, the shared parent reduces the comparison to a one-bucket timestamp advantage together with the validity interval $[1,900)$. For later withheld blocks, the same argument is applied recursively to successive public/private block pairs; after canceling the parent terms, the condition becomes a non-negative bucket advantage together with the same validity interval. Thus, the theorem gives the necessary and sufficient success conditions for SUUM. The detailed proof is given in \appref{Proof of theorem 6}.

\heading{SUUM Successful Condition with Minimal Risk.}
\thmref{theorem7} refines the general SUUM success condition to the minimal post-mining risk case. The first withheld block uses the smallest advantage that still guarantees success, namely one-bucket, and every later withheld block uses zero additional bucket advantage while remaining valid.

\begin{theorem}[\textbf{SUUM Successful Condition with Minimal Risk}]\label{theorem7}
    The SUUM attack is successful with minimal risk if and only if the following conditions are met:

\begin{enumerate}[(1)]
\item For $i=1$, the conditions $\lfloor \frac{t_{i}^{p_{h}} - t_{i - 1}^{p_{h}}}{9} \rfloor - \lfloor \frac{t_{i}^{p_{a}} - t_{i - 1}^{p_{a}}}{9} \rfloor = 1$ and $t_{i}^{p_{a}} - t_{i - 1}^{p_{a}} = t_{1}^{p_{a}} - t_{0}^{p_{a}} \in \lbrack 1,900)$ must be satisfied.

\item For $i \geq 2$, the conditions $
\lfloor \frac{t_{i}^{p_{h}} - t_{i - 1}^{p_{h}}}{9} \rfloor - \lfloor \frac{t_{i}^{p_{a}} - t_{i - 1}^{p_{a}}}{9} \rfloor = 0$ and $
t_{i}^{p_{a}} - t_{i - 1}^{p_{a}} \in \lbrack 1,900)$ must be satisfied.
\end{enumerate}
\end{theorem}

Among all timestamp choices that make SUUM succeed, minimal post-mining risk is achieved by taking the smallest admissible bucket advantage in each step: equality to $1$ for the first withheld block and equality to $0$ for every later withheld block. The validity interval $[1,900)$ is unchanged. Hence, these conditions are necessary and sufficient for SUUM to succeed with minimal post-mining risk. The detailed proof is given in \appref{Proof of theorem 7}.

\subsection{State Space}
We categorize the state space of SUUM into three types: deployment state, downgrade state, and attack state. We further divide the attack state into withholding state and releasing state. Specifically, the deployment state captures the scenario where the SUUM adversary waits to kick-start the attack, the downgrade state captures the situation where SUUM is downgraded to UUM, and the attack state captures the execution of attacks by the SUUM adversary. The details of each state can be found in \appref{Section5.2}.

\subsection{Reward Analysis}
\heading{SUUM Adversary Rewards.}
Relative to UUM, SUUM further improves the adversary's normalized realized main-chain coinbase reward share through block withholding and staircase release. By the steady-state reward relation established in the Preliminaries, this also yields a strictly larger absolute realized main-chain coinbase reward rate per unit physical time. We defer the formal statement and proof to \appref{Proof of theorem 8}.

The appendix derivation uses the steady-state probabilities of the SUUM states together with the same reward notions $\widetilde{\mathbb{R}}_{\mathcal{P}}$, $\widehat{\mathbb{R}}_{\mathcal{P}}$, and $\mathbb{R}_{\mathcal{P}}$. The detailed derivation is given in \appref{Proof of theorem 8}.

\section{Comparison}\label{Comparison}
We compare honest mining (HM), RUM, UUM, and SUUM under the same adversarial hash power, blockchain parameters, coinbase-only realized main-chain accounting, and steady-state model.

\heading{Reward Comparison.} Under both the normalized realized main-chain coinbase reward share and the absolute realized reward rate, the adversarial reward follows \(\mathrm{SUUM}>\mathrm{UUM}>\mathrm{RUM}>\mathrm{HM}\). UUM improves upon RUM by enlarging the attack-initiation window, while SUUM further benefits from block withholding and staircase release; the formal statement and proof are given in \appref{Proof of Theorem 9}.

\heading{Post-mining Risk Comparison.} RUM and UUM each have one successful completion pattern that contributes post-mining risk, whereas SUUM has two such patterns because an attack can complete through either its Downgrade or Attack state. This comparison concerns cumulative post-mining risk across successful attack rounds and is formalized in \appref{Proof of Theorem 10}.

\heading{Forking Rate Comparison.} The induced forking rates satisfy \({\mathbb{F}\mathbb{R}}^{SUUM}>{\mathbb{F}\mathbb{R}}^{UUM}>{\mathbb{F}\mathbb{R}}^{RUM}>{\mathbb{F}\mathbb{R}}^{HM}=0\). The ordering follows because UUM removes RUM's restrictive timestamp window and SUUM additionally creates forks from its Downgrade state; see \appref{Proof of Theorem 11}.

\heading{Pre-mining Risk Comparison.} RUM has zero pre-mining risk, whereas UUM and SUUM incur the same non-zero but negligible reduction in block-generation rate, approximately \(\mu_{\mathcal{A}}/(T\cdot 2048)\). Under Ethereum 1.x parameters, this corresponds to a relative efficiency loss of about \(1/2048\approx0.0488\%\), as established in \appref{Proof of Theorem 12}.

The attack-specific reward derivations for UUM and SUUM appear in \appref{Proof of theorem 5} and \appref{Proof of theorem 8}, respectively.

\begin{figure*}[t]
  \centering
  \includegraphics[width=1\linewidth]{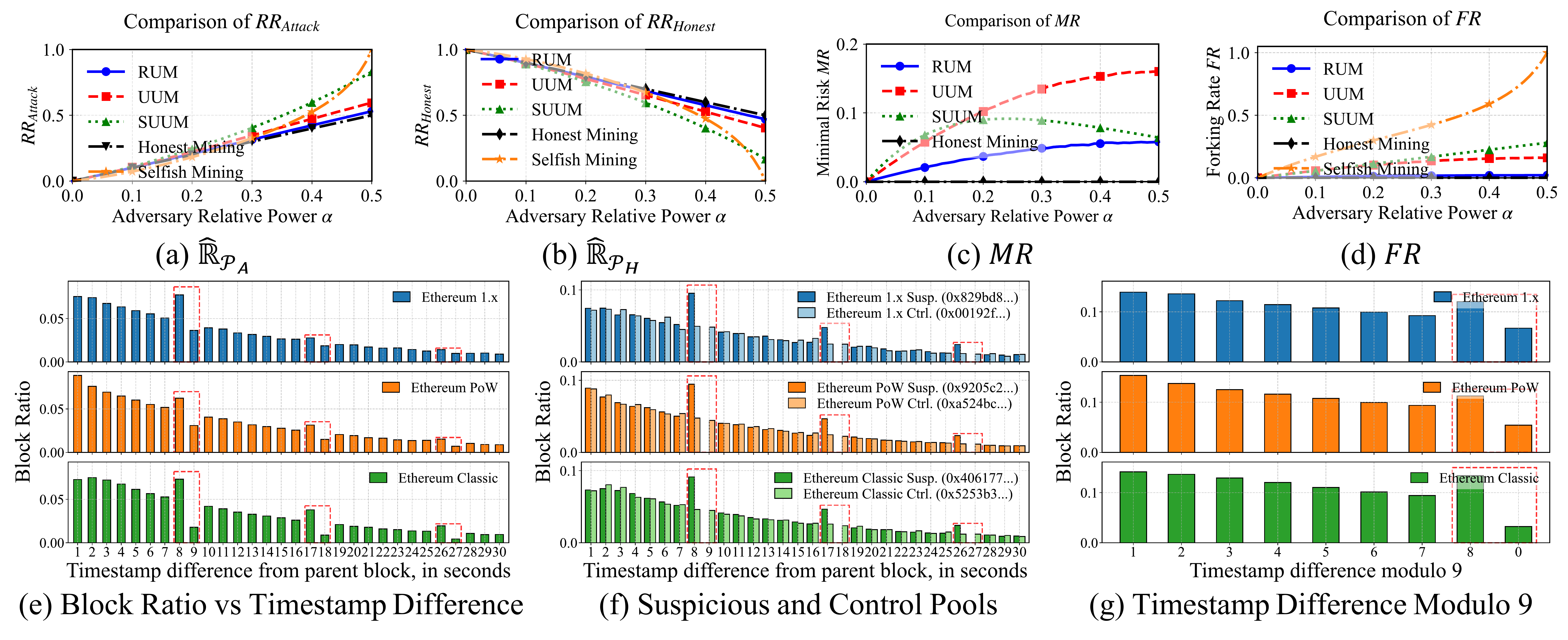}
    \caption{Results of our experiments and on-chain analysis. (a) Adversary normalized realized main-chain coinbase reward share, $\widehat{\mathbb{R}}_{\mathcal{P}_A}$. (b) Honest participants' normalized realized main-chain coinbase reward share, $\widehat{\mathbb{R}}_{\mathcal{P}_H}$. (c) Minimal post-mining risk. (d) Forking rate. (e) Distribution of main-chain blocks across ETH, ETHW, and ETC. (f) Block ratios of suspicious and control pools. (g) Block ratios by timestamp-difference residue modulo 9.}

  \label{Exp}
\end{figure*}

\begin{table*}[t]
\caption{Run-length distribution of consecutive main-chain blocks for representative flagged pools}
\label{tab:consecutive_blocks_0x406177}
\centering
\setlength{\tabcolsep}{3pt}
\resizebox{\textwidth}{!}{%
\begin{tabular}{@{}|c|c|c|c|c|c|c|c|c|c|c|c|c|c|c|c|c|c|c|c|c|c|@{}}
\toprule
Consecutive Blocks & 2 & 3 & 4 & 5 & 6 & 7 & 8 & 9 & 10 & 11 & 12 & 13 & 14 & 15 & 16 & 17 & 18 & 19 & 20 & 21 \\
\midrule
\texttt{0x406177...} & 
\fbox{5600} & 
\fbox{3356} & 
\fbox{2032} & 
\fbox{1207} & 
\fbox{756} & 
\fbox{467} & 
\fbox{246} & 
\fbox{177} & 
\fbox{99} & 
\fbox{67} & 
\fbox{43} & 
\fbox{22} & 
\fbox{17} & 
\fbox{9} & 
\fbox{2} & 
\fbox{7} & 
0 & 
0 & 
\fbox{1} & 
\fbox{1} \\
\texttt{0x829bd8...} & 
\fbox{1956} & 
\fbox{327} & 
\fbox{49} & 
\fbox{4} & 
0 & 
\fbox{1} & 
0 & 
0 & 
0 & 
0 & 
0 & 
0 & 
0 & 
0 & 
0 & 
0 & 
0 & 
0 & 
0 & 
0 \\
\texttt{0x9205c2...} & 
\fbox{4770} & 
\fbox{1497} & 
\fbox{532} & 
\fbox{161} & 
\fbox{61} & 
\fbox{22} & 
\fbox{11} & 
\fbox{3} & 
\fbox{2} & 
0 & 
0 & 
0 & 
0 & 
0 & 
0 & 
0 & 
0 & 
0 & 
0 & 
0 \\
\bottomrule
\end{tabular}%
}
\begin{tablenotes}
\footnotesize
\setlength{\itemindent}{0pt}
\setlength{\rightskip}{10pt}
\item* \noindent
This table reports the run-length distribution of consecutive main-chain blocks for representative flagged pools on ETC, ETH, and ETHW. Unlike the modulo-9 anomaly statistics in Table~\ref{tab:unified_pool_statistics}, this table focuses on \emph{withholding-related sequence evidence}. In particular, the ETC pool \texttt{0x406177...4e95} exhibits a much heavier tail in the run-length distribution than the ETH and ETHW pools, including one streak of length 21 and another of length 20. In the ETC dataset of 100,000 sampled main-chain blocks, the 21-block streak occurs at heights \(23438117 \rightarrow 23438137\) (2025/11/20 15:32:51 \(\rightarrow\) 2025/11/20 15:37:18), and the 20-block streak occurs at heights \(23393481 \rightarrow 23393500\) (2025/11/13 15:38:16 \(\rightarrow\) 2025/11/13 15:42:35). In total, this pool produces nine streaks of length at least 17, including seven distinct 17-block streaks, one 20-block streak, and one 21-block streak. Such repeated long-streak behavior is difficult to reconcile with RUM-style one-step timestamp pruning alone and is instead more consistent with SUUM-style persistent multi-block withholding.
\end{tablenotes}
\end{table*}

\begin{table*}[t]
\caption{Statistical results for the full datasets of three Ethereum 1.x-style blockchains}
\label{tab:chi_square_full_chain}
\centering
\setlength{\tabcolsep}{3pt}
\begin{tabular}{@{}|c|c|c|c|c|c|c|c|@{}}
\toprule
Chain & Blocks (Total) & 0 mod 9 (E/O) & 8 mod 9 (E/O) & $\chi^2(8)$ & Exact binom. & $G(8)$ & Shift exact \\
\midrule
ETH & 111{,}185 & \fbox{8851.3 / 7495} $\mathbf{\downarrow}$ \textbf{1356.3} & \fbox{12306.8 / 13385} $\mathbf{\uparrow}$ \textbf{1078.2} & 303.9 & 53.1 & 312.8 & 68.4 \\
ETHW & 100{,}000 & \fbox{8002.9 / 5423} $\mathbf{\downarrow}$ \textbf{2579.9} & \fbox{8562.2 / 11256} $\mathbf{\uparrow}$ \textbf{2693.8} & 1682.6 & 221.3 & 1712.8 & 372.9 \\
ETC & 100{,}000 & \fbox{7921.6 / 3221} $\mathbf{\downarrow}$ \textbf{4700.6} & \fbox{8639.9 / 13395} $\mathbf{\uparrow}$ \textbf{4755.1} & 5408.5 & 835.8 & 5843.2 & 1270.3 \\
\bottomrule
\end{tabular}
\begin{tablenotes}
\footnotesize
\setlength{\itemindent}{0pt}
\setlength{\rightskip}{10pt}
\item* \noindent
This table reports goodness-of-fit results for the $\Delta t \bmod 9$ distributions in the full sampled datasets on ETH, ETHW, and ETC. The expected counts are computed from the same-chain empirical baseline after excluding the flagged staircase-anomalous pool(s), and are then rescaled to the original number of valid $\Delta t$ samples on that chain. Specifically, the ETH baseline excludes pool \texttt{0x829bd8...}, the ETHW baseline excludes pool \texttt{0x9205c2...}, and the ETC baseline excludes pools \texttt{0x406177...} and \texttt{0x35aa26...}. Here, \textbf{E/O} denotes expected/observed counts. For each annotated entry, $\mathbf{\downarrow}$ indicates a deficit relative to expectation, while $\mathbf{\uparrow}$ indicates an enrichment relative to expectation. \textbf{Exact binom.} reports $-\log_{10} p$ for the one-sided exact binomial test of deficit at $(\Delta t \bmod 9)=0$. $\boldsymbol{G(8)}$ reports the likelihood-ratio statistic for the full 9-category distribution. \textbf{Shift exact} reports $-\log_{10} p$ for a one-sided exact conditional test of whether residue 8 is abnormally enriched relative to residue 0 under the attacker-excluded same-chain baseline. Larger values indicate stronger significance.
\end{tablenotes}
\end{table*}

\begin{table*}[t]
\caption{Unified view of flagged pool statistics, timestamp manipulation evidence, and estimated reward}
\label{tab:unified_pool_statistics}
\centering
\scriptsize
\setlength{\tabcolsep}{3pt}
\resizebox{\textwidth}{!}{
\begin{tabular}{@{}|c|c|c|c|c|c|c|c|c|c|c|c|c|@{}}
\toprule
\multirow{2}{*}{Pool ID} & \multirow{2}{*}{Chain} & \multicolumn{3}{c|}{Observed Pool Statistics} & \multicolumn{7}{c|}{Timestamp Manipulation Evidence} & \multicolumn{1}{c|}{Reward Impact} \\
\cmidrule(lr){3-5} \cmidrule(lr){6-12} \cmidrule(lr){13-13}
& & $HP$ & \begin{tabular}[c]{@{}c@{}}Obs.\\Share\end{tabular} & Blocks (Pool/Total) & $n$ & 0 mod 9 (E/O) & 8 mod 9 (E/O) & $\chi^2(8)$ & \begin{tabular}[c]{@{}c@{}}Exact\\binom\end{tabular} & $G(8)$ & \begin{tabular}[c]{@{}c@{}}Shift\\exact\end{tabular} & \begin{tabular}[c]{@{}c@{}}Extra\\ Profits (\$)\end{tabular} \\
\midrule
\texttt{0x829bd8...} & ETH & 0.11563 & 0.15323 & 17{,}037 / 111{,}186 & 17{,}037 & \fbox{1356.3 / 0} $\mathbf{\downarrow}$ \textbf{1356.3} & \fbox{1885.8 / 2964} $\mathbf{\uparrow}$ \textbf{1078.2} & 1983.3 & 613.8 & 3247.1 & 630.7 & \fbox{2,122,778} \\
\texttt{0x00192f...} & ETH & -- & 0.07272 & 8{,}086 / 111{,}186 & 8{,}086 & 640.4 / 679 $\mathbf{\uparrow}$ \textbf{38.6} & 911.2 / 723 $\mathbf{\downarrow}$ \textbf{188.2} & 50.5 & 0.02 & 53.2 & 0.00 & -- \\
\midrule
\texttt{0x9205c2...} & ETHW & 0.19066 & 0.32237 & 32{,}237 / 100{,}000 & 32{,}237 & \fbox{2579.9 / 0} $\mathbf{\downarrow}$ \textbf{2579.9} & \fbox{2760.2 / 5454} $\mathbf{\uparrow}$ \textbf{2693.8} & 5219.4 & 1167.8 & 7211.6 & 1192.1 & \fbox{5,373} \\
\texttt{0xa524bc...} & ETHW & -- & 0.15823 & 15{,}823 / 100{,}000 & 15{,}823 & 1267.3 / 1263 $\mathbf{\downarrow}$ \textbf{4.3} & 1346.5 / 1382 $\mathbf{\uparrow}$ \textbf{35.5} & 5.8 & 0.34 & 5.7 & 0.64 & -- \\
\midrule
\texttt{0x406177...} & ETC & 0.30960 & 0.59339 & 59{,}339 / 100{,}000 & 59{,}339 & \fbox{4704.0 / 0} $\mathbf{\downarrow}$ \textbf{4704.0} & \fbox{5128.9 / 9883} $\mathbf{\uparrow}$ \textbf{4754.1} & 9114.4 & 2128.4 & 12868.5 & 1524.4 & \fbox{832,927} \\
\texttt{0x5253b3...} & ETC & -- & 0.09586 & 9{,}586 / 100{,}000 & 9{,}586 & 749.4 / 793 $\mathbf{\uparrow}$ \textbf{43.6} & 823.1 / 845 $\mathbf{\uparrow}$ \textbf{21.9} & 14.0 & 0.02 & 14.1 & 0.13 & -- \\
\bottomrule
\end{tabular}
}
\begin{tablenotes}
\footnotesize
\setlength{\itemindent}{0pt}
\setlength{\rightskip}{10pt}
\item* \noindent This table unifies the representative flagged pool statistics, the statistical evidence for staircase-style timestamp manipulation, and the estimated reward impact under the SUUM benchmark. Here, $\boldsymbol{HP}$ denotes the external honest share benchmark of the mining pool, while \textbf{Obs. Share} denotes the realized main-chain ratio of the pool in our sampled window. \textbf{Blocks (Pool/Total)} reports the corresponding raw counts on the sampled main-chain. We emphasize that \textbf{Obs. Share} is an outcome variable in the attacked sample and should not be interpreted as the external honest share benchmark. The column $n$ denotes the number of valid timestamp differences used in the modulo-9 statistical tests. \textbf{E/O} denotes expected/observed counts. For each annotated entry, $\mathbf{\downarrow}$ indicates a deficit relative to expectation, while $\mathbf{\uparrow}$ indicates an enrichment relative to expectation. For the three flagged pools, the same-chain baseline is formed by all other miners on the same blockchain. For the three comparison pools, the baseline is re-computed after excluding the corresponding flagged pool on the same chain, so that the expected counts are not contaminated by the most prominent flagged pool on that chain. \textbf{Exact binom} reports $-\log_{10} p$ for the one-sided exact binomial test of deficit at $(\Delta t \bmod 9)=0$. $\boldsymbol{G(8)}$ reports the likelihood-ratio statistic for the full 9-category distribution. \textbf{Shift exact} reports $-\log_{10} p$ for a one-sided exact conditional test of whether residue 8 is abnormally enriched relative to residue 0 under the same-chain baseline formed by all other miners. Finally, \textbf{Extra Profits (\$)} reports the corresponding model-based estimate under that SUUM benchmark over the observation window, compared to honest mining.
\end{tablenotes}
\end{table*}

\section{Experimentation}\label{Simulated Estimation}

We conducted extensive experiments to assess the UUM and SUUM attacks. The setup and implementation details can be found in \appref{Setup and Implementation Details}.

\heading{The Steady-state Probability.}
We analyze the steady-state probabilities of the RUM, UUM, and SUUM attacks under varying adversarial power ratios, i.e., how frequently each attack can be launched. The steady-state probability is crucial as it is directly related to the profit of the adversary. 

The steady-state probability of being in the attack state increases with the adversary's relative power $\alpha$ for all three attacks. SUUM exhibits the highest attack state probability at higher values of $\alpha$. This result directly validates the theoretical design of SUUM described in \secref{SUUM Method}: by allowing the adversary to withhold blocks and build a private chain, SUUM expands the adversary's strategy space and enables persistent capture of attack windows. 

The results show that SUUM's design achieves superior persistence in maintaining the attack state, enabling adversaries to exert sustained influence on the blockchain. Taking $\alpha=0.25$ as an example, SUUM maintains an attack state probability of 0.61, while UUM and RUM achieve only 0.47 and 0.17. So SUUM remains in active attack windows approximately 61\% of the time, an increase of 14\% and 44\% compared to UUM and RUM (see \tabref{tab:State_probability}). This result shows the attack persistence advantage achieved by block withholding mechanism. This aligns with \thmref{theorem7}, where SUUM's reward advantage stems from its higher steady-state probability of attack execution. The findings underscore the need for countermeasures to mitigate such persistent attacks, as discussed in \secref{Discussion}.

\heading{Normalized Realized Reward Share.}
To evaluate the economic impact of the proposed attacks, we analyze the normalized realized main-chain coinbase reward shares obtained by the adversary and honest participants under RUM, UUM, SUUM, and honest mining strategies.

We use the normalized realized main-chain coinbase reward share defined in the Preliminaries, measured as a participant's realized main-chain coinbase rewards divided by the total realized main-chain coinbase rewards in the observation window. Panels~(a) and~(b) in \figref{Exp} report $\widehat{\mathbb{R}}_{\mathcal{P}_A}$ and $\widehat{\mathbb{R}}_{\mathcal{P}_H}$, respectively. By the steady-state reward relation established there, a strict increase in $\widehat{\mathbb{R}}_{\mathcal{P}_A}$ also implies a strict increase in the adversary's absolute realized main-chain coinbase reward rate.

As shown in \figref{Exp}.(a), SUUM consistently yields the highest $\widehat{\mathbb{R}}_{\mathcal{P}_A}$ among the compared strategies. For instance, at $\alpha=0.25$, the adversary reaches $\widehat{\mathbb{R}}_{\mathcal{P}_A}=33.30\%$ in SUUM, whereas the corresponding values are 28.41\% and 26.12\% for UUM and RUM, respectively.

\figref{Exp}.(b) shows the corresponding degradation in the honest participants' normalized realized main-chain coinbase reward share. Consistent with panel~(a), the honest participants obtain the lowest $\widehat{\mathbb{R}}_{\mathcal{P}_H}$ under SUUM.

\heading{The Minimal Post-mining Risk.}
As mentioned throughout the paper, our crucial observation is that RUM attack ignores the post-mining risk. Thus, we evaluate the minimal post-mining risk. 
The minimal post-mining risk is quantified as $MR=\mathcal{D}_{Attack}-\mathcal{D}_{Honest},$
where $\mathcal{D}_{Attack}$ and $\mathcal{D}_{Honest}$ represent the difficulty of adversarial and honest blocks at the same height. 

\figref{Exp}.(c) shows the post-mining risk for each attack. Honest mining maintains a baseline risk of zero, as no difficulty manipulation occurs. RUM introduces non-negligible risk, although its strict timestamp constraints limit difficulty fluctuations. UUM exhibits marginally higher risk due to relaxed initiation conditions (\thmref{theorem1}), allowing occasional difficulty spikes. SUUM has the highest risk among all three attacks, as its block withholding strategy (\secref{SUUM Method}) disrupts difficulty adjustment continuity. 

Nevertheless, through our multi-layered minimal post-mining risk control, SUUM effectively mitigates the risk of rising mining difficulty while ensuring an advantage in the forking race (\thmref{theorem7}). Simulation results show that the maximum difficulty increase from a single round of SUUM attack does not exceed 0.09 (\figref{Exp}.(c)). The upper bound 0.09 is the \textit{expected normalized increase} in difficulty per attack round under minimal risk calibration. According to the difficulty adjustment rule, a more noticeable impact on subsequent mining will only occur when the cumulative change in the difficulty exceeds a threshold of 2048 (Definition \ref{def:difficulty}). Therefore, it would require approximately $2048 \div 0.09 > 22755$ rounds of SUUM to cause one significant effect. This further indicates that the risk incurred by a single round of SUUM attack is minimal.

\heading{The Forking Rate.}
We also assess the forking rate $FR$. $FR$ is defined as the number of intentionally induced forks divided by the total number of blocks mined in the system. We vary the adversarial power $\alpha$ from 0 to 0.5 for the attacks. 

We show the results in \figref{Exp}.(d). Honest mining maintains a 0\% forking rate, as all participants adhere to the main-chain. RUM triggers occasional forks, occurring only when adversaries mine blocks with timestamps $t<9$ (\thmref{Forking Rate Comparison of Uncle Maker-based attack and honest mining}). UUM exhibits higher fork rates, as its relaxed constraints (\secref{section4.1}) permit more frequent adversarial interventions. SUUM maximizes forks by combining withheld block releases (\secref{SUUM Method}) and timestamp manipulation.

\section{Analysis of Real-World Behavior, Mitigation, and Discussion}
In this section, we further assess on-chain behavior using real-world data. Our goal is to assess whether there exists any real-world behavior that is similar to our attacks.

Note that RUM detects possible on-chain attack behavior using Ethereum 1.0 data mainly
through \textit{modulo-9 timestamp residues}. For two consecutive main-chain blocks,
let $\Delta t_i=t_i-t_{i-1}$. Since the Ethereum-style difficulty adjustment rule
depends on $\lfloor \Delta t_i/9 \rfloor$, an abnormal enrichment of
$\Delta t_i \equiv 8 \pmod 9$ together with a deficit of
$\Delta t_i \equiv 0 \pmod 9$ suggests timestamp calibration.

We provide a more complete evidence chain for SUUM on three blockchain systems as mentioned in the introduction. In terms of the attack evidence, we consider four components. First, we compare each pool's realized main-chain ratio with an external honest share benchmark (Table~\ref{tab:unified_pool_statistics}). The goal is to test whether block withholding behavior exists. Second, we evaluate whether multiple blocks are released simultaneously, the main difference of SUUM compared to other attacks (Table~\ref{tab:consecutive_blocks_0x406177}). Third, we perform an analysis over disjoint 10{,}000-block windows to test whether repeated multi-block release patterns persist across time windows rather than arising from a single coincidence (Table~\ref{tab:window_stability_summary}). Finally, after excluding suspicious pools from the same-chain baseline, we quantify the deficit at residues \(9m\) and the enrichment at residues \(9m-1\), which captures both the extent and direction of timestamp calibration (Table~\ref{tab:unified_pool_statistics}).

\subsection{Analysis Results}
\heading{Dataset.} We obtain data from three ETH 1.x-style blockchains: Ethereum (ETH, block heights 15424591--15535776, 111,186 main-chain blocks), Ethereum Classic (ETC, block heights 23353196--23453195, 100,000 main-chain blocks), and EthereumPoW (block heights 22989157--23089156, 100,000 main-chain blocks). As mentioned in the introduction, our empirical analysis covers 311,186 main-chain blocks in total.

\heading{Timestamp Difference in Main-Chain Blocks.}
We measure the timestamp gap $\Delta t$ between each main-chain block and its parent on the three blockchains; the corresponding distributions are shown in \figref{Exp}.(e). The residue distribution of $\Delta t \bmod 9$ is clearly non-uniform. This chain-level anomaly is statistically significant across all three full datasets; the corresponding goodness-of-fit, exact binomial, likelihood-ratio, and directional shift tests are reported in Table~\ref{tab:chi_square_full_chain}.

We believe this deviation is unlikely to be coincidental as the timestamp differences of blocks on the spikes and their preceding blocks residues sharply at $9m-1$. For example, in the ETC pool \texttt{0x406177...}, among 59,339 valid timestamp gaps, the observed count for $\Delta t \bmod 9 = 0$ is 0, whereas the count for $\Delta t \bmod 9 = 8$ is 9,883. We view the repeated $9m \rightarrow 9m-1$ skew as evidence consistent with timestamp calibration for post-mining risk control.

\heading{Pools with the Strongest Manipulation Signatures.}
For each blockchain, we sort mining pools by the total number of mined blocks and inspect the largest pools first. In the dataset, the strongest staircase-style timestamp manipulation signatures appear in Ethereum 1.x pool \texttt{0x829bd8...}, EthereumPoW pool \texttt{0x9205c2...}, and Ethereum Classic pool \texttt{0x406177...}. These pools do not merely show a mild reduction of \(\Delta t \equiv 0 \pmod 9\); instead, they exhibit a near-deterministic ladder pattern in which the probability mass expected at \(9,18,27,36,\ldots\) is systematically shifted to \(8,17,26,35,\ldots\). This behavior matches the timestamp control rule needed to preserve the attacker's advantage while limiting difficulty growth.

Among these pools, \texttt{0x406177...} is the most informative case because multiple blocks are released simultaneously and the same pattern repeats several times on the main-chain. Such a pattern is more consistent with SUUM rather than RUM. Namely, since RUM omits post-mining risks, the timestamps of the withheld blocks in the private chain are not carefully manipulated and thus expect to be less \textit{detectable}. In contrast, SUUM requires more careful manipulation on the timestamps and thus exhibits a more clear pattern. 

\heading{A Case Study on Ethereum Classic (F2Pool).} Based on the observation above, we provide a closer look at the ETC pool \texttt{0x406177...} (publicly labeled as F2Pool \cite{Thepool0x406177...4e95ispubliclylabeledasF2Pool}).

First, the pool's timestamp calibration signature is extremely strong. Among 59,339 valid timestamp differences associated with this pool, the observed count for \(\Delta t \bmod 9 = 0\) is exactly 0, whereas the count for \(\Delta t \bmod 9 = 8\) is 9,883. 

Second, and more importantly, this pool exhibits repeated behaviors with the same pattern and a large number of blocks seem to be withheld and released together. Namely, the pool releases up to 21 consecutive blocks, occurring at heights \(23438117 \rightarrow 23438137\). Also, there is another streak of length 20 at heights \(23393481 \rightarrow 23393500\). In total, there are nine streaks of length at least 17. 

Note that under the null hypothesis that the pool \texttt{0x406177...} mined honestly with an external hash-share benchmark of \(\alpha=30.96\%\) \cite{ethereumclassichashrate}, the exact probability that a Bernoulli sequence of length 100,000 contains even one streak of length at least 21 is only
\begin{align}
\Pr(M_{100000}\ge 21 \mid \alpha=0.3096) \approx 1.399\times 10^{-6},
\end{align}
where \(M_n\) denotes the maximum run length. Hence, even a single 21-block streak is already highly implausible under honest mining with this share. Observing one 21-block streak, another 20-block streak, and nine streaks of length at least 17 makes a no-withholding explanation even less plausible.

\heading{Extra Profit.}
We calculated the extra profit obtained by these pools through timestamp manipulation during the statistical period, which can be expressed as
\(
\text{Extra Profit} = HP \cdot N \cdot V \cdot \left(\widehat{\mathbb{R}}_{\mathcal{P}_A}^{SUUM} - \widehat{\mathbb{R}}_{\mathcal{P}_A}^{HM}\right),
\)
where $HP$ denotes the mining pool's relative hash power, $N$ is the total number of blocks during the statistical period, $V$ is the market value of a single block, based on the closing price on the last day of the statistical period, and $\widehat{\mathbb{R}}_{\mathcal{P}_A}^{SUUM} - \widehat{\mathbb{R}}_{\mathcal{P}_A}^{HM}$ is the modeled increase in the adversary's normalized realized main-chain coinbase reward share relative to honest mining. By the steady-state reward relation established in the Preliminaries, this also corresponds to a proportional increase in absolute realized main-chain coinbase reward rate. The results show that pool \texttt{0x829bd8...} makes a total extra profit of \$2,122,778, as shown in \tabref{tab:unified_pool_statistics}.

\subsection{Impact of Uncle/Nephew Mechanism}
\heading{Impact on Profitability.} Historical statistics from Ethereum 1.0 indicate that uncle and nephew rewards were a non-trivial component of PoW issuance, accounting for roughly 6.6\% of total mining rewards \cite{bitmex2021ethsupply, eip649, eip1234}. This historical evidence suggests that SUUM can derive additional economic benefit from systematically inducing and harvesting uncle/nephew rewards, beyond ordinary main-chain block rewards.

\heading{Impact on $pu$.} Honest miners may reference uncles which can include orphaned blocks by setting \(pu = 1\) (\secref{Threat model}, Definition \ref{def:difficulty}). This increases the difficulty increment of their block by \(\lfloor \mathcal{D}_{i}^{p} / 2048 \rfloor\). Nevertheless, the attacker’s victory is still guaranteed under the precise timestamp calibration defined in Theorems~\ref{theorem2} and~\ref{SUUM Successful Condition}. The first adversarial block achieves a strict per-block advantage of at least \(\lfloor \mathcal{D}_{0}^{p} / 2048 \rfloor\). For later blocks, the calibration ensures that the attacker’s difficulty increment remains no lower than that of honest blocks, even when \(pu = 1\). Therefore, the additional advantage from an uncle reference cannot offset the attacker’s initial lead and continued calibration.

\subsection{Relative Loss in Mining Speed}
For an adversary with relative power \(\mu_{\mathcal{A}}\), a successful SUUM attack temporarily increases the difficulty of its own chain. Under minimal risk calibration, the difficulty increases by exactly one adjustment unit, i.e., \(\Delta\mathcal{D} = \lfloor \mathcal{D}_{\text{parent}} / 2048 \rfloor\) (Definition~\ref{def:difficulty}). Consequently, the adversary’s block generation rate is reduced by a factor of approximately \(\Delta\mathcal{D} / \mathcal{D}_{\text{parent}} \approx 1/2048\). This corresponds to a relative increase in the expected time to find the next block of about \(1/2048 \approx 0.0488\%\), independent of \(\mu_{\mathcal{A}}\). The absolute reduction in block generation rate is proportional to \(\mu_{\mathcal{A}}\) and, for \(\mu_{\mathcal{A}}=0.25\), amounts to roughly \(\mu_{\mathcal{A}} \cdot (1/2048) \approx 0.0122\%\) of the network’s total block production rate. This negligible slowdown is the inherent post-mining risk of the attack and is overwhelmingly outweighed by the substantial reward gain from the successful block reorganization.

\subsection{Stability Across Time Windows}
\label{sec:window_stability}
We partition the sampled chain segments into non-overlapping windows of 10,000 consecutive main-chain blocks and recompute the residue-level statistics within each window.

For each window, we define a \textit{directional staircase score} 
\(
S_w =\frac{N_{8,w}-N_{0,w}}{n_w},
\)
where \(N_{0,w}\) and \(N_{8,w}\) denote the counts of \((\Delta t \bmod 9)=0\) and \((\Delta t \bmod 9)=8\) in window \(w\), respectively, and \(n_w\) is the number of valid timestamp differences contributed by the pool in that window. A larger positive value indicates a stronger staircase-style distortion toward residue 8 and away from residue 0.

Table~\ref{tab:window_stability_summary} shows that the three suspicious pools exhibit consistently positive staircase scores across all windows: ETH is positive in \(12/12\) windows, ETHW in \(10/10\), and ETC in \(10/10\). Moreover, their median scores remain stably high, namely 0.179, 0.170, and 0.165, respectively. In contrast, the matched control pools stay much closer to zero, with substantially smaller median scores of 0.006, 0.010, and 0.002, and with several windows becoming negative. Thus, the staircase anomaly is not confined to a single burst or a narrow sub-interval, but persists repeatedly across time for the suspicious pools.

\subsection{Mitigation Measures} \label{Discussion}
We now analyze the essence of the observed timestamp manipulation behaviors, discuss their attribution limits, and propose some mitigation measures to prevent such attacks and their advanced variants. These measures mainly focus on adjusting the consensus mechanism and incentive mechanism of the blockchain, aiming to reduce the profit space of adversaries and enhance the security and fairness of the timestamp-based Nakamoto-style blockchain. We propose four primary mitigation measures: adjusting the difficulty adjustment algorithm, strengthening timestamp verification, introducing an incentive compatible response for uncles, and improving monitoring and response. The details can be found in \appref{Appendix Discussion}. 

\subsection{Generality of Our Attacks}\label{Generality}
\heading{Generality to Other Blockchains.} The SUUM attack is specifically enabled by the Ethereum 1.x difficulty adjustment mechanism, which updates difficulty per block using a deterministic function of the timestamp difference. This allows precise timestamp calibration to win forking races by achieving higher per-block difficulty while controlling long-term risk. Thus, our attacks are directly applicable to Ethereum 1.x and its derivatives (e.g., Ethereum Classic, EthereumPoW). In contrast, Bitcoin and other chains with periodic difficulty retargeting (e.g., every 2016 blocks) follow a fundamentally different model. Their future difficulty is calculated from the total mining time of the previous epoch, making individual block timestamp manipulation ineffective and delayed. Moreover, Bitcoin resolves forks by selecting the chain with the greatest cumulative proof-of-work—a metric independent of timestamp values. Consequently, the specific mechanics and profitability of UUM/SUUM attacks are unique to protocols featuring timestamp-based, per-block difficulty adjustment like Ethereum 1.x.

\heading{Generality to PoW using the Heaviest Chain Rule.}
An interesting fact is that our SUUM attack can be applied to the heaviest chain rule used in the PoW-based protocols of Expanse \cite{Expanse} and Musicoin \cite{Musicoin}. The heaviest chain rule selects the chain with the greatest cumulative proof-of-work (i.e., the highest total difficulty) as the main-chain. Upon release, the withheld private chain $\mathcal C_a$ has a strictly higher total difficulty than the competing public chain $\mathcal C_h$ of the same length. 

The proof leverages two properties derived from our minimal risk timestamp calibration. For the first block ($i=1$), the calibration ensures a strict per-block difficulty advantage $\mathcal D_1^{\mathcal P_a} - \mathcal D_1^{\mathcal P_h} = \lfloor \mathcal D_0 / 2048 \rfloor > 0$. For each subsequent block ($i \geq 2$), the calibration ensures $\mathcal D_i^{\mathcal P_a} \geq \mathcal D_i^{\mathcal P_h}$. Crucially, because $\mathcal D_{i-1}^{\mathcal P_a} > \mathcal D_{i-1}^{\mathcal P_h}$, the term $\lfloor \mathcal D_{i-1}^{\mathcal P_a} / 2048 \rfloor$ in the difficulty adjustment formula is greater than or equal to its counterpart on the honest chain. This means the per-block difficulty on the attacker's chain never falls behind, and the initial advantage is preserved and compounded. 

Therefore, by induction: $\sum_{i=1}^{k} \mathcal D_i^{\mathcal P_a} = \mathcal D_1^{\mathcal P_a} + \sum_{i=2}^{k} \mathcal D_i^{\mathcal P_a} > \mathcal D_1^{\mathcal P_h} + \sum_{i=2}^{k} \mathcal D_i^{\mathcal P_h} = \sum_{i=1}^{k} \mathcal D_i^{\mathcal P_h}$.
This inequality holds strictly because of the guaranteed positive difference at $i=1$. Consequently, under a heaviest-chain fork-choice rule, all honest miners will adopt the attacker's chain immediately upon its release.

\section{Conclusion}\label{Conclusion}

We study timestamp-based Nakamoto-style blockchains and present two new attacks: unrestricted uncle maker (UUM) and staircase-unrestricted uncle maker (SUUM). Compared to prior attacks such as riskless uncle maker (RUM) attack, our attacks expand the scope of risk in the sense that we not only ensure the success of the attacks but also make sure that they are more profitable. Our experimentation and analysis of on-chain data from three blockchain systems validate that it is highly likely that some mining pools are launching SUUM-style attacks and earning an estimated additional profit of \$2,961,078 in total, compared to adhering to the protocol. We responsibly disclosed our findings to the Ethereum Foundation team.

\cleardoublepage
\appendix
\section*{Ethical Considerations}
\textbf{We are committed to complying with all relevant research ethics guidelines, and this study involves no ethical risks. The research content is not associated with animal experimentation, human subjects, environmental protection, medical health, or military applications. We confirm full adherence to all ethical principles outlined in the Call for Papers (CFPs). We pledge to proactively address any ethical concerns that may arise during the review process or post-publication, welcome academic feedback regarding research ethics norms, and will refine the research protocol based on such input.}

\section*{Open Science}
\textbf{We fully endorse and strictly adhere to the Open Science Policy outlined in the Call for Papers (CFP). The formal proofs for all theorems and lemmas presented in this paper are provided in the Appendix. To facilitate reproducibility and independent validation, we provide an anonymized implementation of our analysis and evaluation framework at:
\url{https://anonymous.4open.science/r/Timestamp-Manipulation-796A}. We hereby grant the scientific community unrestricted rights to review, validate, and build upon this research work.}

\cleardoublepage

\bibliographystyle{plainurl}
\bibliography{Ref}

\section{Related Work}\label{Related work}
This appendix reviews related work on incentive and consensus attacks in Nakamoto-style and Ethereum-2.x-style blockchains, primarily focusing on three aspects: attacks targeting vulnerabilities in incentive models and consensus protocols of Nakamoto-like and Ethereum 2.x-like blockchains.

\subsection{Attacks on Incentive Model and Consensus Protocol of Nakamoto-like Blockchains}
\heading{Withholding Attack.}
Withholding attacks are prevalent malicious behaviors in Nakamoto-like blockchains \cite{Majorityisnotenough, OptimalSelfishMiningStrategiesinBitcoin, AnanalyticevaluationfortheimpactofuncleblocksbyselfishandstubbornmininginanimperfectEthereumnetwork, TheMiner'sDilemma, BeSelfishandAvoidDilemmas:ForkAfterWithholding(FAW)AttacksonBitcoin, OnSubversiveMinerStrategiesandBlockWithholdingAttackinBitcoinDigitalCurrency, TheGapGame, SelfishMininginEthereum, thehaltgame, greedy-mine, HowtoBeatNakamotointheRace, hu2026blockchainattacksdefenseslayered}. Most consensus-layer attacks (e.g., selfish mining, stubborn mining) rely on block withholding: adversaries withhold blocks to control release timing or continue mining on lagging selfish forks \cite{AnanalyticevaluationfortheimpactofuncleblocksbyselfishandstubbornmininginanimperfectEthereumnetwork, StubbornMining:GeneralizingSelfishMiningandCombiningwithanEclipseAttack}, undermining network fairness, wasting honest nodes’ computing resources, and reducing network security and reliability.

\heading{Bribing Attack.}
Bribing attacks pose severe threats to the security and stability of Nakamoto-like blockchains \cite{PaytoWin:CheapCross-ChainBribingAttacksonPoWCryptocurrencies, WhyBuyWhenYouCanRent, PowerAdjustingandBriberyRacing:NovelMiningAttacksintheBitcoinSystem, IfYouCantBeatThemPayThemBitcoinProtectionRacketisProfitable, BlockchainBribingAttacksandtheEfficacyofCounterincentives, novalbriberyminingattacks, BM-PAW, 11316192}. First proposed by Bonneau, these attacks include multiple mechanisms (e.g., inbound/outbound bribery via smart contracts). Their motives range from double-spending and consensus disruption to censorship and HTLC sabotage \cite{SuboptimalityinDeFi, SoKDecentralizedFinance(DeFi)Attacks}, significantly disrupting blockchain operations.

\heading{DoS attack.}
Denial of Service (DoS) attacks disrupt system availability by consuming network resources \cite{BDoS:BlockchainDenial-of-Service, OptimalSelfishMining-BasedDenial-of-ServiceAttack, SDoS:SelfishMining-BasedDenial-of-ServiceAttack, DETER:DenialofEthereumTxpoolsERvices, ding2025asymmetric}. At the network layer, adversaries often launch traffic flooding attacks (e.g., fake requests) to exhaust bandwidth, hindering legitimate transaction processing, increasing confirmation latency, and weakening blockchain decentralization.

\heading{Timestamp Attack.}
Timestamp manipulation attacks exploit configurable block timestamps in Nakamoto-like blockchains \cite{BlockchainStretchingSqueezingManipulatingTimeforYourBestInterest,Unclemakertimestampingoutthecompetitioninethereum}, interfering with block generation/confirmation or manipulating difficulty adjustments. Research on timestamp attacks at the consensus layer remains limited: while Stretch identified timestamp vulnerabilities in Geth (e.g., reducing honest mining difficulty), the impact on consensus mechanisms was not explored. Note that the UUM attack proposed in this paper belongs to timestamp attacks, and SUUM combines withholding attacks with timestamp attacks.

\heading{Consensus, Network, and Economic Studies.}
Related work studies Nakamoto consensus under asynchronous networks, deterministic PoW safety, volatile block rewards, and MEV-aware consensus design~\cite{HashSplit,FastDeterministicPoW2026,Bitcoinundervolatileblockrewards,MADDAG2026}. Other studies examine Ethereum's builder market, elastic restaking, execution uncertainty, synchronization attacks, and validator-network privacy~\cite{EthereumBuilderMarket2025,ElasticRestakingNetworks2025,PerilsParallelism2026,SnappingSnapSync2023,DeanonymizingEthereumValidators2025}.

\subsection{Attacks on Incentive Model and Consensus Protocol of Ethereum 2.x-like Blockchains}
\heading{Withholding Attack.}
In Ethereum 2.x-like PoS blockchains, withholding attacks threaten network stability \cite{MaxAttestationMattersMakingHonestPartiesLoseTheirIncentivesinEthereumPoS, Ebb-and-FlowProtocols:AResolutionoftheAvailabilityFinalityDilemma, TwoAttacksOnProof-of-StakeGHOST/Ethereum, SelfishBehaviorintheTezosProof-of-StakeProtocol}. Adversaries withhold blocks to delay block propagation and release them opportunistically, exploiting staking/voting mechanisms to disrupt block confirmation, affect block height/forks, and undermine network consistency.

\heading{Nothing-at-stake Attack.} 
Nothing-at-stake attacks severely threaten the activity and performance of Ethereum 2.x-like PoS blockchains \cite{FormalBarrierstoLongestChainProofofStakeProtocols, Ouroboros, SecuringProof-of-StakeBlockchainProtocols}. Leveraging PoS staking/validation flaws, adversaries participate in multiple forks without risk, causing fork chaos, hindering consensus unification, and reducing transaction processing efficiency.

\heading{Long-range Attack.}
Long-range attacks are potential threats to Ethereum 2.x-like PoS blockchains \cite{KeepYourTransactionsOnShortLeashes, Ouroboros, Pikachu}. Adversaries secretly construct alternative chains from historical blockchain points (with low resource consumption), undermining data consistency (e.g., altering past transactions) and eroding the trust foundation of PoS blockchains.

\heading{DoS Attack.}
DoS attacks on Ethereum 2.x-like PoS blockchains take multiple forms \cite{TowardsAutomaticDiscoveryofDenialofServiceWeaknessesinBlockchainResourceModels, UnderstandingEthereumMempoolSecurityunderAsymmetricDoSbySymbolizedStatefulFuzzing, SpeculativeDenial-of-ServiceAttacksInEthereum, Nurgle}. Network-layer attacks involve traffic flooding, while application-layer attacks target smart contracts (e.g., infinite loops) and Txpool (e.g., low-gas high-complexity transactions), causing resource exhaustion or congestion.

\heading{Reorg Attack.}
Reorg attacks threaten the security and stability of Ethereum 2.x-like PoS blockchains \cite{MaxAttestationMattersMakingHonestPartiesLoseTheirIncentivesinEthereumPoS, ThreeAttacksonProof-of-StakeEthereum, li2025bunnyfinder, li2025bunnyfinder, 10.5555/3766078.3766144, nagy2025forking}. Adversaries increase the proportion of Byzantine validator blocks via controlled nodes (including Short/Long Reorg Attacks), replacing the main-chain, making transaction states uncertain, and weakening user trust.

\heading{Liveness and Attestation Defenses.}
Recent work studies accountable liveness, attestation incentives,
reorganization defenses, and liveness attacks that require no additional
adversarial cost~\cite{AccountableLiveness2025,MaxAttestationMattersMakingHonestPartiesLoseTheirIncentivesinEthereumPoS,10.5555/3766078.3766144,EthereumPoSLivenessNoCost2026}.
These studies show that attestation rules affect not only validator
incentives but also the protocol's resistance to chain reorganizations
and prolonged loss of liveness. They further motivate accountable and
reorganization-resilient mechanisms that can identify misbehavior,
preserve honest participation incentives, and restore progress after
adversarial disruption.


\section{Different types of rewards}\label{Different types of rewards}
\begin{definition}[\textbf{Block Reward}]\label{Block Reward}
    Blocks are classified into main-chain blocks and non-main-chain blocks, as shown in \figref{fig:1}.
\begin{enumerate}[(1)]
    \item A participant $p$ will receive mining rewards $\mathbb{R}_{c}$, nephew rewards $\mathbb{R}_{n}( \cdot )$, transaction fee rewards $\mathbb{R}_{g}( \cdot )$, and whale rewards $\mathbb{R}_{w}( \cdot )$ \cite{10179444, thehaltgame} when they successfully publish a block $\mathcal{B}_{i}$ with a timestamp $t_{i}$ that is ultimately selected as a main-chain block. The calculation of each type of reward is as follows:

\begin{enumerate}
\item[$\bullet$] $\mathbb{R}_{c} \overset{def}{=} 2$;
\item[$\bullet$] $\mathbb{R}_{n}(m) \overset{def}{=} m \cdot \frac{1}{32} \cdot
    \mathbb{R}_{c} = m \cdot \frac{1}{16}$, where $m \leq 2$ indicates the number of uncle blocks referenced by block $\mathcal{B}_{i}$;
\item[$\bullet$] $\mathbb{R}_{g}\left( t_{i} \right) \overset{def}{=} \lambda \cdot
    \mathrm{\Delta}t_{i}$, where $\lambda$ represents the accumulation rate of transaction fee rewards and $\mathrm{\Delta}t_{i} \overset{def}{=} t_{i} - t_{i}^{p} = t_{i} - t_{i - 1}$.
\end{enumerate}
    \item A participant $p$ will receive uncle rewards $\mathbb{R}_{u}( \cdot )$ when they successfully publish a block that is ultimately not selected as a main-chain block but is referenced as an uncle block by a main-chain block. The calculation of uncle rewards $\mathbb{R}_{u}( \cdot )$ is as follows:
\begin{enumerate}
\item[$\bullet$] $\mathbb{R}_{u}\left( d_{i} \right) \overset{def}{=} \left\{ \begin{matrix}
{\frac{8 - d_{i}}{8} \cdot 2,} & {d_{i} \in N^{+} ~\text{and}~1 \leq d_{i} \leq 6} \\
{0,} & {otherwise}
\end{matrix} \right.$, where $d_{i}$ represents the generational distance between the uncle block and the main-chain block that references it.
\end{enumerate}
\end{enumerate}
\end{definition}

\section{Methods of RUM Attacks}\label{Methods of RUM Attacks}
\begin{enumerate}
    \item[$\bullet$] In the Deployment Phase, the adversary monitors blocks generated by honest participants and waits for the timestamp difference between a new honest block $\mathcal{B}_{1}^{{p}_{{h}}}$ and the previous main-chain block $\mathcal{B}_{0}^{{p}_{{h}}}$ to fall within the interval \(t_1^{p_h} - t_0^{p_h} \in [9, 18)\) seconds (i.e., \(\lfloor (t_{1}^{p_h} - t_{0}^{p_h}) / 9 \rfloor = 1\). By the difficulty adjustment formula, the term $\lfloor \frac{t_i - t_{i-1}}{9} \rfloor$ evaluates the deviation of block time from the target block time (Definition \ref{def:difficulty}). If the timestamp difference is less than 9 seconds, the block difficulty increases. If it exceeds 18 seconds, the difficulty decreases. Thus, the interval of $[9, 18)$ ensures that \(\lfloor (t_{1}^{p_h} - t_{0}^{p_h}) / 9 \rfloor = 1\), i.e., the honest block's difficulty equals that of its parent block. In this way, the adversary gains an advantage in the forking race, and the attack has zero pre-mining risk.
    \item[$\bullet$] In the Execution Phase, the adversary mines locally on the same parent block \(\mathcal{B}_{0}^{p_h}\), attempting to generate the next block \(\mathcal{B}_{1}^{p_a}\) before honest miners. If the adversary succeeds in mining \(\mathcal{B}_{1}^{p_a}\) first, the timestamp \(t_{1}^{p_a}\) is set as:
\(
1 \leq t_{1}^{p_a} - t_{0}^{p_h} < 9 \text{ seconds}\), i.e., \( \lfloor (t_{1}^{p_a} - t_{0}^{p_h}) / 9 \rfloor = 0).
\)
By Definition~\ref{def:difficulty}, when \(\lfloor (t_{1}^{p_a} - t_{0}^{p_h}) / 9 \rfloor = 0\), the block difficulty increases.
Meanwhile, the timestamp of the competing honest block \(\mathcal{B}_{1}^{p_h}\) satisfies \(\lfloor (t_{1}^{p_h} - t_{0}^{p_h}) / 9 \rfloor = 1\), i.e., the difficulty is the same as its parent block \(\mathcal{B}_{0}^{p_h}\).
Consequently, the adversary's block \(\mathcal{B}_{1}^{p_a}\) has a higher difficulty than the honest block, i.e., \(\mathcal{D}_{1}^{p_a} > \mathcal{D}_{1}^{p_h}\).
According to the fork selection rule (Example~\ref{example1}), honest miners select the chain led by block \(\mathcal{B}_{1}^{p_a}\), and block \(\mathcal{B}_{1}^{p_h}\) is then orphaned. 
\end{enumerate}

\begin{figure}[t]
  \centering
  \includegraphics[width=\linewidth]{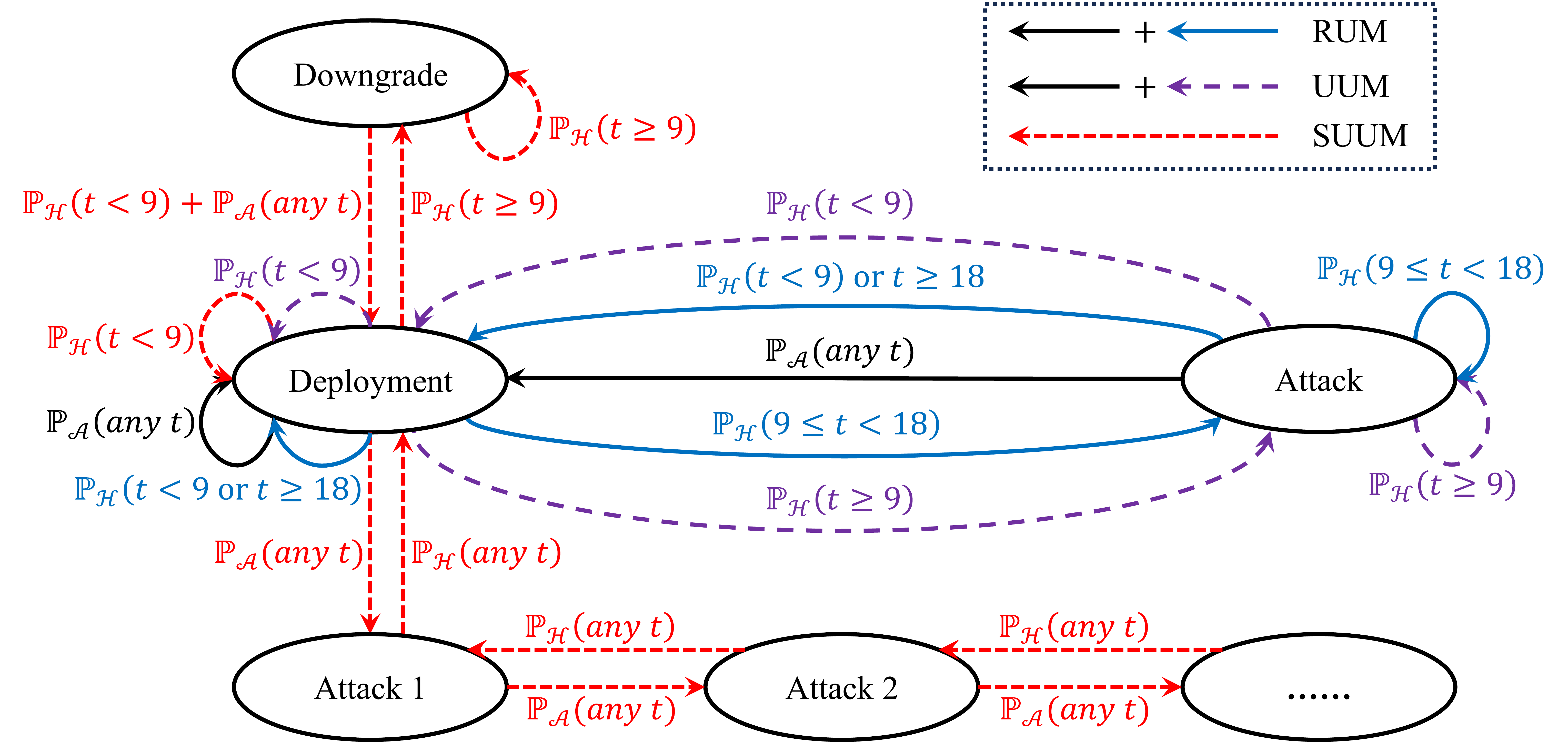}
  \caption{State Transition Process. This figure illustrates the state transition process under different mining strategies. It delineates the dynamic transformation relationships among different states in the blockchain system. \textcolor{black}{Black} + \textcolor{cyan}{Cyan} transitions denote RUM \cite{Unclemakertimestampingoutthecompetitioninethereum}. \textcolor{black}{Black} + \textcolor{violet}{Violet} transitions denote UUM. \textcolor{red}{Red} transitions denote SUUM. Compared to RUM, UUM relaxes the triggering condition to increase the probability of launching the attack and make the attack more profitable. Compared to UUM, SUUM further provides a more fine-grained control on whether more blocks can be withheld. Via SUUM, more blocks can be withheld while incurring only negligible additional pre-mining risk, which makes the attack even more profitable.
  }
  \label{State Transition Process}
\end{figure}

\section{Methods of UUM Attacks}\label{section4.1}
\begin{enumerate}
    \item[$\bullet$] In the Deployment Phase, similar to the RUM attack, the adversary still monitors the timestamp difference between a newly generated honest block $\mathcal{B}_1^{p_h}$ and its preceding main-chain block $\mathcal{B}_0^{p_h}$. Different from the RUM attack that sets a condition as $t_1^{p_h} - t_0^{p_h} \in [9, 18)$, our UUM attack relaxes it to $t_1^{p_h} - t_0^{p_h} \in [9, +\infty)$, which corresponds to \(\lfloor (t_{1}^{p_h} - t_{0}^{p_h}) / 9 \rfloor \geq 1\). As briefly mentioned in \secref{Our Insight}, such relaxation can greatly improve the probability of launching an attack.
    

    
    \item[$\bullet$] In the Execution Phase, the adversary mines on the parent block of the latest main-chain block. If the adversary generates a valid block $\mathcal{B}_{1}^{p_a}$ before honest participants, and the timestamp difference $t_{1}^{p_a} - t_{0}^{p_a}$ lies in $[1, 900)$, the attack progresses successfully to the Risk Control Phase. The range $[1,900)$, as we later show in \thmref{theorem2}, ensures $\mathcal{D}_{1}^{p_a} > \mathcal{D}_{0}^{p_h}$, i.e., the adversarial chain has an advantage in the forking race. 

    \item[$\bullet$] In the Minimal Risk Control Phase, the adversary adjusts the timestamp $t_{a}^{p_1}$ of its block. The goal is to obtain the smallest difficulty advantage sufficient to win the fork race while minimizing subsequent difficulty growth. Specifically, $\lfloor (t_1^{p_h} - t_0^{p_h})/9 \rfloor-\lfloor (t_1^{p_a} - t_0^{p_h})/9 \rfloor = 1~\text{and}~t_1^{p_a} - t_0^{p_a} \in [1, 900)$. This is because the difficulty of the withheld block can be made the same as the parent block $\mathcal{B}_0^{p_h}$. In this way, the difficulty of subsequent blocks will not be affected, according to Definition~\ref{def:difficulty}. The details can be found in Figure \ref{State Transition Process} and Table \ref{table2}.
\end{enumerate}

\section{Proof of Theorem \ref{theorem1}}\label{Proof of Theorem 1}
The UUM attack is initiated if and only if the difficulty of the newly mined honest block \(\mathcal B_1^{p_h}\) is no greater than the difficulty of its parent block \(\mathcal B_0^{p_h}\) (i.e., \(\mathcal D_1^{p_h} \leq \mathcal D_0^{p_h}\)). This is because a smaller or equal difficulty of \(\mathcal B_1^{p_h}\) creates a window for the adversary to manipulate timestamps and gain a difficulty advantage in the forking race. We prove the necessity and sufficiency separately, building on Definition \ref{def:difficulty}.

For consistency, define: \(\mathcal D_0 = \mathcal D_0^{p_h}\) (difficulty of the parent block \(\mathcal B_0^{p_h}\)); \(k_h = \lfloor\frac{t_1^{p_h} - t_0^{p_h}}{9}\rfloor\) (floor division of the timestamp difference between \(\mathcal B_1^{p_h}\) and \(\mathcal B_0^{p_h}\) by 9). From Definition \ref{def:difficulty}, the difficulty of \(\mathcal B_1^{p_h}\) is:
  \(
  \mathcal D_1^{p_h} = \mathcal D_0 + \max\left\{1 - k_h, -99\right\} \cdot \lfloor\frac{\mathcal D_0}{2048}\rfloor.
  \)
Note that for Ethereum 1.x blocks after height 15, \(\mathcal D_0 > 2^{17}\), so \(\lfloor\frac{\mathcal D_0}{2048}\rfloor > 0\) (this positive scaling factor preserves inequality directions in derivations).

\noindent\textit{Necessity}: Assume the UUM attack is initiated, so \(\mathcal D_1^{p_h} \leq \mathcal D_0\). Substitute the difficulty formula into this inequality:
\begin{align}
\mathcal D_0 + \max\left\{1 - k_h, -99\right\} \cdot \lfloor\frac{\mathcal D_0}{2048}\rfloor \leq \mathcal D_0.
\end{align}
Subtract \(\mathcal D_0\) from both sides:
\begin{align}
\max\left\{1 - k_h, -99\right\} \cdot \lfloor\frac{\mathcal D_0}{2048}\rfloor \leq 0.
\end{align}
Since \(\lfloor\frac{\mathcal D_0}{2048}\rfloor > 0\), we can divide both sides by this term without changing the inequality direction:
\begin{align}
\max\left\{1 - k_h, -99\right\} \leq 0.
\end{align}
This inequality holds if and only if \(1 - k_h \leq 0\) (because \(-99 \leq 0\), but \(1 - k_h\) is the relevant term for practical block timestamps—extremely large timestamp differences leading to \(1 - k_h = -99\) are rare but still satisfy the condition). Solving \(1 - k_h \leq 0\) gives \(k_h \geq 1\), i.e., \(\lfloor\frac{t_1^{p_h} - t_0^{p_h}}{9}\rfloor \in [1, +\infty)\).

\begin{table}[t]
\caption{UUM state transition.}
\label{table2}
\centering
\begin{tabular}{@{}|c|c|c|cc|@{}}
\toprule
\multirow{2}{*}{State} & \multirow{2}{*}{Destination} & \multirow{2}{*}{Transition}  & \multicolumn{2}{c|}{Reward} \\ \cmidrule(l){4-5} 
                       &                                                           &                              & \multicolumn{1}{c|}{$
\mathcal{P}_{\mathcal{H}}$}   &  $
\mathcal{P}_{\mathcal{A}}$ \\ \midrule
Deployment & Deployment & $\mathcal{P}_{\mathcal{A}}~any~t$ &  \multicolumn{1}{c|}{$0$}  & $1$  \\ \midrule
Deployment & Deployment & $\mathcal{P}_{\mathcal{H}}~t < 9$ &  \multicolumn{1}{c|}{$1$}  & $0$  \\ \midrule
Deployment & Attack & $\mathcal{P}_{\mathcal{H}}~t \geq 9$ &  \multicolumn{1}{c|}{$1$}  & $0$  \\ \midrule
Attack & Attack & $\mathcal{P}_{\mathcal{H}}~t \geq 9$ &  \multicolumn{1}{c|}{$1$}  & $0$  \\ \midrule
Attack & Deployment & $\mathcal{P}_{\mathcal{H}}~t < 9$ &  \multicolumn{1}{c|}{$1$}  & $0$  \\ \rowcolor{gray!40} \midrule
Attack & Deployment & $\mathcal{P}_{\mathcal{A}}~any~t$ &  \multicolumn{1}{c|}{$-1$} & $1$  \\ \bottomrule
\end{tabular}
\begin{tablenotes} 
\setlength{\itemindent}{0pt} 
\setlength{\rightskip}{10pt}   
\item* \noindent 
\footnotesize
The state transition represented by the gray row (the sixth row) is the fundamental cause of UUM adversaries gaining an advantage over RUM and honest participants.
\end{tablenotes}
\end{table}

\noindent\textit{Sufficiency}: Assume \(\lfloor\frac{t_1^{p_h} - t_0^{p_h}}{9}\rfloor \in [1, +\infty)\), so \(k_h \geq 1\). Substitute \(k_h \geq 1\) into the difficulty formula of \(\mathcal B_1^{p_h}\):
\begin{align}
\max\left\{1 - k_h, -99\right\} = 1 - k_h.
\end{align}
Thus:
\begin{align}
\mathcal D_1^{p_h} = \mathcal D_0 + (1 - k_h) \cdot \lfloor\frac{\mathcal D_0}{2048}\rfloor.
\end{align}
Since \(k_h \geq 1\), \(1 - k_h \leq 0\). Multiplying this non-positive term by the positive scaling factor \(\lfloor\frac{\mathcal D_0}{2048}\rfloor\) yields a non-positive value:
\begin{align}
(1 - k_h) \cdot \lfloor\frac{\mathcal D_0}{2048}\rfloor \leq 0.
\end{align}
Adding this to \(\mathcal D_0\) gives:
\begin{align}
\mathcal D_1^{p_h} = \mathcal D_0 + \text{non-positive term} \leq \mathcal D_0.
\end{align}
This satisfies the core requirement for UUM initiation (\(\mathcal D_1^{p_h} \leq \mathcal D_0\)), creating a valid attack window for the adversary to manipulate timestamps and gain a difficulty advantage.

Thus, the UUM attack is initiated if and only if \(\lfloor\frac{t_1^{p_h} - t_0^{p_h}}{9}\rfloor \in [1, +\infty)\).

\section{Proof of Theorem \ref{theorem2}}\label{Proof of Theorem 2}
We prove the necessity and sufficiency separately:

\noindent\textit{Necessity}: For the UUM attack to succeed, two core requirements must be satisfied: (1) the adversary’s block \(\mathcal B_1^{p_a}\) is valid; (2) the difficulty of \(\mathcal B_1^{p_a}\) is strictly greater than that of the honest block \(\mathcal B_1^{p_h}\) (i.e., \(\mathcal D_1^{p_a} > \mathcal D_1^{p_h}\)) to win the forking race (Example \ref{example1}).

1. \textit{Validity of } \(\mathcal B_1^{p_a}\): By the blockchain protocol, a block is valid only if its timestamp is strictly greater than its parent block’s timestamp. Since \(\mathcal B_1^{p_a}\) shares the same parent block \(\mathcal B_0^{p_h}\) as \(\mathcal B_1^{p_h}\) (i.e., \(t_0^{p_a} = t_0^{p_h}\)), we require \(t_1^{p_a} > t_0^{p_a}\), which implies \(t_1^{p_a} - t_0^{p_a} \geq 1\) (timestamps are integer-valued in practice).

2. \textit{Difficulty Superiority } \(\mathcal D_1^{p_a} > \mathcal D_1^{p_h}\): From Definition \ref{def:difficulty}, the difficulty of a block is given by:
\begin{align}
\mathcal{D}_i = \mathcal{D}_i^p + \max\left\{1 + pu_i - \lfloor\frac{t_i - t_{i-1}}{9}\rfloor, -99\right\} \cdot \lfloor\frac{\mathcal{D}_i^p}{2048}\rfloor.
\end{align}
For \(\mathcal B_1^{p_h}\) and \(\mathcal B_1^{p_a}\), \(pu_1 = 0\) (no uncle blocks referenced in the first attack round), and they share the same parent difficulty \(\mathcal D_0^{p_h} = \mathcal D_0^{p_a}\) (denoted as \(\mathcal D_0\) for simplicity). Substituting into the difficulty formula:
\begin{align}
\mathcal D_1^{p_h} = \mathcal D_0 + \max\left\{1 - \lfloor\frac{t_1^{p_h} - t_0^{p_h}}{9}\rfloor, -99\right\} \cdot \lfloor\frac{\mathcal D_0}{2048}\rfloor,
\end{align}
\begin{align}
\mathcal D_1^{p_a} = \mathcal D_0 + \max\left\{1 - \lfloor\frac{t_1^{p_a} - t_0^{p_h}}{9}\rfloor, -99\right\} \cdot \lfloor\frac{\mathcal D_0}{2048}\rfloor.
\end{align}
Let \(k_h = \lfloor\frac{t_1^{p_h} - t_0^{p_h}}{9}\rfloor\) and \(k_a = \lfloor\frac{t_1^{p_a} - t_0^{p_h}}{9}\rfloor\). Since \(\mathcal D_0 > 2^{17}\) (Ethereum 1.x blocks after height 15), \(\lfloor\frac{\mathcal D_0}{2048}\rfloor > 0\). For \(\mathcal D_1^{p_a} > \mathcal D_1^{p_h}\), we need:
\begin{align}
\max\left\{1 - k_a, -99\right\} > \max\left\{1 - k_h, -99\right\}.
\end{align}
This inequality holds if and only if \(1 - k_a > 1 - k_h\) (since \(k_h, k_a\) are non-negative integers, \(1 - k_h\) and \(1 - k_a\) are greater than -99 for practical block generation times). Simplifying gives \(k_h - k_a \geq 1\), i.e., \(\lfloor\frac{t_1^{p_h} - t_0^{p_h}}{9}\rfloor - \lfloor\frac{t_1^{p_a} - t_0^{p_h}}{9}\rfloor \in [1, +\infty)\).

3. \textit{Upper Bound on Timestamp Difference}: To avoid excessive difficulty reduction in subsequent blocks (Definition \ref{def:difficulty}), the protocol implicitly restricts the timestamp difference \(t_1^{p_a} - t_0^{p_a}\) to be less than 900 seconds. If \(t_1^{p_a} - t_0^{p_a} \geq 900\), \(\lfloor\frac{t_1^{p_a} - t_0^{p_a}}{9}\rfloor \geq 100\), leading to \(\max\left\{1 - k_a, -99\right\} = -99\) and an uncontrollable drop in difficulty, which invalidates the attack’s stability. Thus, \(t_1^{p_a} - t_0^{p_a} < 900\).

Combining the above, the necessary conditions are \(\lfloor\frac{t_1^{p_h} - t_0^{p_h}}{9}\rfloor - \lfloor\frac{t_1^{p_a} - t_0^{p_h}}{9}\rfloor \in [1, +\infty)\) and \(t_1^{p_a} - t_0^{p_a} \in [1, 900)\).

\noindent\textit{Sufficiency}: Assume the two conditions hold. We need to show \(\mathcal D_1^{p_a} > \mathcal D_1^{p_h}\).

From \(k_h - k_a \geq 1\), we have \(k_a \leq k_h - 1\). Substituting into the difficulty formula:
\begin{align}
1 - k_a \geq 1 - (k_h - 1) = 2 - k_h.
\end{align}
Since \(k_h = \lfloor\frac{t_1^{p_h} - t_0^{p_h}}{9}\rfloor \geq 1\) (UUM initiation condition, Theorem \ref{theorem1}), \(2 - k_h \geq 1 - k_h\). Thus:
\begin{align}
\max\left\{1 - k_a, -99\right\} \geq 2 - k_h > 1 - k_h = \max\left\{1 - k_h, -99\right\}.
\end{align}
Since \(\lfloor\frac{\mathcal D_0}{2048}\rfloor > 0\), multiplying both sides by this term preserves the inequality:
\begin{align}
\max\left\{1 - k_a, -99\right\} \cdot \lfloor\frac{\mathcal D_0}{2048}\rfloor > \max\left\{1 - k_h, -99\right\} \cdot \lfloor\frac{\mathcal D_0}{2048}\rfloor.
\end{align}
Adding \(\mathcal D_0\) to both sides gives \(\mathcal D_1^{p_a} > \mathcal D_1^{p_h}\). By Example \ref{example1}, honest miners will select \(\mathcal B_1^{p_a}\) as the main-chain block, and the UUM attack succeeds.

\section{Proof of Theorem \ref{theorem3}}\label{Proof of Theorem 3}
We build on Theorem \ref{theorem2} and formalize the minimal post-mining risk criterion, proving the necessity and sufficiency separately:

\noindent\textit{Necessity}: Assume the UUM attack succeeds with minimal post-mining risk. By Theorem \ref{theorem2}, the attack’s success requires \(\lfloor\frac{t_{1}^{p_h} - t_{0}^{p_h}}{9}\rfloor - \lfloor\frac{t_{1}^{p_a} - t_{0}^{p_h}}{9}\rfloor \in [1, +\infty)\) and \(t_{1}^{p_a} - t_{0}^{p_a} \in [1, 900)\). We now show \(\lfloor\frac{t_{1}^{p_h} - t_{0}^{p_h}}{9}\rfloor - \lfloor\frac{t_{1}^{p_a} - t_{0}^{p_h}}{9}\rfloor = 1\) is mandatory for minimal risk.

First, define \(k_h = \lfloor\frac{t_1^{p_h} - t_0^{p_h}}{9}\rfloor\) and \(k_a = \lfloor\frac{t_1^{p_a} - t_0^{p_h}}{9}\rfloor\). Post-mining risk (denoted \(MR\)) refers to the long-term growth of blockchain difficulty induced by the attack, which depends directly on the difficulty of the adversary’s block \(\mathcal B_1^{p_a}\) (as \(\mathcal B_1^{p_a}\) becomes the main-chain block and influences subsequent difficulty adjustments). From Definition \ref{def:difficulty}, the difficulty of \(\mathcal B_1^{p_a}\) is:
\begin{align}
\mathcal D_1^{p_a} = \mathcal D_0 + \max\left\{1 - k_a, -99\right\} \cdot \lfloor\frac{\mathcal D_0}{2048}\rfloor,
\end{align}
where \(\mathcal D_0 = \mathcal D_0^{p_h} = \mathcal D_0^{p_a}\) and \(\max\left\{1 - k_a, -99\right\} = 1 - k_a\) for practical \(k_a\) (since block generation times rarely exceed 891 seconds, \(k_a \leq 99\)). A smaller \(\mathcal D_1^{p_a}\) reduces the incremental difficulty passed to subsequent blocks, thus minimizing \(MR\).

For the attack to succeed, \(k_h - k_a = m \geq 1\) (Theorem \ref{theorem2}). If \(m \geq 2\), then \(k_a = k_h - m\), and substituting into \(\mathcal D_1^{p_a}\)’s formula gives:
\begin{align}
\mathcal D_1^{p_a} &= \mathcal D_0 + \left(1 - (k_h - m)\right) \cdot \lfloor\frac{\mathcal D_0}{2048}\rfloor \\
&= \mathcal D_0 + \left((1 - k_h) + m\right) \cdot \lfloor\frac{\mathcal D_0}{2048}\rfloor.
\end{align}
Since \(m \geq 2\), \(\mathcal D_1^{p_a}\) is strictly larger than when \(m = 1\), leading to a higher \(MR\) (as subsequent block difficulty \(\mathcal{D}_2\) increases with \(\mathcal D_1^{p_a}\)). This contradicts the minimal risk requirement. When \(m = 1\), \(k_a = k_h - 1\), and:
\(
\mathcal D_1^{p_a} = \mathcal D_0 + \left(1 - (k_h - 1)\right) \cdot \lfloor\frac{\mathcal D_0}{2048}\rfloor = \mathcal D_0 + \left(2 - k_h\right) \cdot \lfloor\frac{\mathcal D_0}{2048}\rfloor,
\)
which is the smallest possible \(\mathcal D_1^{p_a}\) that still satisfies \(\mathcal D_1^{p_a} > \mathcal D_1^{p_h}\). Additionally, the validity condition \(t_1^{p_a} - t_0^{p_a} \in [1, 900)\) is retained, as it ensures block validity and controlled difficulty adjustment, which are prerequisites for minimal risk (uncontrolled difficulty drops would introduce additional risk).

\noindent\textit{Sufficiency}: Assume \(\lfloor\frac{t_{1}^{p_h} - t_{0}^{p_h}}{9}\rfloor - \lfloor\frac{t_{1}^{p_a} - t_{0}^{p_h}}{9}\rfloor = 1\) and \(t_{1}^{p_a} - t_{0}^{p_a} \in [1, 900)\). We first confirm the attack succeeds, then verify minimal post-mining risk.

1. \textit{Attack Success}: By Theorem \ref{theorem2}, these conditions satisfy the success criteria: \(k_h - k_a = 1 \geq 1\) ensures \(\mathcal D_1^{p_a} > \mathcal D_1^{p_h}\) (difficulty superiority), and \(t_1^{p_a} - t_0^{p_a} \in [1, 900)\) ensures \(\mathcal B_1^{p_a}\) is valid. Honest miners will select \(\mathcal B_1^{p_a}\) as the main-chain block (Example \ref{example1}), so the UUM attack succeeds.

2. \textit{Minimal post-mining risk}: As shown in the necessity proof, \(k_h - k_a = 1\) yields the smallest possible \(\mathcal D_1^{p_a}\) among all valid \(k_h - k_a \geq 1\). From Definition \ref{def:difficulty}, the difficulty of the next block \(\mathcal{D}_2\) is:
\begin{align}
\mathcal{D}_2 = \mathcal D_1^{p_a} + \max\left\{1 - \lfloor\frac{t_2 - t_1^{p_a}}{9}\rfloor, -99\right\} \cdot \lfloor\frac{\mathcal D_1^{p_a}}{2048}\rfloor.
\end{align}
A smaller \(\mathcal D_1^{p_a}\) reduces the base difficulty for \(\mathcal{D}_2\) and the scaling factor \(\lfloor\frac{\mathcal D_1^{p_a}}{2048}\rfloor\), minimizing the long-term difficulty growth rate (i.e., \(MR\)). No other valid \(k_h - k_a\) value can produce a smaller \(\mathcal D_1^{p_a}\), so the risk is minimal.

Thus, the UUM attack is successful with minimal post-mining risk if and only if the two conditions hold.

\section{Proof of Theorem \ref{theorem4}}\label{Proof of Theorem 4}
We build on Definition \ref{def:difficulty}, Theorem \ref{theorem1}, and the core properties of the RUM attack. UUM’s initiation condition is \(\lfloor\frac{t_{1}^{p_h} - t_{0}^{p_h}}{9}\rfloor \in [1, +\infty)\) (Theorem \ref{theorem1}), with attack logic relying on \(\mathcal D_1^{p_h} \leq \mathcal D_0^{p_h}\). RUM’s defining condition is that it has zero pre-mining risk under our metric, which requires the honest block \(\mathcal B_1^{p_h}\) to have the same difficulty as its parent block \(\mathcal B_0^{p_h}\) (i.e., \(\mathcal D_1^{p_h} = \mathcal D_0^{p_h}\)); this occurs when \(t_{1}^{p_h} - t_{0}^{p_h} \in [9, 18)\) seconds (equivalent to \(\lfloor\frac{t_{1}^{p_h} - t_{0}^{p_h}}{9}\rfloor = 1\)).

UUM downgrades to RUM if and only if UUM’s initiation condition reduces to RUM’s defining condition. We prove necessity and sufficiency separately:

\noindent\textit{Necessity}: Assume UUM downgrades to RUM. By definition of RUM, the attack must satisfy the zero pre-mining risk condition, which requires \(\mathcal D_1^{p_h} = \mathcal D_0^{p_h}\). From Definition \ref{def:difficulty}, the difficulty of \(\mathcal B_1^{p_h}\) is:
\begin{align}
\mathcal D_1^{p_h} = \mathcal D_0 + \max\left\{1 - k_h, -99\right\} \cdot \lfloor\frac{\mathcal D_0}{2048}\rfloor,
\end{align}
where \(\mathcal D_0 = \mathcal D_0^{p_h}\) and \(k_h = \lfloor\frac{t_{1}^{p_h} - t_{0}^{p_h}}{9}\rfloor\). For \(\mathcal D_1^{p_h} = \mathcal D_0\), substitute into the formula:
\begin{align}
\mathcal D_0 + \max\left\{1 - k_h, -99\right\} \cdot \lfloor\frac{\mathcal D_0}{2048}\rfloor = \mathcal D_0.
\end{align}
Subtract \(\mathcal D_0\) from both sides:
\begin{align}
\max\left\{1 - k_h, -99\right\} \cdot \lfloor\frac{\mathcal D_0}{2048}\rfloor = 0.
\end{align}
Since \(\mathcal D_0 > 2^{17}\) (Ethereum 1.x blocks after height 15), \(\lfloor\frac{\mathcal D_0}{2048}\rfloor > 0\). Thus, we must have:
\begin{align}
\max\left\{1 - k_h, -99\right\} = 0.
\end{align}
This equality holds if and only if \(1 - k_h = 0\) (since \(-99 \neq 0\)), so \(k_h = 1\). Translating back to the timestamp difference:
\begin{align}
\lfloor\frac{t_{1}^{p_h} - t_{0}^{p_h}}{9}\rfloor = 1.
\end{align}

\noindent\textit{Sufficiency}: Assume \(\lfloor\frac{t_{1}^{p_h} - t_{0}^{p_h}}{9}\rfloor = 1\), so \(k_h = 1\). Substitute \(k_h = 1\) into the difficulty formula of \(\mathcal B_1^{p_h}\) (Definition \ref{def:difficulty}):
\begin{align}
\max\left\{1 - k_h, -99\right\} = \max\left\{1 - 1, -99\right\} = 0.
\end{align}
Thus:
\begin{align}
\mathcal D_1^{p_h} = \mathcal D_0 + 0 \cdot \lfloor\frac{\mathcal D_0}{2048}\rfloor = \mathcal D_0.
\end{align}
This satisfies RUM’s core condition (\(\mathcal D_1^{p_h} = \mathcal D_0\)), meaning the adversary can launch an attack with zero pre-mining risk—consistent with RUM’s design~\cite{Unclemakertimestampingoutthecompetitioninethereum}. Additionally, \(k_h = 1\) satisfies UUM’s initiation condition (\(k_h \in [1, +\infty)\), Theorem \ref{theorem1}), so UUM’s attack logic collapses to RUM’s logic.

Thus, UUM downgrades to RUM if and only if \(\lfloor\frac{t_{1}^{p_h} - t_{0}^{p_h}}{9}\rfloor = 1\).

\section{State Space of UUM Attacks}\label{Section4.2}
\begin{enumerate}
    \item[$\bullet$] \textbf{Deployment State.} In this state, the adversary follows the specification of the protocol. It monitors the blocks generated by honest miners. The critical condition is that the timestamp difference between the latest honest block and its parent block must be less than 9 seconds (i.e., \( t_1^{p_h} - t_0^{p_h} < 9 \)). This condition indicates that the difficulty of the honest chain increases, and the adversary currently lacks a favorable attack window. Because no matter how the adversary adjusts its block timestamp, it cannot guarantee that its block will win the forking race (Definition~\ref{def:difficulty}). In this case, the adversary does not kick-start the attack. 
    \item[$\bullet$] \textbf{Attack State.} When a newly generated block by an honest miner has a timestamp difference greater than or equal to 9 seconds from its parent block (i.e., \( t_1^{p_h} - t_0^{p_h} \geq 9 \)), an attack can be kick-started. According to Definition~\ref{def:difficulty}, this means that the difficulty of this honest block is less than or equal to that of its parent block (\( \mathcal{D}_1^{p_h} \leq \mathcal{D}_0^{p_h} \)), satisfying the initiation condition of UUM (Theorem~\ref{theorem1}). 
\end{enumerate}

\begin{figure}[t]
  \centering
  \includegraphics[width=\linewidth]{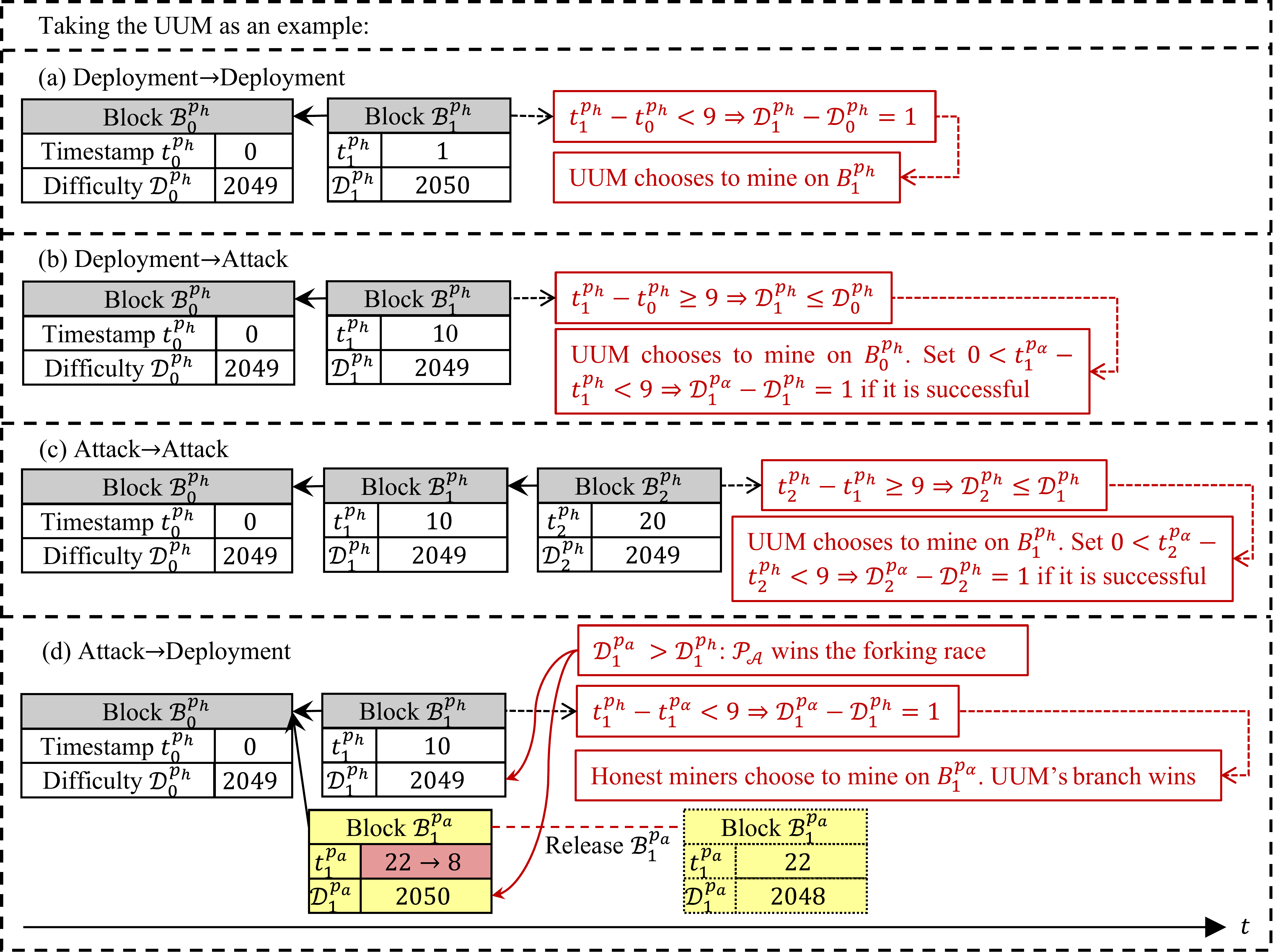}
  \caption{Details of UUM State Transition Process. UUM operates in two states: Deployment (adversary monitors honest block timestamp differences) and Attack (triggered when $t_1^{p_h} - t_0^{p_h} \geq 9$ seconds). In Attack State, the adversary mines on the same parent block, calibrating its block timestamp to $\lfloor (t_1^{p_h} - t_0^{p_h}) / 9 \rfloor - \lfloor (t_1^{p_a} - t_0^{p_h}) / 9 \rfloor = 1$ for higher difficulty (forking race victory) and minimal long-term risk. A valid block ($t_1^{p_a} - t_0^{p_a} \in [1, 900)$) secures success, transitioning back to Deployment for the next window.}
  \label{UUM State Transition Process}
\end{figure}

After an attack is successfully launched, the state transitions back to the deployment state. In UUM, one block is released. The details can be found in \figref{State Transition Process} and \figref{UUM State Transition Process}.

\section{Proof of Theorem \ref{theorem5}}\label{Proof of theorem 5}
We build on Definition \ref{def:fairness}, Theorem \ref{theorem1}--\ref{theorem4}, and the state-transition properties of UUM (Deployment and Attack states). Let \(\mathcal P = \mathcal P_H \cup \mathcal P_A\) be the set of all participants, where \(\mathcal P_H\) and \(\mathcal P_A\) denote the honest participants and the adversary, respectively. Let \(\mu_H = \sum_{p_h \in \mathcal P_H} \mu_{p_h}\) and \(\mu_A = \sum_{p_a \in \mathcal P_A} \mu_{p_a}\) denote their total relative power, with \(\mu_H + \mu_A = 1\). The UUM state probabilities satisfy \(\mathbb{P}(Deploy) + \mathbb{P}(Attack) = 1\).

To avoid conflating per-opportunity reward mass with physical-time reward rate, we proceed in two steps. We first analyze the auxiliary quantity \(\widetilde{\mathbb{R}}_{\mathcal{P}}\), namely realized main-chain coinbase reward mass per steady-state mining opportunity. We then convert the resulting normalized share \(\widehat{\mathbb{R}}_{\mathcal{P}}\) into the physical-time rate \(\mathbb{R}_{\mathcal{P}}\) via
\(
\mathbb{R}_{\mathcal{P}}=\frac{R_c}{T}\widehat{\mathbb{R}}_{\mathcal{P}}.
\)

\noindent\textit{Part 1: Normalized realized reward share.}

Under UUM, the adversary does not create an extra block opportunity; rather, successful attack rounds reassign realized main-chain coinbase rewards away from the honest participants. Therefore, the adversary's auxiliary reward mass remains at its honest-mining baseline \(
\widetilde{\mathbb{R}}_{\mathcal{P}_A}^{UUM} = \mu_A.
\)

By contrast, in the Deployment state the honest participants retain their honest-mining baseline reward mass \(\mu_H\), whereas in the Attack state the adversary wins the forking race (Theorem \ref{theorem3}) and reclaims reward mass that would otherwise contribute to the honest participants' realized main-chain coinbase rewards. Hence,
\begin{align}
\widetilde{\mathbb{R}}_{\mathcal{P}_H}^{UUM}
&= \mathbb{P}(Deploy)\cdot \mu_H
 + \mathbb{P}(Attack)\cdot (\mu_H-\mu_A) \\
&= \mu_H - \mathbb{P}(Attack)\cdot \mu_A.
\end{align}
Since \(\mathbb{P}(Attack)>0\) (Theorem \ref{theorem1}) and \(\mu_A>0\), we have
\(
\widetilde{\mathbb{R}}_{\mathcal{P}_H}^{UUM} < \mu_H.
\)
Therefore, the total auxiliary realized main-chain coinbase reward mass is
\begin{align}
\widetilde{\mathbb{R}}_{total}^{UUM}
&=
\widetilde{\mathbb{R}}_{\mathcal{P}_A}^{UUM}
+
\widetilde{\mathbb{R}}_{\mathcal{P}_H}^{UUM} \\
&=
\mu_A + \mu_H - \mathbb{P}(Attack)\cdot \mu_A \\
&=
1-\mathbb{P}(Attack)\cdot \mu_A.
\end{align}

Hence, the adversary's normalized realized main-chain coinbase reward share is
\begin{align}
\widehat{\mathbb{R}}_{\mathcal{P}_A}^{UUM}
&=
\frac{\widetilde{\mathbb{R}}_{\mathcal{P}_A}^{UUM}}
{\widetilde{\mathbb{R}}_{total}^{UUM}} \\
&=
\frac{\mu_A}{1-\mathbb{P}(Attack)\cdot \mu_A}
>
\mu_A.
\end{align}
Similarly, the honest participants' normalized realized main-chain coinbase reward share is
\begin{align}
\widehat{\mathbb{R}}_{\mathcal{P}_H}^{UUM}
&=
\frac{\widetilde{\mathbb{R}}_{\mathcal{P}_H}^{UUM}}
{\widetilde{\mathbb{R}}_{total}^{UUM}} \\
&=
\frac{\mu_H-\mathbb{P}(Attack)\cdot \mu_A}
{1-\mathbb{P}(Attack)\cdot \mu_A}
<
\mu_H.
\end{align}

\noindent\textit{Part 2: Absolute realized reward rate per unit physical time.}

Under the steady-state target-block-time model used throughout the paper, the protocol retargets difficulty toward the target block interval \(T\). Therefore, the realized main-chain block generation rate converges to \(1/T\), and the corresponding coinbase reward rate satisfies
\(
\mathbb{R}_{\mathcal{P}}
=
\frac{R_c}{T}\widehat{\mathbb{R}}_{\mathcal{P}}.
\)
Applying this relation to UUM yields
\begin{align}
\mathbb{R}_{\mathcal{P}_A}^{UUM}
=
\frac{R_c}{T}\widehat{\mathbb{R}}_{\mathcal{P}_A}^{UUM}
>
\frac{R_c}{T}\mu_A
=
\mathbb{R}_{\mathcal{P}_A}^{HM},
\end{align}
and
\begin{align}
\mathbb{R}_{\mathcal{P}_H}^{UUM}
=
\frac{R_c}{T}\widehat{\mathbb{R}}_{\mathcal{P}_H}^{UUM}
<
\frac{R_c}{T}\mu_H
=
\mathbb{R}_{\mathcal{P}_H}^{HM}.
\end{align}

Thus, under UUM, the adversary's normalized realized main-chain coinbase reward share and its absolute realized main-chain coinbase reward rate per unit physical time both increase, while the honest participants' corresponding reward metrics both decrease.

\section{Methods of SUUM Attacks}\label{Methods of SUUM Attacks}
\begin{enumerate}
\item[$\bullet$]  \textbf{Phase One: SUUM Downgrade Phase.} If the adversary $\mathcal{P}_{\mathcal{A}}$ observes that the difference between the timestamp $t_{1}^{p_{h}}$ of a newly generated block $
\mathcal{B}_{1}^{p_{h}}$ by honest participant $\mathcal{P}_{\mathcal{H}}$ in the network and the timestamp $t_{0}^{p_{h}}$ of the previous main-chain block $\mathcal{B}_{0}^{p_{h}}$ is $t_{1}^{p_{h}} - t_{0}^{p_{h}} \in \lbrack 9, + \infty)$, i.e., $\lfloor (t_{1}^{p_{h}} - t_{0}^{p_{h}})/9 \rfloor \in \lbrack 1, + \infty)$, then SUUM downgrades to UUM. Note that when $\lfloor (t_{1}^{p_{h}} - t_{0}^{p_{h}})/9 \rfloor = 1$, SUUM also downgrades to RUM. The subsequent strategies of adversary $\mathcal{P}_{\mathcal{A}}$ in SUUM are identical to those in UUM. If adversary $\mathcal{P}_{\mathcal{A}}$ in SUUM first discovers a valid block, it proceeds to \textbf{Phase Two}. 

\item[$\bullet$] \textbf{Phase Two: SUUM Deployment Phase.}
At this point, adversary $\mathcal{P}_{\mathcal{A}}$ in SUUM first discovers a valid block $\mathcal{B}_{1}^{p_{a}}$. $\mathcal{P}_{\mathcal{A}}$ will withhold this block $\mathcal{B}_{1}^{p_{a}}$, forming a private chain visible only locally to himself, and then transitions to \textbf{Phase Three}.

\item[$\bullet$] \textbf{Phase Three: SUUM Withholding Block Phase.}
The adversary $\mathcal{P}_{\mathcal{A}}$ in SUUM continues mining along the private chain. If $\mathcal{P}_{\mathcal{A}}$ discovers a new valid block $\mathcal{B}_{2}^{p_{a}}$, it appends this block to its local private chain and continues executing \textbf{Phase Three}. Meanwhile, if an honest miner $\mathcal{P}_{\mathcal{H}}$ discovers a new valid block $\mathcal{B}_{1}^{p_{h}}$, the attack transitions to \textbf{Phase Four}.

\item[$\bullet$] \textbf{Phase Four: SUUM Releasing Block Phase.}
The local private chain of adversary $\mathcal{P}_{\mathcal{A}}$ in SUUM has a length of at least $1$, and honest participant $\mathcal{P}_{\mathcal{H}}$ discovers a new valid block $\mathcal{B}_{1}^{p_{h}}$. In this case, $\mathcal{P}_{\mathcal{A}}$ immediately releases its private block $\mathcal{B}_{1}^{p_{a}}$ and transitions to \textbf{Phase Five}. After $\mathcal{B}_{1}^{p_{a}}$ is released, if the length of $\mathcal{P}_{\mathcal{A}}$'s local private chain still remains at least $1$, the attack transitions to \textbf{Phase Three}. If $\mathcal{P}_{\mathcal{A}}$ has no private blocks, the attack transitions to \textbf{Phase One}.

\item[$\bullet$] \textbf{Phase Five: SUUM Minimal Risk Control Phase.}
$\mathcal{P}_{\mathcal{A}}$ carefully selects the timestamp $t_{1}^{p_{a}}$ for each withheld block $\mathcal{B}_{1}^{p_{a}}$ to manage the post-mining risk. Upon completion of this step, $\mathcal{P}_{\mathcal{A}}$ proceeds with the subsequent steps of \textbf{Phase Four}. The details can be found in Figure \ref{State Transition Process}.
\end{enumerate}

\section{Proof of Theorem \ref{SUUM Successful Condition}}\label{Proof of theorem 6}
We prove the necessity and sufficiency separately, building on Definition \ref{def:difficulty}, Example \ref{example1}, and Theorem \ref{theorem2}. For consistency, define: \(\mathcal D_{i}^{p_h}\) denotes the difficulty of the \(i\)-th honest block on the public chain; \(\mathcal D_{i}^{p_a}\) denotes the difficulty of the \(i\)-th adversarial block on the private chain; \(k_{h,i} = \lfloor\frac{t_{i}^{p_h} - t_{i-1}^{p_h}}{9}\rfloor\), \(k_{a,i} = \lfloor\frac{t_{i}^{p_a} - t_{i-1}^{p_a}}{9}\rfloor\); \(\mathcal D_0 = \mathcal D_{0}^{p_h} = \mathcal D_{0}^{p_a}\) (shared genesis/parent block difficulty for \(i=1\)).

SUUM succeeds if and only if: (1) all adversarial blocks on the private chain are valid; (2) for every \(i \geq 1\), \(\mathcal D_{i}^{p_a} > \mathcal D_{i}^{p_h}\) (ensuring the private chain wins the forking race when released); (3) the private chain remains consistent with blockchain protocol rules.

\noindent\textit{Necessity}: Assume the SUUM attack succeeds. We verify conditions (1) and (2) hold.

1. \textit{Validity of Adversarial Blocks}: By protocol rules, a block is valid only if its timestamp exceeds its parent’s. For \(i=1\), \(t_{1}^{p_a} > t_{0}^{p_a}\) implies \(t_{1}^{p_a} - t_{0}^{p_a} \geq 1\); for \(i \geq 2\), the parent of \(\mathcal B_i^{p_a}\) is \(\mathcal B_{i-1}^{p_a}\), so \(t_{i}^{p_a} > t_{i-1}^{p_a}\) implies \(t_{i}^{p_a} - t_{i-1}^{p_a} \geq 1\). To avoid uncontrolled difficulty reduction (Definition \ref{def:difficulty}), \(t_{i}^{p_a} - t_{i-1}^{p_a} < 900\) (otherwise \(\max\{1 - k_{a,i}, -99\} = -99\), destabilizing subsequent difficulty). Thus, \(t_{i}^{p_a} - t_{i-1}^{p_a} \in [1, 900)\) for all \(i \geq 1\).

2. \textit{Difficulty Superiority for} \(i=1\): From Definition \ref{def:difficulty}, the difficulty of \(\mathcal B_1^{p_h}\) and \(\mathcal B_1^{p_a}\) is:
\begin{align}
\mathcal D_1^{p_h} = \mathcal D_0 + \max\left\{1 - k_{h,1}, -99\right\} \cdot \lfloor\frac{\mathcal D_0}{2048}\rfloor,
\end{align}
\begin{align}
\mathcal D_1^{p_a} = \mathcal D_0 + \max\left\{1 - k_{a,1}, -99\right\} \cdot \lfloor\frac{\mathcal D_0}{2048}\rfloor.
\end{align}
Since \(\mathcal D_0 > 2^{17}\) (Ethereum 1.x blocks after height 15), \(\lfloor\frac{\mathcal D_0}{2048}\rfloor > 0\). For \(\mathcal D_1^{p_a} > \mathcal D_1^{p_h}\), we need \(\max\{1 - k_{a,1}, -99\} > \max\{1 - k_{h,1}, -99\}\). This simplifies to \(1 - k_{a,1} > 1 - k_{h,1}\) (practical block times make \(1 - k_{h,1}, 1 - k_{a,1} > -99\)), so \(k_{h,1} - k_{a,1} \geq 1\), i.e., \(\lfloor\frac{t_1^{p_h} - t_0^{p_h}}{9}\rfloor - \lfloor\frac{t_1^{p_a} - t_0^{p_a}}{9}\rfloor \in [1, +\infty)\).

3. \textit{Difficulty Superiority for} \(i \geq 2\): Assume \(\mathcal D_{i-1}^{p_a} > \mathcal D_{i-1}^{p_h}\) (inductive hypothesis, holds for \(i=2\) as \(i=1\) is proven). For \(\mathcal B_i^{p_h}\) and \(\mathcal B_i^{p_a}\):
\begin{align}
\mathcal D_i^{p_h} = \mathcal D_{i-1}^{p_h} + \max\left\{1 - k_{h,i}, -99\right\} \cdot \lfloor\frac{\mathcal D_{i-1}^{p_h}}{2048}\rfloor,
\end{align}
\begin{align}
\mathcal D_i^{p_a} = \mathcal D_{i-1}^{p_a} + \max\left\{1 - k_{a,i}, -99\right\} \cdot \lfloor\frac{\mathcal D_{i-1}^{p_a}}{2048}\rfloor.
\end{align}
Since \(\mathcal D_{i-1}^{p_a} > \mathcal D_{i-1}^{p_h}\), \(\lfloor\frac{\mathcal D_{i-1}^{p_a}}{2048}\rfloor \geq \lfloor\frac{\mathcal D_{i-1}^{p_h}}{2048}\rfloor\). If \(k_{h,i} - k_{a,i} \geq 0\), then \(1 - k_{a,i} \geq 1 - k_{h,i}\), so:
\begin{align}
\max\left\{1 - k_{a,i}, -99\right\} \cdot \lfloor\frac{\mathcal D_{i-1}^{p_a}}{2048}\rfloor \geq \max\left\{1 - k_{h,i}, -99\right\} \cdot \lfloor\frac{\mathcal D_{i-1}^{p_h}}{2048}\rfloor.
\end{align}
Adding the inductive hypothesis \(\mathcal D_{i-1}^{p_a} > \mathcal D_{i-1}^{p_h}\) gives \(\mathcal D_i^{p_a} > \mathcal D_i^{p_h}\). If \(k_{h,i} - k_{a,i} < 0\), then \(1 - k_{a,i} < 1 - k_{h,i}\), and the inequality may fail—contradicting success. Thus, \(k_{h,i} - k_{a,i} \in [0, +\infty)\) for \(i \geq 2\).

\noindent\textit{Sufficiency}: Assume conditions (1) and (2) hold. We prove the SUUM attack succeeds.

1. \textit{Validity of Adversarial Blocks}: For all \(i \geq 1\), \(t_{i}^{p_a} - t_{i-1}^{p_a} \in [1, 900)\) ensures \(\mathcal B_i^{p_a}\) is valid (timestamp exceeds parent’s) and avoids uncontrolled difficulty adjustment (Definition \ref{def:difficulty}).

2. \textit{Difficulty Superiority for} \(i=1\): By condition (1), \(k_{h,1} - k_{a,1} \geq 1\), so \(1 - k_{a,1} > 1 - k_{h,1}\). With \(\lfloor\frac{\mathcal D_0}{2048}\rfloor > 0\), \(\mathcal D_1^{p_a} > \mathcal D_1^{p_h}\) (same derivation as Theorem \ref{theorem2}).

3. \textit{Inductive Proof for} \(i \geq 2\): Assume \(\mathcal D_{k}^{p_a} > \mathcal D_{k}^{p_h}\) for some \(k \geq 1\). For \(i = k+1\), condition (2) gives \(k_{h,k+1} - k_{a,k+1} \geq 0\), so \(1 - k_{a,k+1} \geq 1 - k_{h,k+1}\). Since \(\mathcal D_k^{p_a} > \mathcal D_k^{p_h}\), \(\lfloor\frac{\mathcal D_k^{p_a}}{2048}\rfloor \geq \lfloor\frac{\mathcal D_k^{p_h}}{2048}\rfloor\), leading to:
\(
\max\left\{1 - k_{a,k+1}, -99\right\} \cdot \lfloor\frac{\mathcal D_k^{p_a}}{2048}\rfloor \geq \max\left\{1 - k_{h,k+1}, -99\right\} \cdot \lfloor\frac{\mathcal D_k^{p_h}}{2048}\rfloor.
\)
Adding \(\mathcal D_k^{p_a} > \mathcal D_k^{p_h}\) yields \(\mathcal D_{k+1}^{p_a} > \mathcal D_{k+1}^{p_h}\). By induction, \(\mathcal D_i^{p_a} > \mathcal D_i^{p_h}\) for all \(i \geq 1\).

4. \textit{Forking Race Victory}: When the private chain is released, every adversarial block \(\mathcal B_i^{p_a}\) has higher difficulty than the corresponding \(\mathcal B_i^{p_h}\) (Example \ref{example1}). The cumulative difficulty of the private chain is thus strictly greater than the public chain, ensuring it becomes the main-chain.

Thus, the SUUM attack is successful if and only if the stated conditions hold.

\section{Proof of Theorem \ref{theorem7}}\label{Proof of theorem 7}
We build on Theorem \ref{SUUM Successful Condition}, Definition \ref{def:difficulty}, and the minimal post-mining risk criterion. For consistency, we reuse and extend the notation from Theorem \ref{SUUM Successful Condition}: \(\mathcal D_{i}^{p_h}\) denotes the difficulty of the \(i\)-th honest block on the public chain; \(\mathcal D_{i}^{p_a}\) denotes the difficulty of the \(i\)-th adversarial block on the private chain; \(k_{h,i} = \lfloor\frac{t_{i}^{p_h} - t_{i-1}^{p_h}}{9}\rfloor\), \(k_{a,i} = \lfloor\frac{t_{i}^{p_a} - t_{i-1}^{p_a}}{9}\rfloor\); \(\mathcal D_0 = \mathcal D_{0}^{p_h} = \mathcal D_{0}^{p_a}\) (shared parent block difficulty for \(i=1\)); post-mining risk (\(MR\)) denotes the incremental growth of subsequent block difficulty induced by the attack, directly dependent on \(\mathcal D_{i}^{p_a}\) (adversarial blocks become main-chain blocks and influence future difficulty adjustments).

Minimal \(MR\) requires minimizing the difficulty of each adversarial block \(\mathcal D_{i}^{p_a}\) while preserving the success condition \(\mathcal D_{i}^{p_a} > \mathcal D_{i}^{p_h}\) (Theorem \ref{SUUM Successful Condition}). We prove necessity and sufficiency separately:

\noindent\textit{Necessity}: Assume the SUUM attack succeeds with minimal post-mining risk. By Theorem \ref{SUUM Successful Condition}, the attack’s success requires: For \(i=1\): \(k_{h,1} - k_{a,1} \in [1, +\infty)\) and \(t_{1}^{p_a} - t_{0}^{p_a} \in [1, 900)\); For \(i \geq 2\): \(k_{h,i} - k_{a,i} \in [0, +\infty)\) and \(t_{i}^{p_a} - t_{i-1}^{p_a} \in [1, 900)\).

We now show the strict equalities \(k_{h,1} - k_{a,1} = 1\) and \(k_{h,i} - k_{a,i} = 0\) (for \(i \geq 2\)) are mandatory for minimal \(MR\).

1. \textit{Case \(i=1\)}: From Definition \ref{def:difficulty}, the difficulty of \(\mathcal B_1^{p_a}\) is:
\begin{align}
\mathcal D_1^{p_a} = \mathcal D_0 + \max\left\{1 - k_{a,1}, -99\right\} \cdot \lfloor\frac{\mathcal D_0}{2048}\rfloor,
\end{align}
where \(\max\left\{1 - k_{a,1}, -99\right\} = 1 - k_{a,1}\) (practical block times ensure \(k_{a,1} \leq 99\)). For \(k_{h,1} - k_{a,1} = m \geq 1\), \(k_{a,1} = k_{h,1} - m\), so:
\begin{align}
\mathcal D_1^{p_a} &= \mathcal D_0 + \left(1 - (k_{h,1} - m)\right) \cdot \lfloor\frac{\mathcal D_0}{2048}\rfloor\\
&= \mathcal D_0 + \left((1 - k_{h,1}) + m\right) \cdot \lfloor\frac{\mathcal D_0}{2048}\rfloor.
\end{align}
Since \(\lfloor\frac{\mathcal D_0}{2048}\rfloor > 0\) (\(\mathcal D_0 > 2^{17}\)), \(\mathcal D_1^{p_a}\) increases with \(m\). A larger \(\mathcal D_1^{p_a}\) raises the base difficulty for subsequent blocks (e.g., \(\mathcal D_2\)) and amplifies their difficulty growth (via \(\lfloor\frac{\mathcal D_1^{p_a}}{2048}\rfloor\)), increasing \(MR\). Thus, minimal \(MR\) requires the smallest \(m\) (i.e., \(m=1\)), so \(k_{h,1} - k_{a,1} = 1\). The validity condition \(t_{1}^{p_a} - t_{0}^{p_a} \in [1, 900)\) is retained (Theorem \ref{SUUM Successful Condition}), as it ensures block validity and controlled difficulty adjustment (prerequisites for minimal risk).

2. \textit{Case \(i \geq 2\)}: Assume \(\mathcal D_{i-1}^{p_a} > \mathcal D_{i-1}^{p_h}\) (inductive hypothesis, holds for \(i=2\) as \(i=1\) is proven). From Definition \ref{def:difficulty}:
\begin{align}
\mathcal D_i^{p_h} = \mathcal D_{i-1}^{p_h} + \max\left\{1 - k_{h,i}, -99\right\} \cdot \lfloor\frac{\mathcal D_{i-1}^{p_h}}{2048}\rfloor,
\end{align}
\begin{align}
\mathcal D_i^{p_a} = \mathcal D_{i-1}^{p_a} + \max\left\{1 - k_{a,i}, -99\right\} \cdot \lfloor\frac{\mathcal D_{i-1}^{p_a}}{2048}\rfloor.
\end{align}
For \(k_{h,i} - k_{a,i} = n \geq 0\), \(k_{a,i} = k_{h,i} - n\), so:
\begin{align}
\mathcal D_i^{p_a} = \mathcal D_{i-1}^{p_a} + \left(1 - (k_{h,i} - n)\right) \cdot \lfloor\frac{\mathcal D_{i-1}^{p_a}}{2048}\rfloor.
\end{align}
Since \(\mathcal D_{i-1}^{p_a} > \mathcal D_{i-1}^{p_h}\), \(\lfloor\frac{\mathcal D_{i-1}^{p_a}}{2048}\rfloor \geq \lfloor\frac{\mathcal D_{i-1}^{p_h}}{2048}\rfloor\). If \(n \geq 1\), \(1 - k_{a,i} = 1 - k_{h,i} + n\), leading to a larger \(\mathcal D_i^{p_a}\) than when \(n=0\) (where \(1 - k_{a,i} = 1 - k_{h,i}\)). A larger \(\mathcal D_i^{p_a}\) increases \(MR\) by accelerating subsequent difficulty growth. When \(n=0\), \(k_{a,i} = k_{h,i}\), and the inductive hypothesis ensures:
\(
\mathcal D_{i-1}^{p_a} + \left(1 - k_{h,i}\right) \cdot \lfloor\frac{\mathcal D_{i-1}^{p_a}}{2048}\rfloor > \mathcal D_{i-1}^{p_h} + \left(1 - k_{h,i}\right) \cdot \lfloor\frac{\mathcal D_{i-1}^{p_h}}{2048}\rfloor = \mathcal D_i^{p_h}.
\)
Thus, \(n=0\) (i.e., \(k_{h,i} - k_{a,i} = 0\)) minimizes \(\mathcal D_i^{p_a}\) while preserving \(\mathcal D_i^{p_a} > \mathcal D_i^{p_h}\), ensuring minimal \(MR\). The validity condition \(t_{i}^{p_a} - t_{i-1}^{p_a} \in [1, 900)\) is retained.

\noindent\textit{Sufficiency}: Assume conditions (1) and (2) hold. We first confirm the attack succeeds and then verify minimal post-mining risk.

1. \textit{Attack Success}: For \(i=1\): \(k_{h,1} - k_{a,1} = 1 \in [1, +\infty)\) and \(t_{1}^{p_a} - t_{0}^{p_a} \in [1, 900)\) satisfy Theorem \ref{SUUM Successful Condition}’s condition (1), ensuring \(\mathcal D_1^{p_a} > \mathcal D_1^{p_h}\) (difficulty superiority) and \(\mathcal B_1^{p_a}\) validity. For \(i \geq 2\): \(k_{h,i} - k_{a,i} = 0 \in [0, +\infty)\) and \(t_{i}^{p_a} - t_{i-1}^{p_a} \in [1, 900)\) satisfy Theorem \ref{SUUM Successful Condition}’s condition (2). By induction: Assume \(\mathcal D_{k}^{p_a} > \mathcal D_{k}^{p_h}\) for \(k \geq 1\); for \(i = k+1\), \(k_{h,k+1} = k_{a,k+1}\) and \(\mathcal D_k^{p_a} > \mathcal D_k^{p_h}\) imply:
\(
\mathcal D_{k+1}^{p_a} = \mathcal D_k^{p_a} + \left(1 - k_{h,k+1}\right) \cdot \lfloor\frac{\mathcal D_k^{p_a}}{2048}\rfloor > \mathcal D_k^{p_h} + \left(1 - k_{h,k+1}\right) \cdot \lfloor\frac{\mathcal D_k^{p_h}}{2048}\rfloor = \mathcal D_{k+1}^{p_h}.
\)

Thus, \(\mathcal D_i^{p_a} > \mathcal D_i^{p_h}\) for all \(i \geq 1\), and the private chain wins the forking race (Example \ref{example1}).

2. \textit{Minimal post-mining risk}: For \(i=1\): \(k_{h,1} - k_{a,1} = 1\) yields the smallest possible \(\mathcal D_1^{p_a}\) among all valid \(k_{h,1} - k_{a,1} \geq 1\) (Necessity proof), minimizing the base difficulty for subsequent blocks. For \(i \geq 2\): \(k_{h,i} - k_{a,i} = 0\) yields the smallest possible \(\mathcal D_i^{p_a}\) among all valid \(k_{h,i} - k_{a,i} \geq 0\) (Necessity proof). Since \(\mathcal D_i^{p_a}\) directly influences \(\mathcal D_{i+1}\) (Definition \ref{def:difficulty}), this choice minimizes the long-term difficulty growth rate (i.e., \(MR\)). No other valid values of \(k_{h,i} - k_{a,i}\) can produce smaller \(\mathcal D_i^{p_a}\) while preserving \(\mathcal D_i^{p_a} > \mathcal D_i^{p_h}\), so the risk is minimal.

Thus, the SUUM attack is successful with minimal post-mining risk if and only if the stated conditions hold.

\begin{table}[t]
\caption{SUUM state transition.}
\label{table3}
\centering
\scriptsize
\resizebox{\columnwidth}{!}{%
\begin{tabular}{@{}|c|c|c|c|c|@{}}
\toprule
\multirow{2}{*}{State} & \multirow{2}{*}{Destination} & \multirow{2}{*}{Transition} & \multicolumn{2}{c|}{Reward} \\ \cmidrule(l){4-5} 
                       &                                                           &                              & \multicolumn{1}{c|}{$\mathcal{P}_{\mathcal{H}}$}   &  $\mathcal{P}_{\mathcal{A}}$ \\ \midrule
Deployment          & Attack $1$   & $\mathcal{P}_{\mathcal{A}}~any~t$ &  \multicolumn{1}{c|}{$0$}  & $1$ \\ \midrule
Deployment          & Deployment    & $\mathcal{P}_{\mathcal{H}}~t < 9$ &  \multicolumn{1}{c|}{$1$}  & $0$ \\ \midrule
Deployment          & Downgrade  & $\mathcal{P}_{\mathcal{H}}~t \geq 9$ &  \multicolumn{1}{c|}{$1$}  & $0$ \\ \midrule
Downgrade        & Deployment    & $\mathcal{P}_{\mathcal{H}}~t < 9$ &  \multicolumn{1}{c|}{$1$}  & $0$ \\ \midrule
Downgrade        & Downgrade  & $\mathcal{P}_{\mathcal{H}}~t \geq 9$ &  \multicolumn{1}{c|}{$1$}  & $0$ \\ \rowcolor{gray!40} \midrule
Downgrade        & Deployment    & $\mathcal{P}_{\mathcal{A}}~any~t$ &  \multicolumn{1}{c|}{$-1$} & $1$ \\ \midrule
Attack $1$         & Deployment    & $\mathcal{P}_{\mathcal{H}}~any~t$ &  \multicolumn{1}{c|}{$0$}  & $0$ \\ \rowcolor{gray!40}  \midrule
{ Attack $i-1$, $i \geq 2$} & Attack $i$   & $\mathcal{P}_{\mathcal{A}}~any~t$ &  \multicolumn{1}{c|}{$-1$}  & $1$ \\ \midrule
{ Attack $i+1$, $i \geq 1$} & Attack $i$   & $\mathcal{P}_{\mathcal{H}}~any~t$ &  \multicolumn{1}{c|}{$1$}  & $0$ \\ \bottomrule
\end{tabular}
}
\begin{tablenotes} 
\footnotesize
\setlength{\itemindent}{0pt} 
\setlength{\rightskip}{10pt}   
\item* \noindent 
The state transitions that lead to honest blocks becoming orphaned include the two gray rows (the sixth and eighth rows).
\end{tablenotes}
\end{table}

\section{State Space of SUUM Attacks}\label{Section5.2}
\begin{enumerate}
\item[$\bullet$] \textbf{SUUM Deployment State.}
Three subsequent scenarios may arise: (1) The SUUM adversary continues to wait until a latest block is generated by an honest participant, with a timestamp difference of greater than or equal to 9 compared to the previous main-chain block. Upon fulfillment of this condition, the \textbf{Deployment State} transitions to the \textbf{Downgrade State}. (2) If the SUUM adversary successfully generates the next valid block, they will withhold it and drive the \textbf{Deployment State} to transition to \textbf{Attack State 1}. (3) If the next block is generated by an honest participant and the timestamp difference is less than $9$ compared to the previous main-chain block, the state remains unchanged, remaining in the \textbf{Deployment State}.

\item[$\bullet$] \textbf{SUUM Downgrade State.}
Three subsequent scenarios may arise: (1) If the SUUM adversary successfully generates a valid block, he will strategically set the block's timestamp and drive the downgrade state to transition back to the \textbf{Deployment State}. (2) If the next block is generated by an honest participant and its timestamp difference is less than $9$ compared to the previous block, this drives the \textbf{Downgrade State} to transition back to the \textbf{Deployment State}. (3) If the next block is generated by an honest participant and its timestamp difference is greater than or equal to $9$ compared to the previous block, the state remains unchanged, remaining in the \textbf{Downgrade State}.

\item[$\bullet$] \textbf{SUUM Attack State.}
Three subsequent scenarios may arise: (1) For \textbf{Attack State \bm{$1$}}, if the next valid block is generated by an honest participant, the SUUM adversary immediately releases a withheld block with a strategically set timestamp and drives the \textbf{Attack state} to transition back to the \textbf{Deployment State}. (2) For \textbf{Attack State \bm{$i$}}, \textbf{\bm{$i \geq 1$}}, if the next block is generated by the SUUM adversary, he will withhold it and drive the \textbf{Attack State \bm{$i$}} to the \textbf{Attack State \bm{$i+1$}}. (3) For \textbf{Attack State \bm{$i+1$}}, \textbf{\bm{$i \geq 1$}}, if the next block is generated by an honest participant, the SUUM adversary immediately releases a withheld block and strategically sets its timestamp. The \textbf{Attack State \bm{$i+1$}} transitions back to the \textbf{Attack State \bm{$i$}}. The details can be found in Figure \ref{State Transition Process} and Table \ref{table3}. We present the steady-state probability in \tabref{tab:State_probability}.
\end{enumerate}

\begin{table}[t]
\caption{Steady-state probabilities under RUM, UUM, and SUUM.}
\label{tab:State_probability}
\centering
\scriptsize
\setlength{\tabcolsep}{2.5pt}
\renewcommand{\arraystretch}{1.15}
\begin{threeparttable}
\resizebox{\columnwidth}{!}{%
\begin{tabular}{@{}|c|c|c|c|c|c|c|c|@{}}
\toprule
\multirow{2}{*}{$\alpha$}
& \multicolumn{2}{c|}{RUM}
& \multicolumn{2}{c|}{UUM}
& \multicolumn{3}{c|}{SUUM} \\ \cmidrule(l){2-8}
& \multicolumn{1}{c|}{Deployment} 
& \multicolumn{1}{c|}{Attack}
& \multicolumn{1}{c|}{Deployment} 
& \multicolumn{1}{c|}{Attack}
& \multicolumn{1}{c|}{Deployment} 
& \multicolumn{1}{c|}{Downgrade}
& Attack \\ \midrule
0.1 & 0.7934 & 0.2066 & 0.4232 & 0.5768 & 0.2878 & 0.3900 & 0.3222 \\ \midrule
0.2 & 0.8147 & 0.1853 & 0.4880 & 0.5120 & 0.2202 & 0.2307 & 0.5491 \\ \midrule
0.3 & 0.8373 & 0.1627 & 0.5509 & 0.4491 & 0.1644 & 0.1334 & 0.7022 \\ \midrule
0.4 & 0.8611 & 0.1389 & 0.6171 & 0.3829 & 0.1201 & 0.0745 & 0.8054 \\ \midrule
0.5 & 0.8851 & 0.1149 & 0.6796 & 0.3204 & 0.0874 & 0.0414 & 0.8712 \\ \bottomrule
\end{tabular}
}
\begin{tablenotes} 
\footnotesize
\setlength{\itemindent}{0pt}
\setlength{\rightskip}{10pt}
\item* \noindent
For SUUM, the attack state is the aggregated steady-state probability of all attack states.
\end{tablenotes}
\end{threeparttable}
\end{table}

\begin{theorem}[\textbf{SUUM Adversary Rewards}]\label{SUUM Adversary Rewards}
Under SUUM, the adversary's normalized realized main-chain coinbase reward share strictly exceeds its honest-mining baseline and is strictly higher than under UUM. Under the steady-state difficulty-retargeting relation
\(
\mathbb{R}_{\mathcal{P}}=\frac{R_c}{T}\widehat{\mathbb{R}}_{\mathcal{P}},
\)
this also implies a strictly higher absolute realized main-chain coinbase reward rate per unit physical time. Correspondingly, the honest participants' normalized and absolute realized main-chain coinbase rewards both decrease further.
\end{theorem}

\section{Proof of Theorem \ref{SUUM Adversary Rewards}}\label{Proof of theorem 8}
We build on Definition \ref{def:fairness}, Theorem \ref{SUUM Successful Condition}--\ref{theorem7}, the UUM reward analysis, and the SUUM state transitions (Deployment, Downgrade, and Attack states). Let \(P = \mathcal P_H \cup \mathcal P_A\) be the set of all participants, where \(\mathcal P_H\) and \(\mathcal P_A\) denote the honest participants and the adversary, respectively. Let \(\mu_H = \sum_{p_h \in \mathcal P_H} \mu_{p_h}\) and \(\mu_A = \sum_{p_a \in \mathcal P_A} \mu_{p_a}\) denote the total relative hash power of the honest participants and the adversary, with \(\mu_H + \mu_A = 1\). The SUUM state probabilities satisfy
\(
\mathbb{P}(Deploy)+\mathbb{P}(Downgrade)+\mathbb{P}(Attack)=1.
\)

As in the proof of Theorem \ref{theorem5}, we first analyze the auxiliary quantity \(\widetilde{\mathbb{R}}_{\mathcal{P}}\), namely realized main-chain coinbase reward mass per steady-state mining opportunity, and then convert the resulting normalized share \(\widehat{\mathbb{R}}_{\mathcal{P}}\) into the physical-time rate \(\mathbb{R}_{\mathcal{P}}\) via
\(
\mathbb{R}_{\mathcal{P}}=\frac{R_c}{T}\widehat{\mathbb{R}}_{\mathcal{P}}.
\)

\noindent\textit{Part 1: Normalized realized reward share.}

Under SUUM, block withholding and staircase release enlarge the reward-reclaiming state space, but they still do not create an extra mining opportunity in a single step. Therefore, the adversary's auxiliary realized main-chain coinbase reward mass remains
\(
\widetilde{\mathbb{R}}_{\mathcal{P}_A}^{SUUM} = \mu_A.
\)

The honest participants, however, lose realized main-chain coinbase reward mass whenever SUUM enters either the Downgrade state or the Attack state. Hence,
\begin{align}
\widetilde{\mathbb{R}}_{\mathcal{P}_H}^{SUUM}
&=
\mathbb{P}(Deploy)\cdot \mu_H\\
&+
\big(\mathbb{P}(Downgrade)+\mathbb{P}(Attack)\big)\cdot(\mu_H-\mu_A) \\
&=
\mu_H-\big(\mathbb{P}(Downgrade)+\mathbb{P}(Attack)\big)\cdot \mu_A.
\end{align}
Since \(\mathbb{P}(Downgrade)+\mathbb{P}(Attack)>0\) (Theorem \ref{SUUM Successful Condition}) and \(\mu_A>0\), we obtain
\(
\widetilde{\mathbb{R}}_{\mathcal{P}_H}^{SUUM} < \mu_H.
\)

Therefore, the total auxiliary realized main-chain coinbase reward mass is
\begin{align}
\widetilde{\mathbb{R}}_{total}^{SUUM}
&=
\widetilde{\mathbb{R}}_{\mathcal{P}_A}^{SUUM}
+
\widetilde{\mathbb{R}}_{\mathcal{P}_H}^{SUUM} \\
&=
1-\big(\mathbb{P}(Downgrade)+\mathbb{P}(Attack)\big)\cdot \mu_A.
\end{align}

Hence, the adversary's normalized realized main-chain coinbase reward share is
\begin{align}
\widehat{\mathbb{R}}_{\mathcal{P}_A}^{SUUM}
&=
\frac{\widetilde{\mathbb{R}}_{\mathcal{P}_A}^{SUUM}}
{\widetilde{\mathbb{R}}_{total}^{SUUM}} \\
&=
\frac{\mu_A}
{1-\big(\mathbb{P}(Downgrade)+\mathbb{P}(Attack)\big)\cdot \mu_A}
>
\mu_A.
\end{align}
Similarly, the honest participants' normalized realized main-chain coinbase reward share is
\begin{align}
\widehat{\mathbb{R}}_{\mathcal{P}_H}^{SUUM}
&=
\frac{\widetilde{\mathbb{R}}_{\mathcal{P}_H}^{SUUM}}
{\widetilde{\mathbb{R}}_{total}^{SUUM}} \\
&=
\frac{
\mu_H-\big(\mathbb{P}(Downgrade)+\mathbb{P}(Attack)\big)\cdot \mu_A
}{
1-\big(\mathbb{P}(Downgrade)+\mathbb{P}(Attack)\big)\cdot \mu_A
}
<
\mu_H.
\end{align}

Compared with UUM, SUUM strictly enlarges the reward-reclaiming region by adding withholding and staircase release on top of the downgrade path. Therefore, the denominator contraction term under SUUM is strictly larger than under UUM, which yields
\(
\widehat{\mathbb{R}}_{\mathcal{P}_A}^{SUUM}
>
\widehat{\mathbb{R}}_{\mathcal{P}_A}^{UUM}.
\)

\noindent\textit{Part 2: Absolute realized reward rate per unit physical time.}

Under the same steady-state target-block-time model, we again have
\(
\mathbb{R}_{\mathcal{P}}
=
\frac{R_c}{T}\widehat{\mathbb{R}}_{\mathcal{P}}.
\)
Therefore,
\begin{align}
\mathbb{R}_{\mathcal{P}_A}^{SUUM}
=
\frac{R_c}{T}\widehat{\mathbb{R}}_{\mathcal{P}_A}^{SUUM}
>
\frac{R_c}{T}\mu_A
=
\mathbb{R}_{\mathcal{P}_A}^{HM},
\end{align}
and, moreover,
\(
\mathbb{R}_{\mathcal{P}_A}^{SUUM}
>
\mathbb{R}_{\mathcal{P}_A}^{UUM}.
\)
Similarly,
\begin{align}
\mathbb{R}_{\mathcal{P}_H}^{SUUM}
=
\frac{R_c}{T}\widehat{\mathbb{R}}_{\mathcal{P}_H}^{SUUM}
<
\frac{R_c}{T}\mu_H
=
\mathbb{R}_{\mathcal{P}_H}^{HM}.
\end{align}

Thus, under SUUM, the adversary's normalized realized main-chain coinbase reward share and its absolute realized main-chain coinbase reward rate per unit physical time both increase, and both are strictly larger than under UUM. Correspondingly, the honest participants' realized main-chain reward metrics both decrease further.

\begin{theorem}[\textbf{Reward Comparison}] \label{Reward Comparison of Uncle Maker-based attack and honest mining}
Under both the normalized realized main-chain coinbase reward-share metric $\widehat{\mathbb{R}}_{\mathcal{P}_A}$ and the absolute realized main-chain coinbase reward-rate metric $\mathbb{R}_{\mathcal{P}_A}$, SUUM outperforms UUM, which in turn outperforms RUM, all of which surpass honest mining.
\end{theorem}

\section{Proof of Theorem \ref{Reward Comparison of Uncle Maker-based attack and honest mining}}\label{Proof of Theorem 9}
We compare the four strategies under two reward metrics: the adversary's normalized realized main-chain coinbase reward share \(\widehat{\mathbb{R}}_{\mathcal{P}_A}\) and the corresponding absolute realized main-chain coinbase reward rate per unit physical time \(\mathbb{R}_{\mathcal{P}_A}\).

We first establish the ordering under \(\widehat{\mathbb{R}}_{\mathcal{P}_A}\). The ordering under \(\mathbb{R}_{\mathcal{P}_A}\) then follows immediately from the steady-state difficulty-retargeting relation
\(
\mathbb{R}_{\mathcal{P}}=\frac{R_c}{T}\widehat{\mathbb{R}}_{\mathcal{P}}.
\)

First, prior work has already established that RUM yields a larger adversarial normalized realized reward share than honest mining \cite{Unclemakertimestampingoutthecompetitioninethereum}. Thus,
\(
\widehat{\mathbb{R}}_{\mathcal{P}_A}^{RUM} > \widehat{\mathbb{R}}_{\mathcal{P}_A}^{HM}.
\)

Next, we compare UUM with RUM. RUM enters the reward-reclaiming attack state only when an honest block satisfies
\(
\mathcal{P}_{\mathcal{H}}(9 \le t < 18),
\)
whereas UUM enters the corresponding state under the weaker condition
\(
\mathcal{P}_{\mathcal{H}}(t \ge 9).
\)
Therefore, UUM has a strictly larger feasible reward-reclaiming region than RUM, which strictly reduces the honest participants' auxiliary realized reward mass \(\widetilde{\mathbb{R}}_{\mathcal{P}_H}\) and hence strictly increases the adversary's normalized realized reward share. It follows that
\(
\widehat{\mathbb{R}}_{\mathcal{P}_A}^{UUM}
>
\widehat{\mathbb{R}}_{\mathcal{P}_A}^{RUM}.
\)

Finally, SUUM further enlarges the adversary's strategy space by introducing withholding and staircase release on top of the downgrade path. This expands the set of states in which the adversary can reclaim realized main-chain coinbase rewards that would otherwise contribute to the honest participants' realized rewards. Therefore,
\(
\widehat{\mathbb{R}}_{\mathcal{P}_A}^{SUUM}
>
\widehat{\mathbb{R}}_{\mathcal{P}_A}^{UUM}.
\)

Combining the above inequalities yields
\begin{align}
\widehat{\mathbb{R}}_{\mathcal{P}_A}^{SUUM}
>
\widehat{\mathbb{R}}_{\mathcal{P}_A}^{UUM}
>
\widehat{\mathbb{R}}_{\mathcal{P}_A}^{RUM}
>
\widehat{\mathbb{R}}_{\mathcal{P}_A}^{HM}.
\end{align}

Applying
\(
\mathbb{R}_{\mathcal{P}}=\frac{R_c}{T}\widehat{\mathbb{R}}_{\mathcal{P}}
\)
to all four strategies yields
\begin{align}
\mathbb{R}_{\mathcal{P}_A}^{SUUM}
>
\mathbb{R}_{\mathcal{P}_A}^{UUM}
>
\mathbb{R}_{\mathcal{P}_A}^{RUM}
>
\mathbb{R}_{\mathcal{P}_A}^{HM}.
\end{align}
This proves Theorem \ref{Reward Comparison of Uncle Maker-based attack and honest mining}.

To present the post-mining risk in a compact form, we use \(\Sigma(\tau)\) to denote the cumulative post-mining risk contributed by all successful attack rounds whose completion pattern is \(\tau\). Each occurrence contributes one Definition~\ref{def:post_mining_risk} risk term; thus, \(\Sigma(\tau)\) is a shorthand for a sum of single-round post-mining risks, not merely a count of transition events.

\begin{theorem}[\textbf{Post-mining Risk Comparison}]
\label{Risk Comparison of Uncle Maker-based attack and honest mining}
Compared to honest mining, the increased post-mining risks posed by RUM, UUM, and SUUM attacks are as follows:
\begin{enumerate}[(1)]
    \item The increased post-mining risk associated with the RUM attack is denoted by
    \(
    \Sigma\!\left(
    Attack\overset{\mathcal{P}_{\mathcal{A}}~~any~t}{\Rightarrow}Deploy
    \right).
    \)
    \item The increased post-mining risk associated with the UUM attack is denoted by
    \(
    \Sigma\!\left(
    Attack\overset{\mathcal{P}_{\mathcal{A}}~~any~t}{\Rightarrow}Deploy
    \right).
    \)
    \item The increased post-mining risk associated with the SUUM attack is denoted by
    $\sum\begin{pmatrix}{Downgrade\overset{\mathcal{P}_{\mathcal{A}}~~any~t}{\Rightarrow}Deploy} \\ {+ Attack~1\overset{\mathcal{P}_{\mathcal{H}}~~any~t}{\Rightarrow}Deploy}
\end{pmatrix}$.
\end{enumerate}
\end{theorem}

\section{Proof of Theorem \ref{Risk Comparison of Uncle Maker-based attack and honest mining}}\label{Proof of Theorem 10}
We discuss the post-mining risk posed by each of the three types of attacks as follows.

First, we analyze the post-mining risk posed by the RUM attack. Recalling Theorems \ref{theorem2} and \ref{theorem3} from \ref{The design of rum attack}, we understand that each successful execution of a RUM attack increases the post-mining risk of the blockchain by $1$. We use the notation $Attack\overset{\mathcal{P}_{\mathcal{A}}~~any~t}{\Rightarrow}Deploy$ to represent this situation. That is, every time a RUM adversary successfully executes an attack and triggers the state change from the attack state to the deployment state, it leads to an increment in the blockchain's post-mining risk. Consequently, the increased risk associated with the RUM attack can be quantified as the number of successful attacks by a RUM adversary, denoted by $\sum\left( {Attack\overset{\mathcal{P}_{\mathcal{A}}~~any~t}{\Rightarrow}Deploy} \right)$.

The proof of UUM follows a methodology analogous to that of RUM.

Finally, we analyze the post-mining risk posed by SUUM attack. Recalling Theorems \ref{SUUM Successful Condition} and \ref{theorem7} from \secref{The design of suum attack}, we understand that each successful execution of a SUUM attack also increases the post-mining risk of the blockchain by $1$. This increase in risk is associated with two specific types of state transitions, represented by $Downgrade\overset{\mathcal{P}_{\mathcal{A}}~~any~t}{\Rightarrow}Deploy$ and $Attack~1\overset{\mathcal{P}_{\mathcal{H}}~~any~t}{\Rightarrow}Deploy$. In other words, when an SUUM adversary successfully conducts an attack and makes the state change happen in either of these two ways, it adds to the overall risk related to the blockchain's difficulty. Therefore, the increased risk associated with SUUM attack can be quantified as the number of successful attacks by a SUUM adversary, denoted by $\sum\begin{pmatrix}{Downgrade\overset{\mathcal{P}_{\mathcal{A}}~~any~t}{\Rightarrow}Deploy} \\
{+ Attack~1\overset{\mathcal{P}_{\mathcal{H}}~~any~t}{\Rightarrow}Deploy}
\end{pmatrix}$.

\begin{theorem}[\textbf{Forking Rate Comparison}]
\label{Forking Rate Comparison of Uncle Maker-based attack and honest mining}
    For an adversary $\mathcal{A}$ with the same computational power, the block forking rates induced by adopting RUM, UUM, SUUM, and honest mining satisfy the following relationship: ${\mathbb{F}\mathbb{R}}^{SUUM} > {\mathbb{F}\mathbb{R}}^{UUM} > {\mathbb{F}\mathbb{R}}^{RUM} > {\mathbb{F}\mathbb{R}}^{HM}$.
    Specifically,
\begin{enumerate}[(1)]
\item For honest mining, ${\mathbb{F}\mathbb{R}}^{HM} = 0$.
\item For RUM, ${\mathbb{F}\mathbb{R}}^{RUM} = \mathbb{P}_{Attack}^{RUM} \cdot \mathcal{P}_{\mathcal{A}}(t < 9)$.
\item For UUM, ${\mathbb{F}\mathbb{R}}^{UUM} = \mathbb{P}_{Attack}^{UUM} \cdot \mathcal{P}_{\mathcal{A}}(any~t)$.
\item For SUUM, ${\mathbb{F}\mathbb{R}}^{SUUM} = \mathbb{P}_{Attack}^{SUUM} + \mathbb{P}_{Downgrade}^{SUUM} \cdot \mathcal{P}_{\mathcal{A}}(any~t)$.
\end{enumerate}
\end{theorem}

\section{Proof of Theorem \ref{Forking Rate Comparison of Uncle Maker-based attack and honest mining}}\label{Proof of Theorem 11}
We compare the attack-induced forking rate under honest mining, RUM, UUM, and SUUM.

For honest mining, no deliberate fork is induced. Thus,
\(
{\mathbb{F}\mathbb{R}}^{HM} = 0.
\)

For RUM, a deliberate fork occurs only when the adversary is already in the attack state and mines the next block with \(t<9\). Therefore,
\begin{equation}
{\mathbb{F}\mathbb{R}}^{RUM} =
\mathbb{P}_{Attack}^{RUM}\cdot \mathcal{P}_{\mathcal{A}}(t<9).
\end{equation}

For UUM, the timestamp constraint is relaxed to any valid \(t\). Hence,
\begin{equation}
{\mathbb{F}\mathbb{R}}^{UUM} =
\mathbb{P}_{Attack}^{UUM}\cdot \mathcal{P}_{\mathcal{A}}(any~t).
\end{equation}

For SUUM, forks arise from two sources: occupying the attack state itself, and successful adversarial mining from the downgrade state. Thus,
\begin{align}
{\mathbb{F}\mathbb{R}}^{SUUM}
&= \sum_{i=0}^{n}\mathbb{P}_{Attack~i}^{SUUM}
+\mathbb{P}_{Downgrade}^{SUUM}\cdot \mathcal{P}_{\mathcal{A}}(any~t) \\
&= \mathbb{P}_{Attack}^{SUUM}
+\mathbb{P}_{Downgrade}^{SUUM}\cdot \mathcal{P}_{\mathcal{A}}(any~t).
\end{align}

From the state-transition analyses of RUM, UUM, and SUUM, the steady-state attack-state occupancies satisfy
\begin{equation}
\mathbb{P}_{Attack}^{SUUM} >
\mathbb{P}_{Attack}^{UUM} >
\mathbb{P}_{Attack}^{RUM}.
\end{equation}
This is because UUM relaxes RUM's trigger condition from \(9 \le t < 18\) to \(t \ge 9\), while SUUM further enlarges the set of attack relevant states by introducing withholding and downgrade to attack pathways.
Moreover,
\begin{equation}
1 > \mathcal{P}_{\mathcal{A}}(any~t) > \mathcal{P}_{\mathcal{A}}(t<9).
\end{equation}
Therefore,
\begin{equation}
{\mathbb{F}\mathbb{R}}^{SUUM} >
{\mathbb{F}\mathbb{R}}^{UUM} >
{\mathbb{F}\mathbb{R}}^{RUM} >
{\mathbb{F}\mathbb{R}}^{HM},
\end{equation}
which proves Theorem \ref{Forking Rate Comparison of Uncle Maker-based attack and honest mining}.

\begin{theorem}[\textbf{Pre-mining Risk Comparison}]\label{thm12}
For an adversary $\mathcal{A}$ with relative power $\mu_{\mathcal{A}}$, the pre-mining risk (reductions in block generation rate) of RUM, UUM, and SUUM satisfies:
\begin{enumerate}[(1)]
\item $\mathbb{AC}^{\mathrm{RUM}} = 0,$
\item $\mathbb{AC}^{\mathrm{UUM}} = \mathbb{AC}^{\mathrm{SUUM}}
= \mu_{\mathcal{A}} \cdot \frac{1}{T} \left(1 - \frac{\mathcal{D}_0}{\mathcal{D}_0 + \lfloor \mathcal{D}_0/2048 \rfloor}\right)
\approx \mu_{\mathcal{A}} \cdot \frac{1}{T} \cdot \frac{1}{2048}.$
\end{enumerate}
Therefore, UUM and SUUM incur negligible, but non-zero, additional pre-mining risk under Ethereum 1.x parameters.
\end{theorem}

\section{Proof of Theorem \ref{thm12}}\label{Proof of Theorem 12}
We begin by formalizing the mining probability model in Ethereum 1.x.
Let $T = 13$ seconds be the target average block time.
The probability per hash for a given difficulty $\mathcal{D}$ is $1/\mathcal{D}$, because the target value $\mathit{target}$ satisfies $\mathit{target} = \mathit{max\_target}/\mathcal{D}$ and $\mathit{max\_target}=2^{256}$.
For an adversary with relative hash power $\mu_{\mathcal{A}}$, the hashrate (hashes per second) owned by the adversary is $\mu_{\mathcal{A}} \cdot H$, where $H$ is the total network hashrate.
In steady state, the network hashrate $H$ adjusts so that the expected block generation time equals $T$, i.e., $H \cdot (1/\mathcal{D}) = 1/T$, or $H = \mathcal{D}/T$.

Hence, the adversary's block generation rate (blocks per second) when mining on a chain of difficulty $\mathcal{D}$ is
\begin{align}
    \mathcal P_{\mathcal{A}}(\mathcal{D}) = \mu_{\mathcal{A}} \cdot H \cdot \frac{1}{\mathcal{D}} = \mu_{\mathcal{A}} \cdot \frac{1}{T}.
\end{align}
Notice that this rate is independent of $\mathcal{D}$ as long as the adversary's hashrate $\mu_{\mathcal{A}} H$ scales proportionally to $\mathcal{D}$ (which is the case in steady state). However, during an attack the adversary temporarily mines on a chain whose difficulty differs from the main-chain. Let $\mathcal{D}_0$ be the difficulty of the parent block (common to both chains) and $\mathcal{D}_1$ be the difficulty of the next block on the adversary's private chain after a successful attack. The adversary's hashrate remains $\mu_{\mathcal{A}} H$, but the effective difficulty it faces changes.

Therefore, the adversary's block generation rate on its own chain becomes
\begin{align}
    \mathcal P_{\mathcal{A}}(\mathcal{D}_1) = \mu_{\mathcal{A}} \cdot H \cdot \frac{1}{\mathcal{D}_1} = \mu_{\mathcal{A}} \cdot \frac{\mathcal{D}_0}{T} \cdot \frac{1}{\mathcal{D}_1},
\end{align}
where we used $H = \mathcal{D}_0/T$ (the main‑chain steady‑state relation before the attack). The \emph{pre-mining risk} $\mathbb{AC}$ is defined as the reduction in the adversary's block generation rate caused by the increased difficulty of its own chain, i.e.,
\begin{align}
    \mathbb{AC} = \mathcal P_{\mathcal{A}}(\mathcal{D}_0) - \mathcal P_{\mathcal{A}}(\mathcal{D}_1)
               = \mu_{\mathcal{A}} \cdot \frac{1}{T} \left(1 - \frac{\mathcal{D}_0}{\mathcal{D}_1}\right).
\end{align}

\paragraph{RUM attack.}
For the RUM attack the initiation condition $\bigl\lfloor (t_1^{p_h} - t_0^{p_h})/9 \bigr\rfloor = 1$ guarantees $\mathcal{D}_1 = \mathcal{D}_0$ (see \thmref{theorem4}). Consequently,
\begin{align}
    \mathbb{AC}^{\text{RUM}} = \mu_{\mathcal{A}} \cdot \frac{1}{T} \left(1 - 1\right) = 0.
\end{align}

\begin{table*}[t]
\caption{Additional matched control pools across ETH, ETHW, and ETC}
\label{tab:matched_controls_appendix}
\centering
\setlength{\tabcolsep}{3pt}
\begin{tabular}{@{}|c|c|c|c|c|c|c|@{}}
\toprule
Chain & Pool ID & Obs. Share & $n$ & 0 mod 9 (E/O) & 8 mod 9 (E/O) & Shift exact \\
\midrule
ETH  & \texttt{0x1ad91e...} & 0.10112 & 11{,}243 & \fbox{890.3 / 930} $\mathbf{\uparrow}$ \textbf{39.7} & \fbox{1275.7 / 1014} $\mathbf{\downarrow}$ \textbf{261.7} & 0.00 \\
ETH  & \texttt{0x2daa35...} & 0.03456 & 3{,}843  & \fbox{305.3 / 320} $\mathbf{\uparrow}$ \textbf{14.7} & \fbox{429.6 / 325} $\mathbf{\downarrow}$ \textbf{104.6} & 0.00 \\
ETHW & \texttt{0x00192f...} & 0.20222 & 20{,}222 & \fbox{1616.4 / 1623} $\mathbf{\uparrow}$ \textbf{6.6} & \fbox{1759.3 / 1666} $\mathbf{\downarrow}$ \textbf{93.3} & 0.03 \\
ETHW & \texttt{0x92d003...} & 0.20192 & 20{,}192 & \fbox{1611.3 / 1627} $\mathbf{\uparrow}$ \textbf{15.7} & \fbox{1720.8 / 1748} $\mathbf{\uparrow}$ \textbf{27.2} & 0.35 \\
ETC  & \texttt{0xa6e43e...} & 0.22013 & 22{,}013 & \fbox{1786.3 / 1709} $\mathbf{\downarrow}$ \textbf{77.3} & \fbox{1901.8 / 1903} $\mathbf{\uparrow}$ \textbf{1.2} & 0.73 \\
ETC  & \texttt{0x981217...} & 0.03719 & 3{,}717  & \fbox{295.1 / 289} $\mathbf{\downarrow}$ \textbf{6.1} & \fbox{322.5 / 309} $\mathbf{\downarrow}$ \textbf{13.5} & 0.21 \\
\bottomrule
\end{tabular}
\begin{tablenotes}
\footnotesize
\setlength{\itemindent}{0pt}
\setlength{\rightskip}{-260pt}
\item* \noindent
This table extends the matched-control analysis beyond the representative controls. For each control pool, the expected counts are computed from the same-chain empirical baseline after excluding the identified suspicious pool(s) and the control pool itself. Here, \textbf{E/O} denotes expected/observed counts. For each annotated entry, $\mathbf{\downarrow}$ indicates a deficit relative to expectation, with magnitude \(\text{E}-\text{O}\), while $\mathbf{\uparrow}$ indicates an enrichment relative to expectation, with magnitude \(\text{O}-\text{E}\). The column \textbf{Shift exact} reports the one-sided exact conditional test for enrichment of residue 8 relative to residue 0. None of these additional controls exhibit the staircase-style directional distortion with the magnitude observed in the suspicious pools.
\end{tablenotes}
\end{table*}

\paragraph{UUM and SUUM attacks.}
For UUM and SUUM the initiation condition is $\bigl\lfloor (t_1^{p_h} - t_0^{p_h})/9 \bigr\rfloor \ge 1$, which by Definition~\ref{def:difficulty} implies $\mathcal{D}_1 \ge \mathcal{D}_0 + \bigl\lfloor \mathcal{D}_0/2048 \bigr\rfloor$. Under the minimal‑risk calibration (\thmref{theorem3} and \thmref{theorem7}) the increase is exactly one adjustment unit:
\begin{align}
    \Delta\mathcal{D} = \mathcal{D}_1 - \mathcal{D}_0 = \bigl\lfloor \mathcal{D}_0/2048 \bigr\rfloor.
\end{align}
Thus,
\begin{align}
    \mathbb{AC}^{\text{UUM}} = \mathbb{AC}^{\text{SUUM}}
        &= \mu_{\mathcal{A}} \cdot \frac{1}{T} \left(1 - \frac{\mathcal{D}_0}{\mathcal{D}_0 + \Delta\mathcal{D}}\right) \\
        &= \mu_{\mathcal{A}} \cdot \frac{1}{T} \cdot \frac{\Delta\mathcal{D}}{\mathcal{D}_0 + \Delta\mathcal{D}}.
\end{align}
Since $\Delta\mathcal{D} \ll \mathcal{D}_0$ for any realistic difficulty, we can approximate
\begin{align}
    \mathbb{AC}^{\text{UUM}} = \mathbb{AC}^{\text{SUUM}}
        &\approx \mu_{\mathcal{A}} \cdot \frac{1}{T} \cdot \frac{\Delta\mathcal{D}}{\mathcal{D}_0}
         = \mu_{\mathcal{A}} \cdot \frac{1}{T} \cdot \frac{\bigl\lfloor \mathcal{D}_0/2048 \bigr\rfloor}{\mathcal{D}_0} \\
        &\approx \mu_{\mathcal{A}} \cdot \frac{1}{T} \cdot \frac{1}{2048}.
\end{align}
Substituting $T = 13$ seconds gives
\begin{align}
    \mathbb{AC}^{\text{UUM}} = \mathbb{AC}^{\text{SUUM}} \approx \mu_{\mathcal{A}} \cdot \frac{1}{13 \times 2048}.
\end{align}
For $\mu_{\mathcal{A}} = 0.25$ this yields an absolute pre-mining risk of about $9.4 \times 10^{-6}$ blocks per second, which corresponds to a relative efficiency loss of roughly $1/2048 \approx 0.0488\%$. This loss is negligible compared to the substantial reward gain from a successful attack.



\section{Setup and Implementation Details}\label{Setup and Implementation Details}
\heading{Setup.} We conducted extensive experiments to assess the UUM and SUUM attacks. Since Nakamoto consensus requires extensive computational power, we follow prior works~\cite{Unclemakertimestampingoutthecompetitioninethereum, LargerscaleNakamotostyleBlockchainsDon’tNecessarilyOfferBetterSecurity} and simulate the attacks instead. These simulations aimed to quantify the probabilities of launching the attacks, reward distributions, post-mining risks, and forking rates. For comparison, we also simulate honest mining, where each miner honestly follows the protocol, and the RUM attack.  

\heading{Implementation Details.} We implemented a discrete-event simulator to model the Ethereum 1.x blockchain protocol, incorporating the fork selection rules and difficulty adjustment mechanisms as described in \secref{The design of rum attack}. Our experiments are implemented in Python, building upon the discrete-event simulation framework used in RUM \cite{Unclemakertimestampingoutthecompetitioninethereum}. The code is structured to simulate the Ethereum 1.x blockchain protocol, including fork selection rules and difficulty adjustment mechanisms.

\begin{table*}[t]
\caption{Stability of the staircase score across non-overlapping 10,000-block windows}
\label{tab:window_stability_summary}
\centering
\setlength{\tabcolsep}{4pt}
\begin{tabular}{@{}|c|c|c|c|c|c|c|@{}}
\toprule
Chain & Susp. $>0$ / all & Susp. median $S_w$ & Susp. range & Ctrl. $>0$ / all & Ctrl. median $S_w$ & Ctrl. range \\
\midrule
ETH  & 12 / 12 & 0.179 & [0.140, 0.189] & 9 / 12 & 0.006 & [-0.013, 0.053] \\
ETHW & 10 / 10 & 0.170 & [0.157, 0.176] & 8 / 10 & 0.010 & [-0.017, 0.025] \\
ETC  & 10 / 10 & 0.165 & [0.160, 0.174] & 5 / 10 & 0.002 & [-0.011, 0.025] \\
\bottomrule
\end{tabular}
\begin{tablenotes}
\footnotesize
\setlength{\itemindent}{0pt}
\setlength{\rightskip}{-260pt}
\item* \noindent
Each chain is partitioned into non-overlapping windows of 10,000 consecutive sampled main-chain blocks. For each pool and each window, the staircase score is defined as \(S_w=(N_{8,w}-N_{0,w})/n_w\), where \(N_{0,w}\) and \(N_{8,w}\) are the counts of residues 0 and 8, and \(n_w\) is the number of valid timestamp differences contributed by that pool in the window. Suspicious pools remain consistently positive across all windows, whereas the matched controls stay substantially closer to zero and are sometimes negative.
\end{tablenotes}
\end{table*}

\begin{table*}[t]
\caption{Window-by-window staircase scores over non-overlapping 10,000-block windows}
\label{tab:window_stability_appendix}
\centering
\scriptsize
\setlength{\tabcolsep}{3pt}
\renewcommand{\arraystretch}{0.92}
\resizebox{\textwidth}{!}{%
\begin{tabular}{@{}|c|cccc|cccc|cccc|cccc|cccc|cccc|@{}}
\toprule
\multirow{3}{*}{Win.}
& \multicolumn{8}{c|}{ETH}
& \multicolumn{8}{c|}{ETHW}
& \multicolumn{8}{c|}{ETC} \\
\cmidrule(lr){2-9}\cmidrule(lr){10-17}\cmidrule(lr){18-25}
& \multicolumn{4}{c|}{\texttt{0x829bd8...}}
& \multicolumn{4}{c|}{\texttt{0x00192f...}}
& \multicolumn{4}{c|}{\texttt{0x9205c2...}}
& \multicolumn{4}{c|}{\texttt{0xa524bc...}}
& \multicolumn{4}{c|}{\texttt{0x406177...}}
& \multicolumn{4}{c|}{\texttt{0x5253b3...}} \\
\cmidrule(lr){2-5}\cmidrule(lr){6-9}\cmidrule(lr){10-13}\cmidrule(lr){14-17}\cmidrule(lr){18-21}\cmidrule(lr){22-25}
& $n_w$ & $N_0$ & $N_8$ & $S_w$
& $n_w$ & $N_0$ & $N_8$ & $S_w$
& $n_w$ & $N_0$ & $N_8$ & $S_w$
& $n_w$ & $N_0$ & $N_8$ & $S_w$
& $n_w$ & $N_0$ & $N_8$ & $S_w$
& $n_w$ & $N_0$ & $N_8$ & $S_w$ \\
\midrule
1  & 1527 & 0 & 255 & 0.167 & 752  & 75 & 65 & -0.013 & 3182 & 0 & 501 & 0.157 & 1807 & 165 & 135 & -0.017 & 5769 & 0 & 1001 & 0.174 & 979  & 67 & 91 & 0.025 \\
2  & 1659 & 0 & 314 & 0.189 & 741  & 68 & 70 & 0.003  & 3051 & 0 & 500 & 0.164 & 1836 & 126 & 157 & 0.017  & 5877 & 0 & 968  & 0.165 & 946  & 76 & 97 & 0.022 \\
3  & 1661 & 0 & 301 & 0.181 & 787  & 73 & 69 & -0.005 & 3185 & 0 & 552 & 0.173 & 1550 & 113 & 150 & 0.024  & 5906 & 0 & 1029 & 0.174 & 990  & 86 & 86 & 0.000 \\
4  & 1523 & 0 & 274 & 0.180 & 738  & 67 & 70 & 0.004  & 3180 & 0 & 539 & 0.169 & 1472 & 126 & 143 & 0.012  & 5831 & 0 & 932  & 0.160 & 942  & 82 & 81 & -0.001 \\
5  & 1512 & 0 & 284 & 0.188 & 722  & 51 & 64 & 0.018  & 3136 & 0 & 528 & 0.168 & 1569 & 115 & 118 & 0.002  & 5966 & 0 & 983  & 0.165 & 924  & 82 & 96 & 0.015 \\
6  & 1510 & 0 & 281 & 0.186 & 706  & 62 & 63 & 0.001  & 3162 & 0 & 545 & 0.172 & 1533 & 131 & 143 & 0.008  & 6220 & 0 & 1064 & 0.171 & 1005 & 84 & 88 & 0.004 \\
7  & 1543 & 0 & 237 & 0.154 & 755  & 61 & 59 & -0.003 & 3210 & 0 & 548 & 0.171 & 1460 & 108 & 144 & 0.025  & 5859 & 0 & 944  & 0.161 & 934  & 79 & 69 & -0.011 \\
8  & 1545 & 0 & 278 & 0.180 & 710  & 57 & 68 & 0.015  & 3379 & 0 & 587 & 0.174 & 1489 & 113 & 117 & 0.003  & 5844 & 0 & 958  & 0.164 & 963  & 78 & 72 & -0.006 \\
9  & 1525 & 0 & 214 & 0.140 & 739  & 50 & 59 & 0.012  & 3374 & 0 & 594 & 0.176 & 1529 & 134 & 120 & -0.009 & 6089 & 0 & 1023 & 0.168 & 954  & 75 & 90 & 0.016 \\
10 & 1483 & 0 & 251 & 0.169 & 673  & 55 & 61 & 0.009  & 3378 & 0 & 560 & 0.166 & 1578 & 132 & 155 & 0.015  & 5978 & 0 & 981  & 0.164 & 946  & 84 & 75 & -0.010 \\
11 & 1367 & 0 & 243 & 0.178 & 687  & 56 & 67 & 0.016  & --   & -- & --  & --     & --   & --  & --  & --     & --   & -- & --   & --     & --   & -- & -- & -- \\
12 & 182  & 0 & 32  & 0.176 & 76   & 4  & 8  & 0.053  & --   & -- & --  & --     & --   & --  & --  & --     & --   & -- & --   & --     & --   & -- & -- & -- \\
\bottomrule
\end{tabular}%
}
\begin{tablenotes}
\footnotesize
\setlength{\itemindent}{0pt}
\setlength{\rightskip}{-260pt}
\item* \noindent
Each blockchain is partitioned into non-overlapping windows of 10,000 consecutive sampled main-chain blocks. For each suspicious pool and its matched control pool, we report the number of valid timestamp differences \(n_w\), the counts of residues 0 and 8, and the directional staircase score \(S_w=(N_{8,w}-N_{0,w})/n_w\). The ETHW and ETC datasets each contain 100,000 sampled blocks and therefore yield 10 windows. The ETH dataset contains 111,186 sampled blocks and therefore yields 12 windows in total; entries beyond the available windows of ETHW and ETC are marked as ``--''.
\end{tablenotes}
\end{table*}

The simulator is written in Python, using NumPy for numerical computations and Matplotlib for visualization. Compared to the original simulator for RUM, our simulator implements four different attack strategies: RUM, UUM, SUUM and selfish mining. We also extend the state transition models to support multiple attack states for SUUM, and incorporate real-world blockchain data analysis using Pandas for data processing. We also quantify the post-mining risk and forking rates, as they are crucial parameters for our attacks. 

Our simulator strictly enforces the timestamp validity rules of the Ethereum 1.x protocol: (1) Monotonicity: A block's timestamp $t_i$ must be strictly greater than its parent's timestamp $t_{i-1}$. (2) Bounded future window: A block's timestamp must not exceed the current system time by more than a protocol-defined tolerance. We implemented the Geth client's rule: a block is rejected if its timestamp is more than 15 seconds into the future relative to the node's local clock. In our synchronized simulation, the current time is defined as the maximum timestamp among all received valid blocks, plus any network propagation delay. (3) Lower bound: As per the difficulty adjustment formula (Definition~\ref{def:difficulty}), the timestamp difference $\Delta t = t_i - t_{i-1}$ is effectively constrained to be $\geq 1$. Values of $\Delta t \geq 900$ lead to the maximum difficulty reduction, which the attack's minimal risk control explicitly avoids. The adversary's timestamp pre-selection strategy (Theorems~\ref{theorem3},~\ref{theorem7}) inherently respects these constraints to ensure the validity of its published blocks.

We adopt the same core parameters as \cite{Unclemakertimestampingoutthecompetitioninethereum} to ensure comparability: (1) Adversarial power ratio $\alpha$: Varied from 0 to 0.5 in increments of 0.05. (2) Block generation: Modeled as a Poisson process with the average block interval parameter aligned with the Ethereum 1.x specification. (3) Timestamp constraints: Followed Ethereum 1.x style blockchain validity rules. (4) Difficulty adjustment: Calculated dynamically based on block timestamps and parent difficulties. In all experiments, each block has 1Mb size, and each experiment lasts for 100,000 blocks.

\section{Additional Experimental Results}
\label{Additional Experimental Results}
Beyond the representative matched control pools shown in \tabref{tab:unified_pool_statistics}, \tabref{tab:matched_controls_appendix} reports additional matched controls across ETH, ETHW, and ETC. These additional controls likewise remain close to the same-chain baseline and do not exhibit staircase-style directional distortion comparable to that of the suspicious pools.

We additionally present the stability of the staircase score and the window-by-window staircase scores across non-overlapping 10,000-block windows in \tabref{tab:window_stability_summary} and \tabref{tab:window_stability_appendix}. The results show that the three suspicious pools have positive staircase scores in all 10{,}000-block windows. By contrast, the matched control pools stay much closer to zero. Several control windows are even negative. This suggests that the residue anomaly is not caused by a short-lived event or temporary noise. The results also show a consistent directional pattern. In the suspicious pools, \(N_0\) is repeatedly suppressed and \(N_8\) is repeatedly enriched. This indicates that the anomaly is stable over time and highly consistent with SUUM-style behavior.

\section{Mitigation Measures}\label{Appendix Discussion}

Our Uncle-Maker variants exploit a structural weakness shared by many PoW protocols: block timestamps are partially miner controlled within a validity window, while the difficulty rule reacts to the parent and child time difference. Consequently, defenses should (1) reduce the sensitivity of difficulty to a single timestamp choice and (2) tighten timestamp validity using consensus enforceable rules derived from on-chain history (rather than relying on trusted external time sources).

\subsection{Adjust the Difficulty Adjustment Algorithm}
\label{app:mitigation-daa}
The core vulnerability is that the difficulty update reacts to a single miner controlled time $\Delta t$. A robust difficulty rule should be stable under bounded timestamp manipulation while still tracking long-term changes in hash rate.

\begin{enumerate}
  \item[$\bullet$] \textbf{Clamp the time delta used by the difficulty function.}
  Replace the raw $\Delta t = t_i - t_{i-1}$ by a clamped version
  $\Delta t'=\min\{\Delta_{\max}$, $\max\{\Delta_{\min},\Delta t\}\}$ before it enters the difficulty update.
  This prevents adversaries from inducing extreme per-block difficulty drops/raises by pushing timestamps to the edge of the validity window.

  \item[$\bullet$] \textbf{Use a windowed statistic instead of a single parent delta.}
  Compute the effective time delta from a short history, e.g.,
  $\Delta t' = \mathrm{median}\{t_i-t_{i-1},\, t_{i-1}-t_{i-2},\ldots,t_{i-k+1}-t_{i-k}\}$,
  or use an average over $k$ blocks.
  Windowing reduces the leverage of manipulating any single block and makes post-mining risk control significantly harder.

  \item[$\bullet$] \textbf{Limit the maximum per-block downward difficulty adjustment.}
  Introduce an explicit bound such as $\mathcal{D}_i \ge (1-\gamma)\cdot \mathcal{D}_{i-1}$ for a small $\gamma$.
  Since UUM/SUUM rely on strategically decreasing difficulty on the attacker controlled branch, bounding downward movement directly reduces the achievable advantage, at the risk of slightly slower adaptation under sudden hashrate loss.
\end{enumerate}

\subsection{Strengthen Timestamp Verification}
\label{app:mitigation-ts}
Current PoW clients typically enforce monotonicity ($t_i>t_{i-1}$) and a bounded future window ($t_i \le \mathrm{Now}+\Delta_{\text{future}}$). Uncle-Maker attacks benefit from repeatedly setting timestamps near the boundary where the difficulty rule becomes most exploitable.

\begin{enumerate}
  \item[$\bullet$] \textbf{Adopt a median-time-past (MTP) lower bound.}
  Require $t_i > \mathrm{median}(t_{i-1},t_{i-2},\ldots,t_{i-m})$ for a small $m$.
  This is a consensus enforceable rule that significantly limits an attacker’s ability to pull timestamps backward/forward across multiple consecutive blocks.

  \item[$\bullet$] \textbf{Reduce and make explicit the effective manipulation window.}
  Tighten $\Delta_{\text{future}}$ and combine it with the MTP rule to shrink the feasible region of timestamps that remain valid and useful for manipulating difficulty.
  Importantly, this does not require any trusted external time source; it only constrains timestamps relative to on-chain history.

  \item[$\bullet$] \textbf{Enforce minimum progression relative to recent history.}
  A conservative rule such as $t_i \ge \mathrm{MTP} + \delta$ (for small $\delta$) can further reduce
  the granularity available to timestamp steering, but must be chosen carefully to avoid rejecting honest blocks under benign clock drift.
\end{enumerate}

\subsection{Introduce an Incentive Compatible Response for Uncles}
If the protocol rewards uncles, then deliberately increasing fork/uncle rates can become a profitable strategy. While uncle rewards were originally introduced to compensate for benign propagation delays, parameters and eligibility can be tuned to reduce the marginal benefit of manufacturing uncles.

\begin{enumerate}
  \item[$\bullet$] \textbf{Tighten uncle eligibility conditions.}
  Add conservative consistency checks for uncle inclusion, for example
  requiring the uncle timestamp to lie in a narrow range relative to its ancestry and the including block.
  This discourages strategically timestamped uncles while preserving inclusion of honest uncles caused by network latency.

  \item[$\bullet$] \textbf{Reduce the marginal reward from additional uncles.}
  Cap uncle derived reward per block/epoch, or reduce the incremental reward for multiple uncles.
  The goal is to make extra uncles unattractive in expectation, while still providing partial compensation for benign stale blocks.
\end{enumerate}

\subsection{Improve Monitoring and Response}
Some defenses are best deployed operationally by pools, clients, and monitoring services. Such measures do not change consensus, but can reduce practical impact and enable faster incident response.

\begin{enumerate}
  \item[$\bullet$] \textbf{Transparent anomaly indicators.}
  Track simple, explainable statistics over time: the empirical distribution of $\Delta t=t_i-t_{i-1}$,
  the fraction of blocks near the validity boundaries, and the correlation between boundary hitting $\Delta t$
  patterns and spikes in uncle/fork rates.

  \item[$\bullet$] \textbf{Conservative reaction policies.}
  If anomalies persist, participants can increase confirmation depth for affected periods,
  treat uncle heavy segments with additional caution, and share indicators across the community.
\end{enumerate}

\end{document}